\documentclass[preprint,12pt]{elsarticle} % preprint/review

\usepackage{lineno}
\modulolinenumbers[1]
\usepackage{color}
\usepackage[usenames,dvipsnames]{xcolor}
\usepackage{fullpage}
\usepackage{amsmath}
\usepackage{amssymb}
\usepackage{mathrsfs}
\usepackage{newfloat}
\usepackage{graphicx}
\usepackage{subfig}
\usepackage{multirow}
\usepackage{stackengine}
\usepackage{caption}
\usepackage{tikz}
\usepackage{float}
\usepackage{multicol}
\usetikzlibrary{arrows,positioning,calc}
\usepackage{pstool}
\usepackage{textcomp}
\usepackage{tcolorbox}
\usepackage[hidelinks]{hyperref}
\hypersetup{colorlinks,bookmarksopen,linkcolor=blue,citecolor=blue,urlcolor=blue}

\newcommand*\patchAmsMathEnvironmentForLineno[1]{%
\expandafter\let\csname old#1\expandafter\endcsname\csname #1\endcsname
\expandafter\let\csname oldend#1\expandafter\endcsname\csname end#1\endcsname
\renewenvironment{#1}%
{\linenomath\csname old#1\endcsname}%
{\csname oldend#1\endcsname\endlinenomath}}% 
\newcommand*\patchBothAmsMathEnvironmentsForLineno[1]{%
\patchAmsMathEnvironmentForLineno{#1}%
\patchAmsMathEnvironmentForLineno{#1*}}%
\AtBeginDocument{%
\patchBothAmsMathEnvironmentsForLineno{equation}%
\patchBothAmsMathEnvironmentsForLineno{align}%
\patchBothAmsMathEnvironmentsForLineno{flalign}%
\patchBothAmsMathEnvironmentsForLineno{alignat}%
\patchBothAmsMathEnvironmentsForLineno{gather}%
\patchBothAmsMathEnvironmentsForLineno{multline}%
}
\definecolor{myred}{RGB}{222,45,38}
\definecolor{myblue}{RGB}{0,115,189}
\definecolor{mygreen}{RGB}{49,156,54}

\newcommand{\bs}[1]{{\boldsymbol{#1}}}
\newcommand{\tr}[1]{\mathrm{tr\,{#1}}}

\DeclareFloatingEnvironment{algorithm}
\DeclareFloatingEnvironment{procedure}
\pgfdeclarelayer{bg} % declare background layer
\pgfsetlayers{bg,main} % set the order of the layers (main is the standard layer)
\graphicspath{{figs/}}

\journal{Int. J. Solids Struct.}

\biboptions{authoryear}
\begin{document}

% \pagecolor{lightgray}

\begin{frontmatter}

\title{Integrated Digital Image Correlation for Micro-Mechanical Parameter Identification in Multiscale Experiments\tnoteref{mytitlenote}}
\tnotetext[mytitlenote]{The post-print version of this article is published in \emph{Int. J. Solids Struct.}, \href{https://doi.org/10.1016/j.ijsolstr.2023.112130}{10.1016/j.ijsolstr.2023.112130}. This manuscript version is made available under the \href{https://creativecommons.org/licenses/by/4.0/}{CC-BY 4.0} license.}

%% Group authors per affiliation:
\author[TUe]{O.~Roko\v{s}\corref{correspondingauthor}} % \fnref{myfootnote}
\ead{O.Rokos@tue.nl}

\author[TUe]{R.H.J.~Peerlings}
\ead{R.H.J.Peerlings@tue.nl}

\author[TUe]{J.P.M.~Hoefnagels}
\ead{J.P.M.Hoefnagels@tue.nl}

\author[TUe]{M.G.D.~Geers}
\ead{M.G.D.Geers@tue.nl}

\address[TUe]{Mechanics of Materials, Department of Mechanical Engineering, Eindhoven University of Technology, P.O.~Box~513, 5600~MB~Eindhoven, The~Netherlands}
\cortext[correspondingauthor]{Corresponding author.}
% \fntext[myfootnote]{Since 1880.}

\begin{abstract}
% Some background information
Micromechanical constitutive parameters are important for many engineering materials, typically in microelectronic applications and material design. Their accurate identification poses a three-fold experimental challenge: (i) deformation of the microstructure is observable only at small scales, requiring SEM or other microscopy techniques; (ii) external loadings are applied at a (larger) engineering or device scale; and~(iii)~material parameters typically depend on the applied manufacturing process, necessitating measurements on material produced with the same process.
% The principal activity of the study and its scope
In this paper, micromechanical parameter identification in heterogeneous solids is addressed through multiscale experiments combined with Integrated Digital Image Correlation (IDIC) in conjunction with various possible computational homogenization schemes.
% Some information about the methods used in the study
To this end, some basic concepts underlying multiscale approaches available in the literature are first reviewed, discussing their respective advantages and disadvantages from the computational as well as experimental point of view. A link is made with recently introduced uncoupled methods, which allow for identification of material parameter ratios at the microscale, still lacking a proper normalization. Two multiscale methods are analysed, allowing to bridge the gap between microstructural kinematics and macroscopically measured forces, providing the required normalization. It is shown that an integrated experimental--computational scheme provides relaxed requirements on scale separation.
% The most important results of the study
The accuracy and performance of the discussed techniques are analysed by means of virtual experimentation under plane strain and large strain assumptions for unidirectional fibre-reinforced composites. The robustness against image noise is also assessed.
% A statement of conclusion or recommendation
The obtained results demonstrate that the expected accuracy is typically within~$10\,\%$ RMS error for all multiscale methods, but decreasing to~$1\,\%$ RMS error for the optimal method.
\end{abstract}


\end{frontmatter}

% \linenumbers

%
%-----------------------------------------------------------------------------
%	SECTION Introduction
%-----------------------------------------------------------------------------
%
\section{Introduction}
\label{Sect:Introduction}
In many engineering applications, accurate knowledge of micro-mechanical parameters is of high relevance, especially for predicting the mechanical response and the lifespan of advanced multi-phase materials or micro-electro-mechanical devices. Through micro-structural numerical simulations, these material parameters furthermore help to better understand individual mechanical processes in solids occurring across the scales such as damage, ductile fracture, or crack initiation and propagation, and thus contribute to the design of new materials, see \cite{Hoc:2003}, \cite{Rupil:2011}, \cite{Blaysat:2015}, \cite{Garoz:2017}, or \cite{Buljac:2017}. 

Multiple aspects complicate the accurate identification of individual micro-structural constituents, such as: 
(i) inherently small dimensions, complicating manufacturing of microscopic specimens for micro-tensile tests and subsequent measurements;
(ii) in numerous situations, such as 3D printing, material parameters depend on the manufacturing process and spatial position in the structure (i.e., final dimensions, shape, and morphology of the entire specimen are relevant and cannot be neglected);
and~(iii) external loadings are applied at a larger engineering or device scale, the effect of which needs to be assessed at the microscale. The last point~(iii) requires a multi-scale setting, in which the specimen length scale~$L$ is significantly larger compared to the length scale of the typical micro-structural features~$\ell$, i.e., the scale ratio~$L/\ell \gg 1$. When mechanical properties of the micro-structural constituents are of interest, their deformation needs to be captured with a sufficiently high resolution using optical or scanning electron microscopy, which implies inability to observe mechanical behaviour of the entire specimen at the same time due to limitations in the Field Of View~(FOV). On the other hand, when deformation of the entire specimen is captured, microstructural features are not resolved with a sufficiently high accuracy due to limitations on image resolution.

One of the options available for non-destructive in-situ testing of heterogeneous microstructures is nano-indentation, cf.~\cite{Frohlich:1977} or~\cite{Cross:2005}. However, inaccuracies stemming from sample surface preparation, surface roughness, precise indenter shape, unknown contact and friction behaviour, and unresolved details on nano-plasticity, etc., significantly affect the resulting accuracy. Typically, this leads to systematic overestimation of elastic stiffness properties, as discussed by~\cite{Hardiman:2017}. As an alternative, direct measurements on micro-mechanical specimens using micro-tensile testing machines can be used, as reported in~\citep{Son:2005,Du:2017}. Such an approach brings, nevertheless, significant challenges in conducting a well-defined uniaxial micro-tensile test~\citep{Saif:I:2010,Saif:II:2010,Du:2018}, and may result in significant changes in mechanical properties due to specimen preparation~\citep{Greer:2011}. Such shortcomings can be circumvented by considering larger specimens, as discussed by~\cite{Heripre:2007} or~\cite{Bertin:2016}, where material properties in polycrystals have been identified within larger samples. Yet, such an approach requires bridging the gap between the microstructure and macroscopically applied loads. Such bridging introduces normalization of the micromechanical constitutive parameters, since when deformation of the microstructure is observed, only material parameter ratios between individual constituents can be identified. Knowledge of acting forces then allows for their normalization to proper levels of magnitude. This is equivalent to a mechanical system subjected solely to Dirichlet boundary conditions, for which its deformation state remains the same as long as the ratios of the material parameters do not change (although the magnitude of the reaction forces changes significantly).

This can be achieved, for instance, through an estimated underlying stress field, as discussed by~\cite{Schmidt:2012}, or by minimizing differences between macroscopically measured and microstructurally estimated homogenized stiffness properties, e.g., \cite{Matzenmiller:2005} or~\cite{Blaheta:2015}. Typically, however, micro-mechanical parameter identification still relies on macroscopic measurements (of displacements and/or forces), that need to be linked to micro-mechanical properties using, e.g., computational homogenization~(FE\textsuperscript{2}). This has been theoretically studied by~\cite{Burczynski:2009}, \cite{Klinge:2012,Klinge:2012a}, \cite{Steinmann:2015}, \cite{Schmidt:2015,Schmidt:2016}, or~\cite{Beluch:2017}. Typically, virtual tests were employed, generating macroscopic measurements by forward evaluation of the multiscale model, assuming validity of (i) the multiscale model (i.e., the homogenization technique) used, and (ii) separation of scales. Problems with uniqueness of micro-mechanical parameters leading to the same homogenized macroscopic response is often mentioned. To remedy this, multiple tests have to be performed or more robust optimization techniques have to be used, seeking the global minimum, cf.~\cite{Zohdi:2003}, \cite{Chaparro:2008}, or~\cite{Unger:2017}. This issue becomes more relevant for a larger number of microstructural constituents to be identified at the microscale, as upper bounds for the maximum number of micromechanical parameters that can be uniquely identified from macroscopic experimental observations exist, see~\cite{Schmidt:2016}. Identification of micromechanical parameters within the multiscale regime has been experimentally performed by~\cite{Fedele:2006}, where differences between the experimentally observed homogenized stress field and predictions from numerical simulations carried out on a representative volume were minimized, assuming certain micromechanical parameters to be known a priori. Finally, parameter identification within the probabilistic realm, e.g., \cite{Rappel2020}, \cite{Fish:2016}, \cite{Fish:2008}, and~\cite{Rosic:2013}, and machine learning realm, e.g., \cite{Flaschel2021,Flaschel2022}, \cite{Joshi2022}, have been discussed as well.

The starting point in this contribution is Digital Image Correlation~(DIC), which is a well-established non-intrusive full-field measurement technique. It is especially suitable for micromechanical identification because it can be readily used in combination with optical microscopy, or with some precautions also with Scanning Electron Microscopy~(SEM), cf.~\cite{Sutton:2007} and~\cite{Maraghechi:2018,Maraghechi:2019}. Its integrated variant, called Integrated DIC~(IDIC), uses prior information on the experimentally measured specimen, typically introduced through an underlying mechanical model, making the method even more versatile. The prior information comprises constitutive laws, morphology of the microstructure, and applied boundary conditions. In effect, this allows for direct identification of all relevant material parameters of the model, as well as high overall accuracy and robustness with respect to measurement uncertainty, as proven in many publications, cf., e.g., \cite{Roux:2006}, \cite{Leclerc:2009}, \cite{Rethore:2009}, \cite{Rethore:2013}, \cite{Neggers:2015}, \cite{Blaysat:2015}, or~\cite{Andre:FEMU}. Other approaches to inverse identification methods can be found in, e.g., \cite{Avril:2008}, or~\cite{Cameron2021} by direct stress identification.

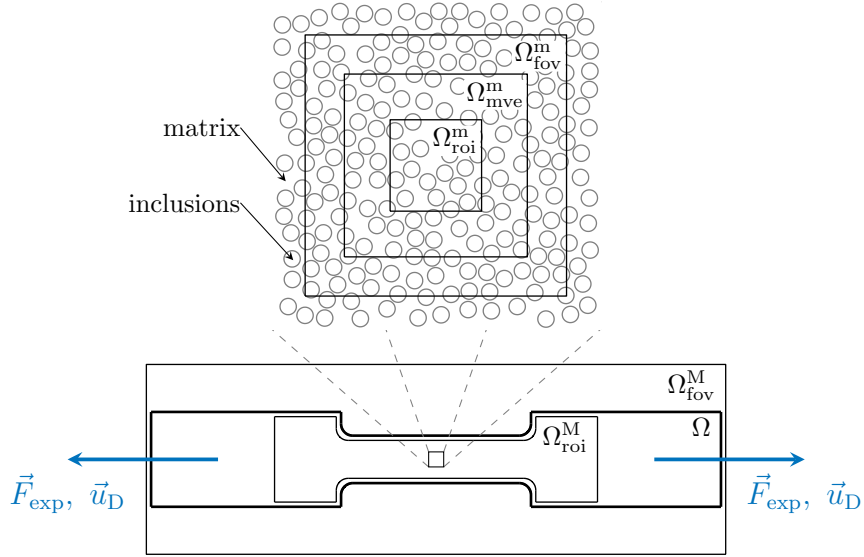
\begin{figure}[t]
	\centering
	% \psfragfig{matfrag/sketchMVE}
	% \psfragfig{matfrag/sketchSpecimen}
	\scalebox{1.0}{
	\begin{tikzpicture}[>=stealth]
		\linespread{1}
		\tikzset{
			mynode/.style={inner sep=0,outer sep=0},
			myarrow/.style={myblue,thick},
		}	
		
		% Specimen and mve
		\begin{pgfonlayer}{bg}
		\node[mynode] (specimen) {
			\includegraphics[scale=1]{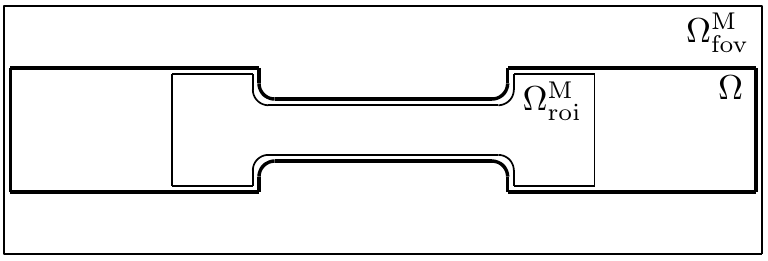}
		};
		\end{pgfonlayer}
		\node[mynode,above=1.0em of specimen] (mve) {
			\includegraphics[scale=1]{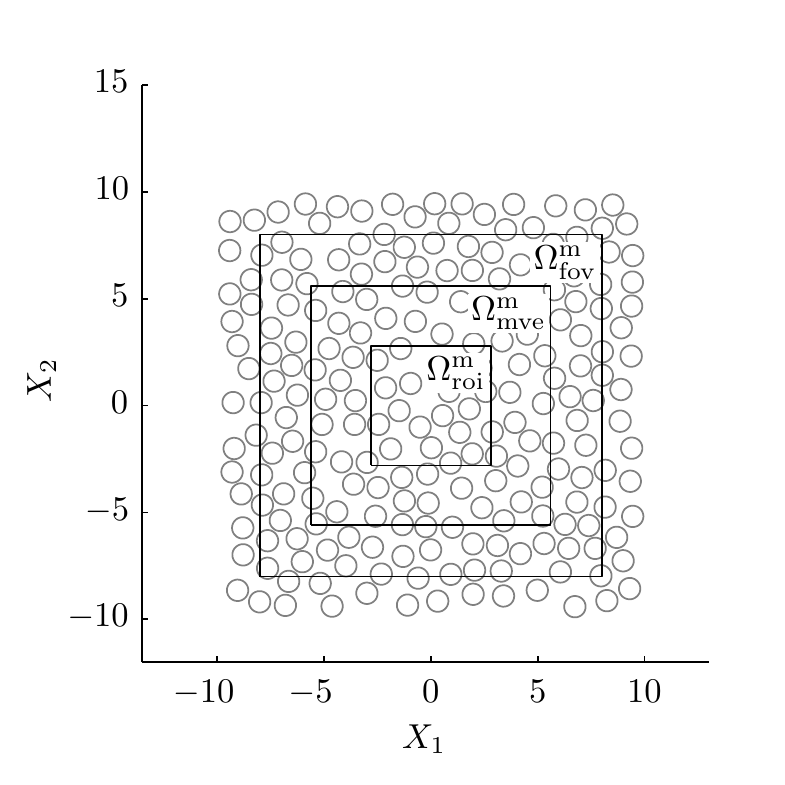}
		};
		\draw (-0.1,-0.1) rectangle (0.1,0.1);
		
		% Forces
		\coordinate []  (Fa) at (-2.9,0);
		\coordinate []  (Fb) at (-4.9,0);
		\node[below=0.1em of Fb,anchor=north] {$\color{myblue}{\vec{F}_\mathrm{exp},\ \vec{u}_\mathrm{D}}$};
		\coordinate []  (Fc) at (2.9,0);
		\coordinate []  (Fd) at (4.9,0);
		\node[below=0.1em of Fd,anchor=north] {$\color{myblue}{\vec{F}_\mathrm{exp},\ \vec{u}_\mathrm{D}}$};
		\draw[->,myblue,line width=0.5mm] (Fa) to (Fb);
		\draw[->,myblue,line width=0.5mm] (Fc) to (Fd);
		
		% Legend
		\node[mynode,left=1.0em of mve, shift={(0.0,0.5)}] (matrix) {\footnotesize matrix~};
		\node[mynode,left=1.0em of mve, shift={(0.0,-0.5)}] (inclusions) {\footnotesize inclusions~};
		\draw[->] (matrix.east) to (-2.0,3.7);
		\draw[->] (inclusions.east) to (-1.9,2.65);
		
		% Zooming lines in the background layer
		\begin{pgfonlayer}{bg}
		\draw[gray,dashed] (0.1,-0.1) to (mve.south east);
		\draw[gray,dashed] (-0.1,-0.1) to (mve.south west);
		\draw[gray,dashed] (0.1,0.1) to (mve.north east);
		\draw[gray,dashed] (-0.1,0.1) to (mve.north west);
		\end{pgfonlayer}
		
		% Plot auxiliary coordinate grid
		% \draw[help lines,step=0.2] (-5,-2) grid (5,6);
		% \draw[help lines,line width=.6pt,step=1] (-5,-2) grid (5,6);
		% \foreach \x in {-5,-4,-3,-2,-1,0,1}
		% \node[anchor=north] at (\x,-2) {\x};
		% \foreach \y in {-2,-1,0,1,2}
		% \node[anchor=east] at (-3,\y) {\y};
		
	\end{tikzpicture}}
	\caption{Sketch of a multiscale experimental set-up in which a macroscopic specimen is subjected to a uni-axial tensile test. Microscopic observations are carried out using a microstructural field of view, $\Omega_\mathrm{fov}^\mathrm{m}$, an underlying mechanical model inside a micro-structural volume element, $\Omega_\mathrm{mve}^\mathrm{m}$, and a microscopic region of interest where the experimental and numerical fields are compared, $\Omega_\mathrm{roi}^\mathrm{m}$. At the macroscale, the deformation field of the entire specimen~$\Omega$, applied boundary conditions, and measured reaction forces are observed through a macroscopic field of view, $\Omega_\mathrm{fov}^\mathrm{M}$, and a macroscopic region of interest, $\Omega_\mathrm{roi}^\mathrm{M}$.}
	\label{Sect:Introduction:Fig1}
\end{figure}

The integration of DIC with a numerical model introduces, nevertheless, several complications because not all of the data required are readily available, or the data is available at different scales. As an example, consider a macroscopic specimen subjected to a uni-axial displacement-controlled tensile test with measured reaction forces, as shown in Fig.~\ref{Sect:Introduction:Fig1}. A microscope used for the observation of microstructural features has a limited FOV, capturing only a small portion of the entire specimen with sufficient resolution and accuracy (note that scanning of the entire specimen at each deformation increment, although theoretically possible, is practically unfeasible). As a consequence, applied boundary conditions (prescribed displacements and corresponding reaction forces) are observable far outside of the FOV, even at a different scale (i.e., at the scale of the entire specimen). As the underlying mechanical model requires combined information from both scales, the gap between the two scales needs to be bridged by a proper multiscale method.

\begin{figure}
	\centering
	\subfloat[Direct Numerical Simulation (DNS), Section~\ref{Sect:DNS}]{\includegraphics[scale=0.7]{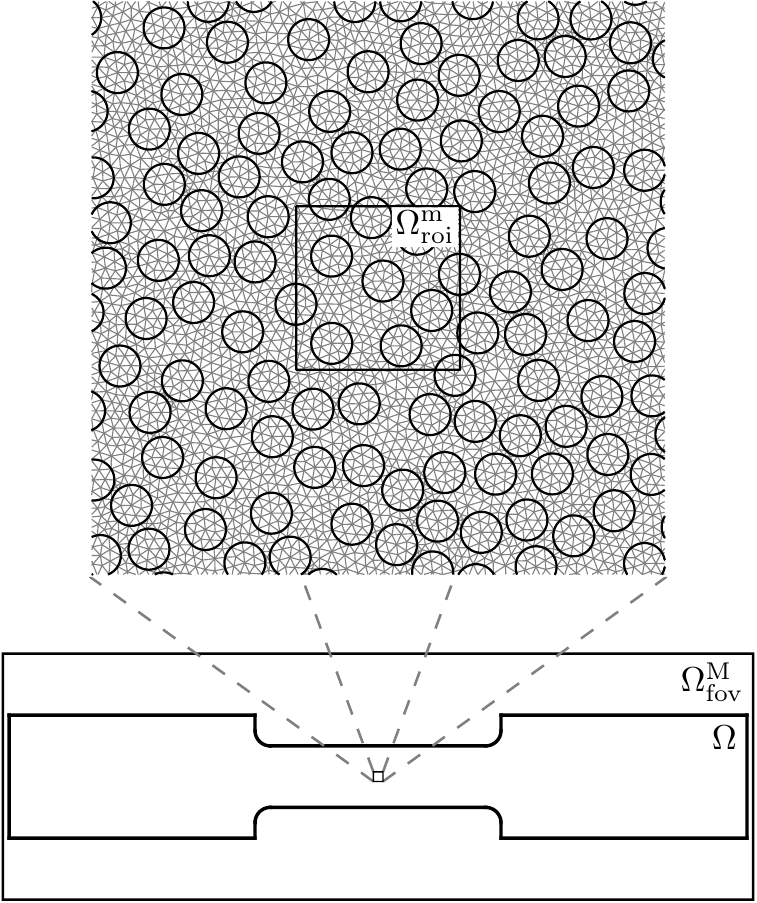}\label{Fig:methods_a}}\hspace{0.0em}
	\subfloat[Concurrent Multiscale Modelling (CMM), Section~\ref{Sect:CMM}]{\includegraphics[scale=0.7]{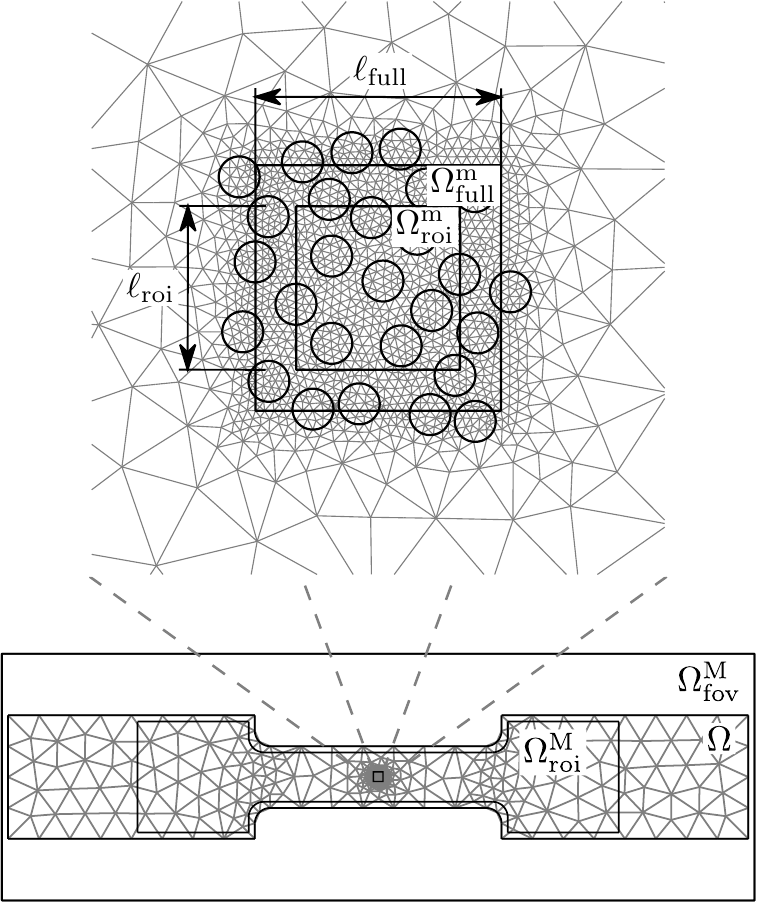}\label{Fig:methods_b}}\hspace{0.0em}
	\subfloat[Bridging Scale Method (BSM), Section~\ref{Sect:BSM}]{\includegraphics[scale=0.7]{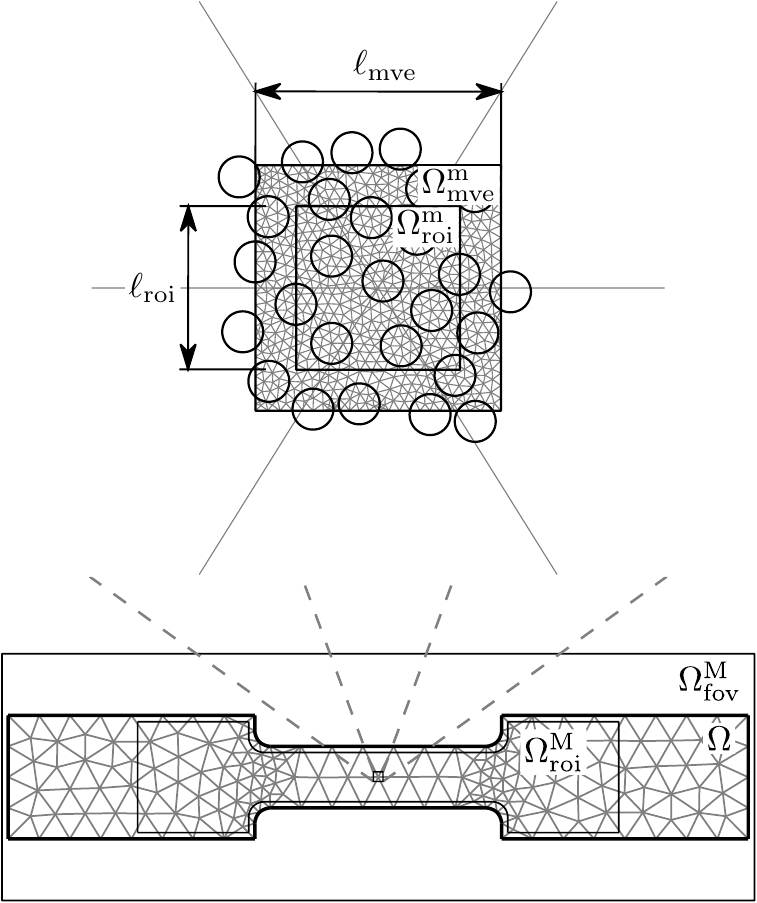}\label{Fig:methods_c}}\\\vspace{2em}
	\subfloat[Stress Integration (SI), Section~\ref{Sect:SI}]{\includegraphics[scale=0.72]{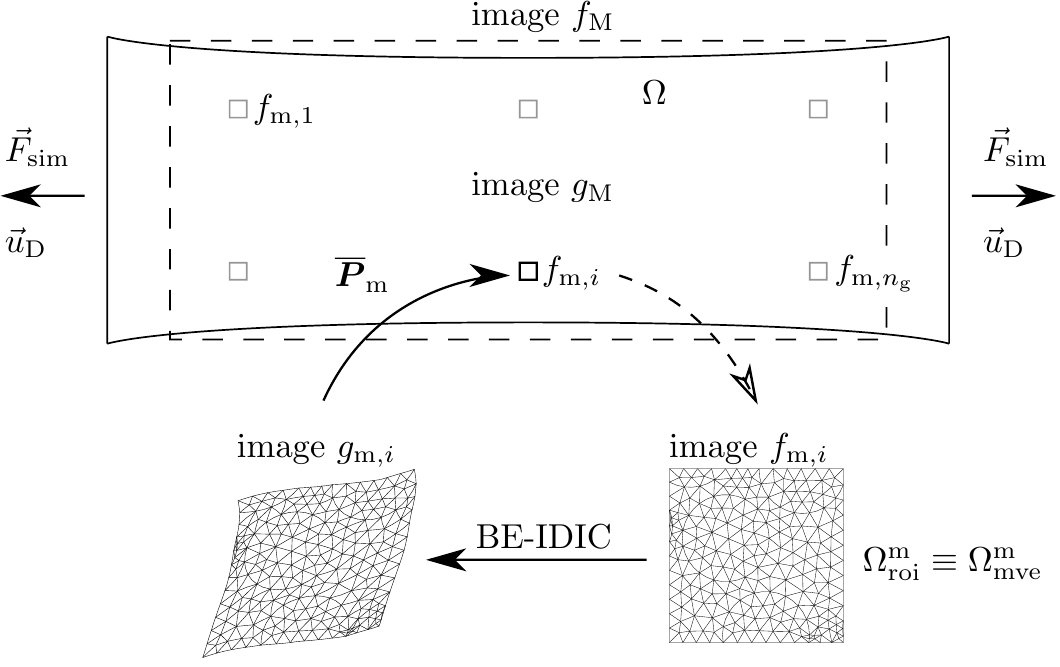}\label{Fig:methods_d}}\hfill			
	\subfloat[Computational homogenization (FE\textsuperscript{2}), Section~\ref{Sect:SIFE2}]{\includegraphics[scale=0.72]{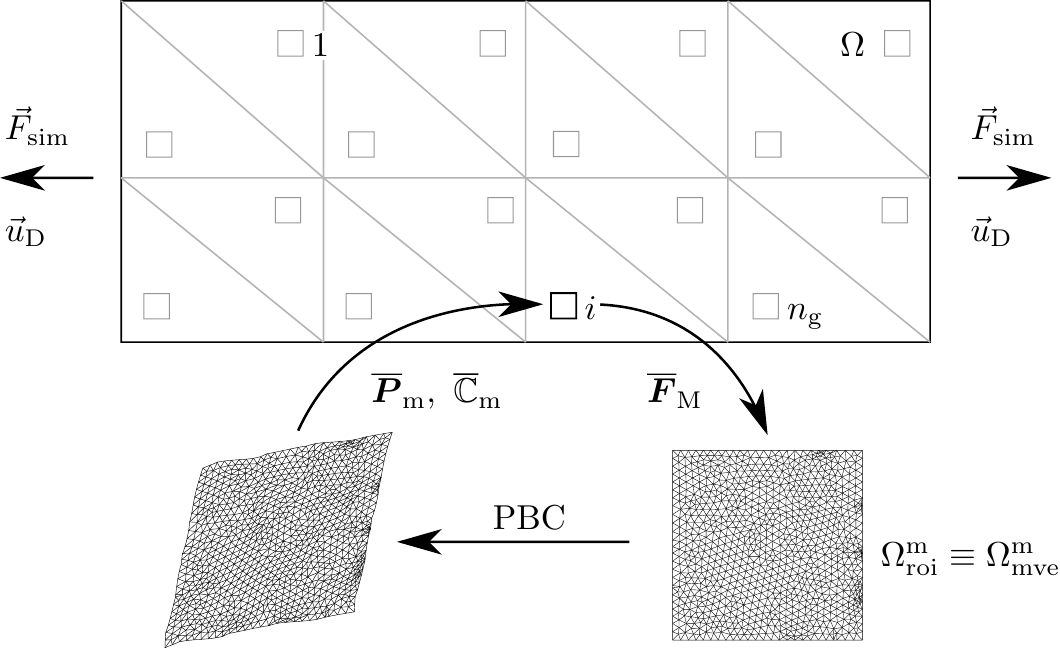}\label{Fig:methods_e}}
	\caption{An overview of the different methodologies considered in this contribution for micromechanical parameter identification. A more detailed description on each methodology and its associated accuracy can be found in the individual sections (listed in the sub-captions).}
	\label{Fig:methods}
\end{figure}

The aim of this paper is to systematically address these challenges for micromechanical parameter identification. To this purpose, experimental measurements will be carried out at two scales, employing IDIC, thereby assuming \emph{all} micromechanical parameters as unknown. An overview of several methods available in the literature is provided in the first step. The basic ideas behind these approaches are briefly described, and their respective advantages and disadvantages are discussed from an experimental as well as a computational perspective. In the second step, two methodologies are proposed to provide the missing normalization of identified material parameter ratios at the microscale by means of uncoupled methods. To a large extent, uncoupled methods can eliminate problems with uniqueness in the identification of mechanical parameters through residual images (indicating global minimum), and reduce effects for a small scale separation by sampling true boundary conditions. However, uncoupled methods only yield material parameter ratios requiring, therefore, a suitable normalization. This is achieved in a first novel method proposed herein by direct stress integration over a number of experimentally observed Microstructural Volume Elements~(MVE), relying on the principle of virtual work. This methodology results in an integrated experimental--computational scheme.  Although satisfactory accuracy is reached, substantial experimental effort is still required. A second novel method is therefore proposed and elaborated, which reduces experimental efforts at the expense of computational resources by making use of FE\textsuperscript{2}. All tested IDIC methodologies are summarized for convenience in Fig.~\ref{Fig:methods}.

The remainder of this paper is divided into five parts. After the introduction, the first part describes the basics of DIC (Section~\ref{Sect:Methods}). The reference virtual experiment, used as a test example for comparison of individual methods, is introduced in the second part (Section~\ref{Sect:VE}). The existing multiscale approaches along with the two newly proposed methodologies based on stress integration and computational homogenization are elaborated in the third part (Sections \ref{Sect:DNS}--\ref{Sect:NM}). In each section of the third part, the theory underlying each method is first briefly outlined, followed by an analysis of its results (such as convergence, typical accuracy, or performance). Mutual comparison and robustness with respect to image noise is summarized in the fourth part (Section~\ref{Sect:comparison}). The paper finally closes with a summary and conclusions in the fifth part (Section~\ref{Sect:Conclusion}).

Throughout this paper, the following notation is adopted, using Cartesian tensors:
\begin{multicols}{2}
\begin{itemize}[\textbf{-}]
\itemsep0em
\item scalars~$ a $,
\item vectors~$ \vec{a} $,
\item second-order tensors~$ \bs{A} $,
\item fourth-order tensors~$ \mathbb{C} $,
\item matrices~$ \bs{\mathsf{A}} $ and column matrices~$ \underline{a} $,
\item $ \vec{a} \cdot \vec{b} = a_i b_i $,
\item $ \bs{A} \cdot \vec{b} =  A_{ij} b_j\vec{e}_i $,
\item $\bs{A}\cdot\bs{B} = A_{ik}B_{kj}\vec{e}_i\vec{e}_j $,
\item $\bs{A}:\bs{B} = A_{ij}B_{ji}$,
\item conjugate~$ \bs{A}^c = \bs{A}^\mathsf{T} $, $ A_{ij}^\mathsf{T} = A_{ji} $,
\item gradient operator~$ \displaystyle \vec{\nabla} \vec{a} = \frac{\partial a_j}{\partial X_i} \vec{e}_i \vec{e}_j $,
\item divergence operator~$ \displaystyle \vec{\nabla} \cdot \vec{a} = \frac{\partial a_i}{\partial X_i} $,
\item derivatives of scalar functions with respect to second-order tensors \\
$\displaystyle \delta\Psi(\bs{F};\delta\bs{F}) = \left.\frac{\mathrm{d}}{\mathrm{d}h}\Psi(\bs{F}+h\delta\bs{F})\right|_{h=0} = \frac{\partial\Psi(\bs{F})}{\partial\bs{F}}:\delta\bs{F} $.
\end{itemize}
\end{multicols}
%
%-----------------------------------------------------------------------------
%	SECTION Methods
%-----------------------------------------------------------------------------
%
\section{Digital Image Correlation}
\label{Sect:Methods}
The guiding method underpinning micromechanical parameter identification is Digital Image Correlation~(DIC). DIC follows the usual scheme of inverse problems, i.e., it minimizes the least squares difference 
\begin{equation}
\underline{\lambda}^\star \in \underset{\underline{\lambda}\in\mathbb{R}^{n_\lambda}}{\text{arg min}}\ 
\frac{1}{2} \| f(\vec{X}) - g(\vec{X}+\vec{u}(\vec{X},\underline{\lambda})) \| ^2_{\Omega_\mathrm{roi}}
\label{SubSect:DIC:Eq1}
\end{equation}
between observation and model prediction. In Eq.~\eqref{SubSect:DIC:Eq1}, $\|\bullet\|_{\Omega_\mathrm{roi}}$ denotes the $L^2$-norm defined over a Region Of Interest~(ROI), $\Omega_\mathrm{roi} \subset \mathbb{R}^2$, while~$\vec{X} = X_1\vec{e}_1 + X_2\vec{e}_2 $ is the position vector associated with the reference configuration. $f$ and~$g$ are observation data, more precisely grey scale images capturing undeformed and deformed configurations of the experimentally tested specimen. The image~$g$ is mapped through a displacement field~$\vec{u}(\vec{X},\underline{\lambda}) = u_1(\vec{X},\underline{\lambda})\vec{e}_1 + u_2(\vec{X},\underline{\lambda})\vec{e}_2$, which is a function of a set of admissible model parameters, stored in a column~$\underline{\lambda} = [\lambda_1, \dots, \lambda_{n_\lambda}]^\mathsf{T} \in \mathbb{R}^{n_\lambda}$. Through this mapping, a back-deformed image is constructed (through the model prediction), which is compared against the reference image~$f$.

Choosing the field~$\vec{u}$ as a linear combination of user-selected basis functions~$ \vec{\psi}_i(\vec{X}) \in \mathbb{R}^2 $, $i = 1, \dots, n_\lambda$, i.e.,
\begin{equation}
\vec{u}(\vec{X},\underline{\lambda}) = \sum_{i = 1}^{n_\lambda}\vec{\psi}_i(\vec{X})\lambda_i, \quad \vec{X} \in \Omega_\mathrm{roi},
\label{SubSect:DIC:Eq2}
\end{equation}
results in a so-called Global DIC scheme. In order to avoid ill-posedness or local minima, the supports of the interpolation functions, $\vec{\psi}_i$, need to span multiple pixels \citep[cf., e.g.,][]{HORN:1981}, be regularized through an additional mechanical system and equilibrium gap method \cite[see, e.g.,][]{tomicevic:2013} or by Tikhonov regularization \citep{Leclerc2011,Yang2014}. Note that the definition of GDIC adopted herein is not unique, since some integrated approaches fulfil the definition of Eq.~\eqref{SubSect:DIC:Eq2} as well, cf., e.g., \cite[Section~5]{Roux:2006}.

On the contrary, in Integrated DIC, $\vec{u}$ is based on the solution of an underlying mechanical model of the boundary value problem at hand, expressed in abstract form as
\begin{equation}
\vec{\mathcal{M}}(\vec{u}(\vec{X},\underline{\lambda}),\underline{\lambda},\vec{X}) = \vec{0}, \quad \vec{X} \in \Omega_\mathcal{M}.
\label{SubSect:DIC:Eq3}
\end{equation}
Obtaining~$\vec{u}$ from Eq.~\eqref{SubSect:DIC:Eq3} usually entails the solution of an associated (system of) partial differential equation(s), carried out typically by means of the Finite Element~(FE) method. The model domain~$\Omega_\mathcal{M}$ differs for the individual methods shown in Fig.~\ref{Fig:methods}, spanning either an inner region of the specimen domain~$\Omega$ or the entire specimen. Dependence on the time-resolved nature of the problem has been omitted for simplicity in Eqs.~\eqref{SubSect:DIC:Eq1}--\eqref{SubSect:DIC:Eq3}, but can easily be taken into account~\citep{Neggers:2015}.

The mechanical solution regularizes the minimization problem of Eq.~\eqref{SubSect:DIC:Eq1} and delivers accurate parameters when an appropriate model~$\vec{\mathcal{M}}$ with no model error is used. As hinted to in both Eqs.~\eqref{SubSect:DIC:Eq2} and~\eqref{SubSect:DIC:Eq3}, $\underline{\lambda}$ may consist of kinematic Degrees Of Freedom~(DOFs), such as kinematic boundary conditions~\citep{Rokos:2018}, material parameters, or topological properties associated with the underlying system. Accuracy of IDIC can further be improved by including the force measurement in the optimization scheme, i.e., by considering
\begin{equation}
\frac{1}{2} \| \vec{F}_\mathrm{exp} - \vec{F}_\mathrm{sim}(\underline{\lambda}) \|^2_\mathrm{E}
\label{SubSect:DIC:Eq4}
\end{equation}
as an additional objective function. The expression in Eq.~\eqref{SubSect:DIC:Eq4} measures the Euclidean norm $ \| \bullet \|_\mathrm{E} $ of the difference between a vector of experimentally observed reaction forces~$\vec{F}_\mathrm{exp}$ and forces predicted by the underlying mechanical model~$\vec{F}_\mathrm{sim}(\underline{\lambda})$. Formally, multi-objective optimization technique should be used for the two objectives, which is nevertheless typically avoided by using weighted sum scalarization, i.e., minimizing a weighted sum of the two objectives (see, e.g., \citealt{Deb:2001} and~\citealt{Neggers:2015}). In this contribution, a slightly different approach will be adopted, as detailed below in Section~\ref{SubSect:DNS}.

The solution of the minimization problem of Eq.~\eqref{SubSect:DIC:Eq1} is usually obtained by a Gauss--Newton algorithm, convergence of which to the global minimum can be readily verified by means of the image residual~$|f(\vec{X})-g(\vec{X}+\vec{u}(\vec{X},\underline{\lambda}))|$, where~$|\bullet|$ denotes the absolute value of a function~$\bullet$. For further details on minimization problem of Eq.~\eqref{SubSect:DIC:Eq1} and its consistent linearisation see~\cite{Neggers:NME:2016}. Solution of the underlying mechanical system in Eq.~\eqref{SubSect:DIC:Eq3} follows standard procedures, discussed in, e.g., \cite{zienkiewicz:Vol2}, \cite{Crisfield:vol1}, or~\cite{Jirasek:2002}.

A two-step Finite Element Model Updating~(FEMU) procedure can be used as an alternative to the presented (one-step) IDIC approach. In FEMU, first the displacement fields are measured through standard DIC, typically Local DIC, which are subsequently used in material parameter identification, carried out in the second step, cf., e.g., \cite{Avril:2008}. As shown by~\cite{Andre:FEMU}, IDIC outperforms FEMU in terms of accuracy and robustness with respect to image noise (especially when displacements are small and/or sensitivity to material parameters is low). Only IDIC is therefore elaborated further in this work.
%
%-----------------------------------------------------------------------------
%	SECTION Reference Model
%-----------------------------------------------------------------------------
%
\section{Reference Model}
\label{Sect:VE}
The performance and accuracy of the different methodologies will be tested by means of virtual experiments, i.e., observations are generated through Direct Numerical Simulations~(DNS) of the full sample with a finely discretized microstructure. To this end, first the underlying model~$\vec{\mathcal{M}}$ of Eq.~\eqref{SubSect:DIC:Eq3}, governing the mechanics of the system, is specified. A standard hyper-elastic mechanical model in absence of body forces, with equilibrium equation
\begin{equation}
\vec{\nabla} \cdot \bs{P}^\mathsf{T}(\vec{X}) = \vec{0}, \quad \vec{X} \in \Omega,
\label{Sect:VE:Eq1}
\end{equation}
under plane strain and large deformation assumptions is adopted. The assumption on plane strain is used to remove any inaccuracies stemming from out-of-plane motion, i.e., subsurface and 3D effects, since our primary objective in this contribution lies in multiscale identification of micromechanical parameters. Although 3D effects are crucial for accurate identification, they are difficult to capture, and are capable of introducing large errors when not treated properly. Methodologies accounting for out-of-plane motion effects should be considered in such cases, including digital height correlation \citep{Kleinendorst:2016,UZUN:2019}, digital volume correlation \citep{Leclerc2011}, and full 3D models. In Eq.~\eqref{Sect:VE:Eq1},
\begin{equation}
\bs{P}(\vec{X}) = \frac{\partial W(\vec{X},\bs{F}(\vec{u}(\vec{X})))}{\partial \bs{F}^\mathsf{T}}, \quad \vec{X} \in \Omega,
\label{Sect:VE:Eq2}
\end{equation}
is the first Piola--Kirchhoff stress tensor, $\vec{\nabla}$ denotes the gradient with respect to the reference configuration, $\bs{F}$ is the deformation gradient, and~$\vec{u}(\vec{X})$ the displacement field. The elastic energy density~$W$ is decomposed as
\begin{equation}
W(\vec{X},\bs{F}(\vec{u}(\vec{X}))) = [1-\chi(\vec{X})]W_1(\bs{F}(\vec{u}(\vec{X}))) + \chi(\vec{X})W_2(\bs{F}(\vec{u}(\vec{X})))
\label{Sect:VE:Eq3}
\end{equation}
according to phases~$\beta = 1,\, 2$, which are geometrically characterized by a phase indicator function of the inclusions
\begin{equation}
\chi(\vec{X}) = \left\{
\begin{array}{ll}
1, & \mbox{if~$\vec{X}$ is inside an inclusion}, \\
0, & \mbox{otherwise}.
\end{array}
\right.
\label{eq:indicator}	
\end{equation}
Individual energy densities are further assumed to follow a compressible hyper-elastic Neo--Hookean constitutive law
\begin{equation}
W_\beta(\bs{F}) = \frac{1}{2}G_\beta(\overline{I}_1-3)+\frac{1}{2}K_\beta(\ln(J))^2,
\label{Sect:VE:Eq4}
\end{equation}
which corresponds to the matrix (for~$\beta = 1$) and inclusions (for~$\beta = 2$). In Eq.~\eqref{Sect:VE:Eq4}, $J = \det{\bs{F}}$, and~$\overline{I}_1 = J^{-2/3}\,\tr(\bs{C})$ is the first isochoric invariant of the right Cauchy--Green deformation tensor~$\bs{C} = \bs{F}^\mathsf{T} \cdot \bs{F}$. For the micro-mechanical parameter identification, the unknown constants are the matrix shear and bulk moduli~$G_1$ and~$K_1$, and the inclusions' shear and bulk moduli~$G_2$ and~$K_2$, for convenience all normalized by the shear modulus of the matrix~$G_\mathrm{m}$. The reference normalized material parameters are summarized in Tab.~\ref{Sect:VE:Tab1} as a function of the material contrast ratio~$\rho$, fixed to a value~$\rho = 4$. Throughout all examples, the micro-structural morphology is assumed to be known exactly for all scans. In practice, however, it should be identified from experimental data, thus introducing additional measurement uncertainty.

\begin{figure}
	\centering
	\def\svgwidth{1.0\textwidth}
	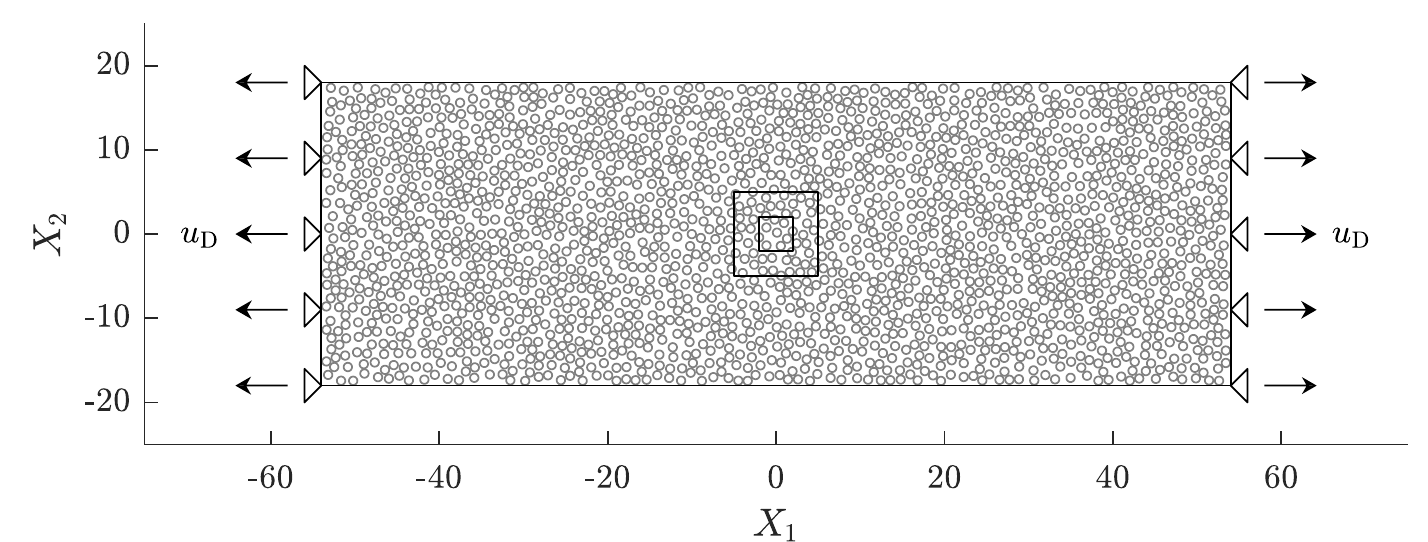
	\caption{A sketch of the specimen's geometry with a random microstructure subjected to~$5\,\%$ overall non-uniform tensile strain. The microstructure (randomly distributed circular inclusions of normalized diameter~$d = 1$) is scanned only in certain parts, depending on the identification method used (cf. Fig.~\ref{Fig:methods}). As an example, two micro ROIs with domains~$\Omega_\mathrm{roi}^\mathrm{m}$ of sizes~$\ell_\mathrm{roi} = 4d$ and~$10d$ are indicated at the specimen's centre.}
	\label{Sect:VE:Fig1}
\end{figure}

For simplicity, the mechanical test adopted is a non-uniform tensile test, sketched in Fig.~\ref{Sect:VE:Fig1}, with a rectangular specimen domain~$\Omega = [-L/2, L/2] \times [-W/2, W/2] = [-54,54]d \times  [-18,18]d$, having randomly distributed inclusions of unit diameter~$d = 1$ with volume fraction~$\phi_\mathrm{dns} = 42.9\,\%$. This means that the typical length scale of the microstructural features, discussed earlier in the introduction, is~$ \ell = d = 1 $. Because the employed constitutive models are size independent, recall Eq.~\eqref{Sect:VE:Eq4}, without any influence on the mechanical behaviour all geometric properties are normalized by the inclusions' diameter~$d = 1$, which can be thought of to be~$50\,\mu\mathrm{m}$, for instance. Macroscopically applied boundary conditions are prescribed to the two vertical edges, cf. Fig.~\ref{Sect:VE:Fig1}, whereas the two horizontal edges are free surfaces. The induced tension corresponds to~$5\,\%$ of overall strain. 
\begin{table}
	\centering
	\caption{Micromechanical parameters adopted in virtual experiment. The shear and bulk moduli~$G_\beta$ and~$K_\beta$ are normalized by the shear modulus of the matrix~$G_\mathrm{m}$. Adopted contrast ratio is fixed to a value~$\rho = 4$.}
	\renewcommand{\arraystretch}{1.2}
	\begin{tabular}{l|r@{}lr@{}l}
		Constitutive parameters                                                     & 	\multicolumn{2}{c}{\renewcommand{\arraystretch}{0.8}\begin{tabular}{@{}c@{}}
				matrix     \\
				{\footnotesize($\beta = 1$)}
		\end{tabular}} & \multicolumn{2}{c}{\renewcommand{\arraystretch}{0.8}\begin{tabular}{@{}c@{}} inclusions \\ {\footnotesize($\beta = 2$)} \end{tabular}}  \\\hline
		Normalized shear modulus, \hfill $G_\beta$                                                  & 1 &                        &  $\rho$ &                        \\
		Normalized bulk modulus, \hfill $K_\beta$                                                   & 3 &                        & $3\rho$ &                        \\
		Poisson's ratio, $\nu_\beta = \frac{3K_\beta-2G_\beta}{2(3K_\beta+G_\beta)}$ & 0 & .35                    &       0 & .35                  
	\end{tabular}
	\label{Sect:VE:Tab1}
\end{table}  

\begin{figure}
	\centering
	\subfloat[micro speckle pattern~$f_\mathrm{m}$]{\includegraphics[scale=1]{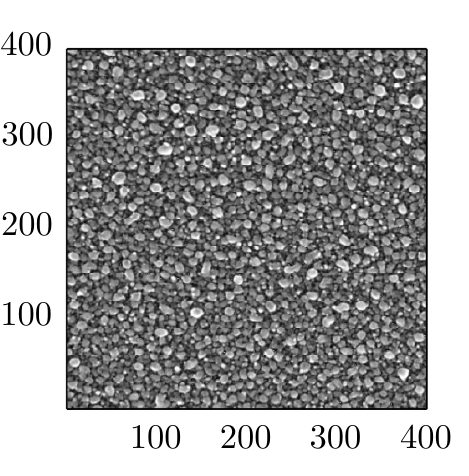}\label{Sect:VE:Fig2a}}\hspace{1em}
	\subfloat[histogram of~$f_\mathrm{m}$]{\includegraphics[scale=1]{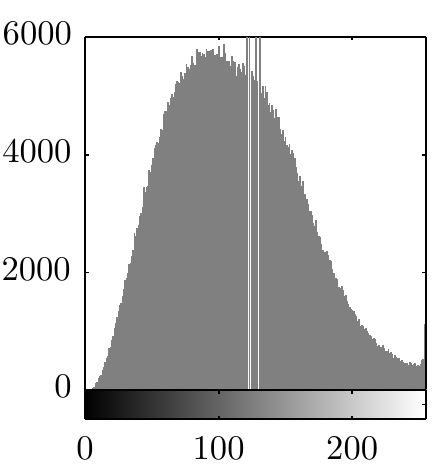}\label{Sect:VE:Fig2b}}\hspace{1em}
	\subfloat[ACF of~$f_\mathrm{m}$, $\ell_\mathrm{c} = 1.8$ pix]{\includegraphics[scale=1]{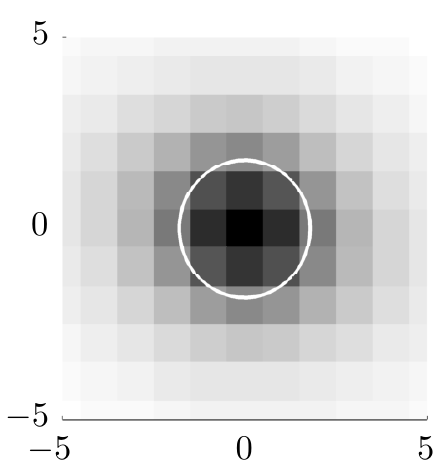}\label{Sect:VE:Fig2c}}\\
	\subfloat[macro speckle pattern~$f_\mathrm{M}$]{\includegraphics[scale=1]{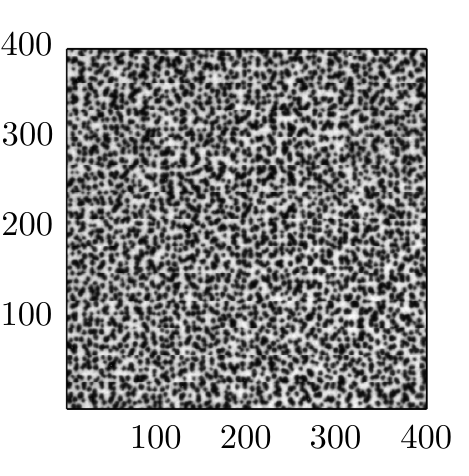}\label{Sect:VE:Fig2d}}\hspace{1em}
	\subfloat[histogram of~$f_\mathrm{M}$]{\includegraphics[scale=1]{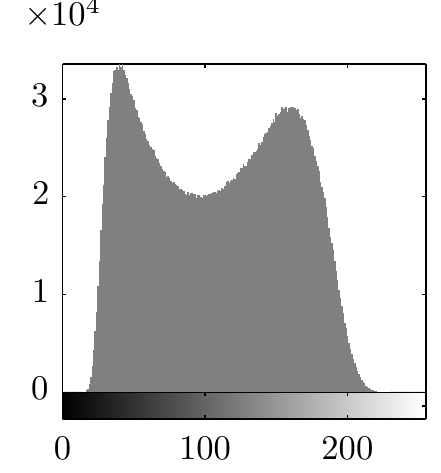}\label{Sect:VE:Fig2e}}\hspace{1em}
	\subfloat[ACF of~$f_\mathrm{M}$, $\ell_\mathrm{c} = 2.8$ pix]{\includegraphics[scale=1]{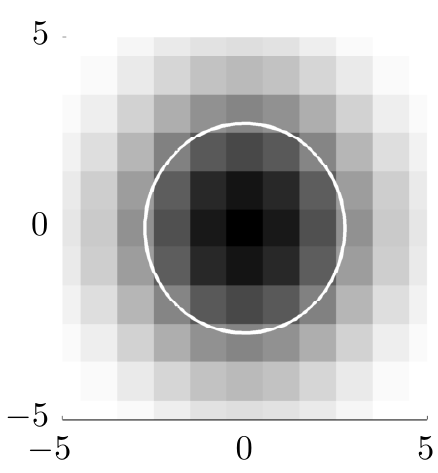}\label{Sect:VE:Fig2f}}
	\caption{Employed speckle patterns (a)~$f_\mathrm{m}$ \cite[Section~4.2.5 of][totally approx.~$ 900 \times 900 $~pixels]{Hoeberichts:2014}, and (d)~$f_\mathrm{M}$ \cite[DIC challenge][Sample~17, totally approx.~$ 2500 \times 1000 $~pixels]{SEM1,SEM2}, along with the corresponding grey value histograms~(b) and~(e), and auto-correlation functions~(c) and~(f).}
	\label{Sect:VE:Fig2}
\end{figure}

Two speckle patterns, $f_\mathrm{m}$ (approx.~$ 900 \times 900 $~pixels) and~$f_\mathrm{M}$ (approx.~$ 2500 \times 1000 $~pixels), are applied at the microscale (inside~$\Omega_\mathrm{fov}^\mathrm{m}$) and at the macroscale (inside~$\Omega_\mathrm{fov}^\mathrm{M}$). Rectangular micro ROIs are considered with edge length~$\ell_\mathrm{roi}$, i.e., $\Omega_\mathrm{roi}^\mathrm{m}$ has the size~$\ell_\mathrm{roi} \times \ell_\mathrm{roi}$. To mimic a fixed camera resolution, $f_\mathrm{m}$ is mapped onto~$\Omega_\mathrm{fov}^\mathrm{m}$ (which is slightly larger than~$\Omega_\mathrm{roi}^\mathrm{m}$), meaning that the micro-pixel size depends on~$\ell_\mathrm{roi}$. Typical pixel densities corresponding to the two patterns thus are~$20 \times 20$~pixels per unit area~$d^2$ for the macro-image, and range from~$50 \times 50$ pixels (corresponding to~$\ell_\mathrm{roi} = 20d$ and scale ratio between the micro ROI size and the specimen size equal to~$5$) to $200 \times 200$ pixels (corresponding to~$\ell_\mathrm{roi} = 4d$ and scale ratio~$25$) per unit area~$d^2$ for the micro-images depending on magnification of the microscopic observation. Figure~\ref{Sect:VE:Fig2} shows both speckle patterns along with their corresponding histograms and Auto-Correlation Functions~(ACF). Several quality parameters are listed in Tab.~\ref{Sect:VE:Tab2}, definitions of which can be found in~\cite{Pan:2008} or~\cite{Dong:2017}. Note that~$\ell_\mathrm{c}$ is defined such that ACF equals~$1/2$, and that in order to closely approximate realistic conditions, physical speckle patterns have been employed, adopted from~\cite[Section~4.2.5]{Hoeberichts:2014} for~$f_\mathrm{m}$, and~\cite[DIC challenge][Sample~17]{SEM1,SEM2} for~$f_\mathrm{M}$. A multiscale pattern is furthermore assumed, i.e., the micro-image~$f_\mathrm{m}$ is applied inside~$\Omega_\mathrm{fov}^\mathrm{m}$ only, whereas the macro-image~$f_\mathrm{M}$ is considered in the remaining part of the domain, i.e., in $\Omega_\mathrm{fov}^\mathrm{M} \subseteq \Omega \backslash \Omega_\mathrm{fov}^\mathrm{m}$. More details on the application of multiscale patterns for \emph{in situ} FE-SEM observations of carbon fibre-reinforced polymer composites can be found, e.g., in~\cite{Tanaka:2011}.

\begin{table}
	\centering
	\caption{Effective parameters of the speckle patterns used at the macroscale~$f_\mathrm{M}$ and microscale~$f_\mathrm{m}$.}
	\renewcommand{\arraystretch}{1.2}
	\begin{tabular}{l@{}l|r@{}l|r@{}l}
		\multicolumn{2}{l|}{Speckle pattern quality parameters}                                & \multicolumn{2}{c|}{image~$f_\mathrm{M}$} & \multicolumn{2}{c}{image~$f_\mathrm{m}$} \\ \hline
		Root-mean-square value,                      & ~$\mathrm{RMS}$                 & 123 & .083                               & 119 & .943                                   \\
		Mean intensity gradient,                     & ~$\delta_f$                     &  30 & .047                               & 25 & .327                                   \\
		Correlation length,                          & ~$\ell_\mathrm{c}$              &   1 & .796  pix                       & 2 & .745 pix                            \\
		Quality factor,                              & ~$Q = \delta_f/\ell_\mathrm{c}$ &  16 & .728                               & 9 & .226
	\end{tabular}
	\label{Sect:VE:Tab2}
\end{table}

Experimental observations~$g_\mathrm{m}$, $g_\mathrm{M}$, and~$\vec{F}_\mathrm{exp}$, are generated through DNS. A Total Lagrangian formulation is used to model the underlying physics of Eq.~\eqref{Sect:VE:Eq1}, employing quadratic isoparametric triangular elements with three-point Gauss integration rule. The mesh generator \emph{gmsh} by~\cite{gmsh} has been used for the triangulation, as shown in Fig.~\ref{Ex:DNS:Fig1a}. Only two images, i.e., the reference~$f$ and deformed~$g$, will be used, although all inverse methods presented below can easily be extended to simultaneous correlation of multiple images in a time-integrated scheme as would be done in real tests to improve accuracy, see~\cite{Neggers:2015}. The synthetic deformed images are generated such that the reference image~$f(\vec{X})$ is pixelized, with discrete pixel positions~$\vec{X}_i$. Since isoparametric elements are used, natural coordinates~$\vec{\eta}_i$ of all pixels are found by inverting the isoparametric mapping. The displacement field is then computed exactly for each of the pixels as $\vec{u}_i = \vec{u}(\vec{X}_i)$, based on the assumed shape of the interpolation functions, and each of the pixels is translated accordingly, i.e., $g(\vec{X}_i) = f(\vec{X}_i - \vec{u}_i)$. Since computed displacements are of non-integer pixel values, the resulting images~$g_\mathrm{m}$ and~$g_\mathrm{M}$ are interpolated at the reference (integer) pixel locations of the underlying grid~$\vec{X}_i$ using cubic interpolation.

\begin{figure}
	\centering
	\def\svgwidth{150px}
	\subfloat[normalized effective material parameters]{\includegraphics[scale=1]{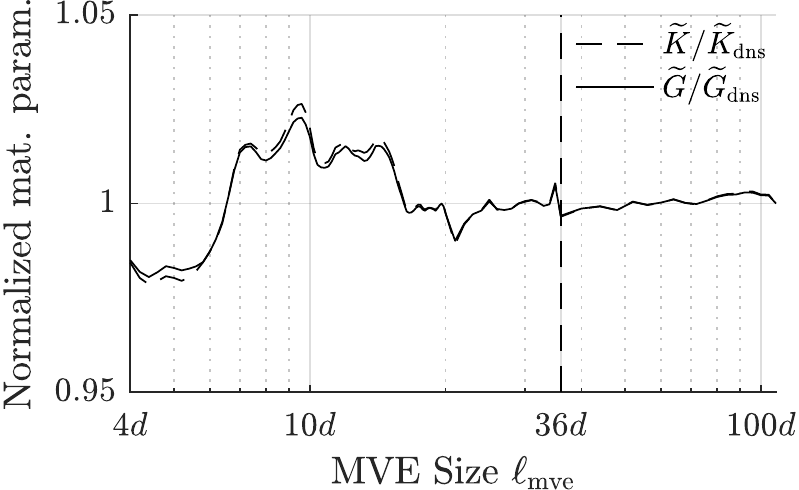}\label{Sect:VE:Fig3a}}\hfill
	\subfloat[normalized volume fraction]{\includegraphics[scale=1]{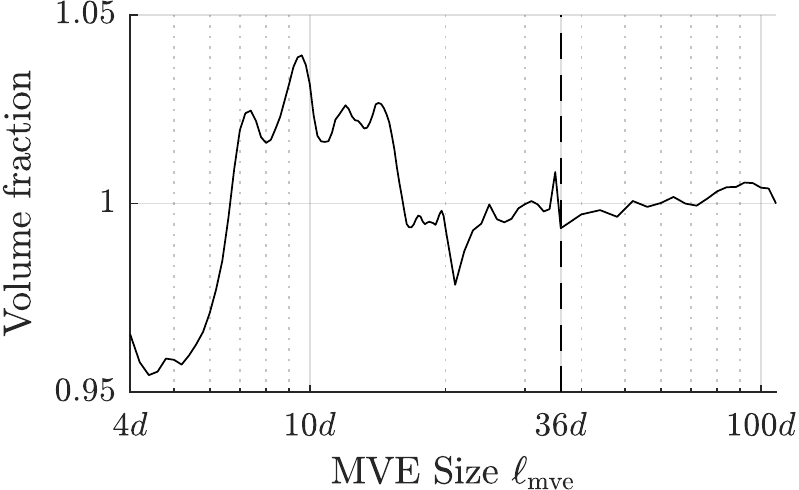}\label{Sect:VE:Fig3b}}
	\caption{(a)~Convergence of the normalized effective material parameters~$\widetilde{K}$ and~$\widetilde{G}$ based on MVE simulations and Eqs.~\eqref{SubSect:FE2:Eq1} and~\eqref{Ex:Comparison:Eq1} for various sizes of~$\Omega_\mathrm{mve}^\mathrm{m}$ ($\ell_\mathrm{mve} \times \ell_\mathrm{mve}$). (b)~Convergence of the normalized volume fraction~$\phi$ as a function of~$\ell_\mathrm{mve}$. Normalization limit values corresponding to~$\ell_\mathrm{mve} = L$ are listed in Tab.~\ref{Sect:VE:Tab3}.}
	\label{Sect:VE:Fig3}
\end{figure}

For further analyses, effective material parameters~$\widetilde{G}$ and~$\widetilde{K}$ obtained for a MVE positioned at the specimen's centre~$\vec{X} = \vec{0}$ (as shown in Fig.~\ref{Sect:VE:Fig1}), are computed to provide a measure for the representativeness of individual MVEs as a function of their size~$\ell_\mathrm{mve}$. To this end, first the homogenized anisotropic constitutive tangent of each MVE, $\overline{\mathbb{C}}_\mathrm{m}$, is computed, considering periodic boundary conditions with exact constitutive models for both constituents in the reference configuration, i.e.,
\begin{equation}
\overline{\bs{\mathsf{C}}}_\mathrm{m} = \frac{1}{|\Omega_\mathrm{mve}^\mathrm{m}|} \bs{\mathsf{B}}^\mathsf{T}(\bs{\mathsf{A}}\bs{\mathsf{K}}_\partial^{-1}\bs{\mathsf{A}}^\mathsf{T})^{-1}\bs{\mathsf{B}}.
\label{SubSect:FE2:Eq1}
\end{equation}
In Eq.~\eqref{SubSect:FE2:Eq1}, $\overline{\bs{\mathsf{C}}}_\mathrm{m}$ is a matrix representation of~$\overline{\mathbb{C}}_\mathrm{m}$, $\bs{\mathsf{K}}_\partial$ is a stiffness matrix condensed to the nodes of the MVE boundary~$\partial\Omega_\mathrm{mve}^\mathrm{m}$, and~$|\Omega_\mathrm{mve}^\mathrm{m}|$ is the size of the micro-model domain~$\Omega_\mathrm{mve}^\mathrm{m}$. The coefficient matrices~$\bs{\mathsf{A}}$ and~$\bs{\mathsf{B}}$ correspond to the discretized version of periodic boundary conditions, expressed as a set of equality constraints
\begin{equation}
\bs{\mathsf{A}}\underline{u}_\partial = \bs{\mathsf{B}}(\underline{\overline{F}}_\mathrm{M} - \underline{I}),
\label{eq:weak_periodicity}
\end{equation}
where~$\underline{u}_\partial$ is a column of MVE boundary displacements, and~$\underline{\overline{F}}_\mathrm{M}$ and~$\underline{I}$ are column representations of the macroscopically prescribed deformation gradient~$\overline{\bs{F}}_\mathrm{M}$ and the identity tensor~$\bs{I}$; for further details see~\citep{Miehe:2003,Miehe:2007,Larsson:2007}. Next, $\widetilde{G}$ and~$\widetilde{K}$ are derived by fitting an isotropic stiffness tensor~$\mathbb{C}_\mathrm{iso}$ to the homogenized anisotropic constitutive tangent~$\overline{\mathbb{C}}_\mathrm{m}$,
\begin{equation}
\{\widetilde{G},\widetilde{K}\} = \underset{G,K}{\text{arg min}}\ \| \mathbb{C}_\mathrm{iso}(G,K) - \overline{\mathbb{C}}_\mathrm{m} \|_\mathrm{T}^2,
\label{Ex:Comparison:Eq1}
\end{equation}
where~$ \| \mathbb{C} \|^2_\mathrm{T} = C_{ijkl} C_{ijkl}$ denotes the tensor norm of fourth-order tensors. More details on constitutive tangent fitting can be found in~\cite{Moakher:2006}.
%
% Analytical solution
%\begin{equation}
%\begin{aligned}
%9\widetilde{K} &= 3\overline{\mathbb{C}}_\mathrm{m} \cddddot \mathbb{I}_\mathrm{v} = c_{1111} + c_{2222} + c_{3333} + 2( c_{1122} + c_{1133} + c_{2233} ), \\
%30\widetilde{G} &= 3\overline{\mathbb{C}}_\mathrm{m} \cddddot \mathbb{I}_\mathrm{d} = 2(c_{1111} + c_{2222} + c_{3333} - c_{1122} - c_{2233} - c_{1133}) + 6(c_{2323} + c_{1313} + c_{1212}),
%\end{aligned}
%\end{equation}
%Introduce double dot product, $\mathbb{I}_\mathrm{v}$, and $\mathbb{I}_\mathrm{d}$ in the notation.
%
The resulting effective parameters (normalized by their limit values) are shown in Fig.~\ref{Sect:VE:Fig3a}. As expected, for small MVE sizes, the homogenized shear and bulk moduli are not representative, whereas the error slowly decreases with increasing MVE size (especially for~$\ell_\mathrm{mve} \geq 16d$). The limit values, corresponding to the entire specimen, $\ell_\mathrm{mve} = L$, which are used for normalization, read as~$\widetilde{G}_\mathrm{dns} = 1.656$, $\widetilde{K}_\mathrm{dns} = 4.656$, see Tab.~\ref{Sect:VE:Tab3}. The representativeness of individual MVEs in terms of homogenized bulk and shear moduli correlates well with the volume fraction of inclusions shown in Fig.~\ref{Sect:VE:Fig3b}. Overall, all effective quantities fluctuate approximately within~$5\,\%$ error relative to their limit values. The small fluctuation observed in both graphs of Fig.~\ref{Sect:VE:Fig3} towards the limit~$\ell_\mathrm{mve} = L$ is explained by the fact that all inclusions are not allowed to intersect specimen's boundary, cf. Fig.~\ref{Sect:VE:Fig1}. For more details on the representativeness of random microstructures see, e.g., \cite{Torquato:2002}.

\begin{table}
	\centering
	\caption{Effective parameters corresponding to the entire specimen~$\Omega$.}
	\renewcommand{\arraystretch}{1.3}
	\begin{tabular}{l|r@{}l}
		Effective parameter & \multicolumn{2}{c}{Value} \\\hline
		Volume fraction, \hfill $\phi_\mathrm{dns}$ & 0 & .429 \\
		Shear modulus, \hfill $\widetilde{G}_\mathrm{dns}$ & 1 & .656 \\
		Bulk modulus, \hfill $\widetilde{K}_\mathrm{dns}$ & 4 & .656
	\end{tabular}
	\label{Sect:VE:Tab3}
\end{table}
%
%-----------------------------------------------------------------------------
%	SECTION Full-Scale Direct Numerical Simulation
%-----------------------------------------------------------------------------
%
\section{Full-Scale Direct Numerical Simulations}
\label{Sect:DNS}
%
%----------------------------------
%	SUBSECTION DNS Methodology
%----------------------------------
%
\subsection{Methodology}
\label{SubSect:DNS}
Full-Scale Direct Numerical Simulations~(DNS), cf. Fig.~\ref{Fig:methods_a} and Proc.~\ref{Sect:DNS:Tab1}, consider a fully resolved microstructure throughout the entire specimen domain, thus requiring considerable experimental as well as computational efforts. The IDIC model~$\vec{\mathcal{M}}$ spans the entire domain (i.e., $\Omega_\mathcal{M} \equiv \Omega$), for which experimentally applied kinematic boundary conditions and reaction forces are known. For the region of interest, typically microscopic images~$\Omega_\mathrm{roi}^\mathrm{m}$ are used, cf. Fig.~\ref{Sect:Introduction:Fig1}. In principle, DNS constitutes a single-scale approach operating at the fully expanded microscale only, and hence is considered as the reference solution.

The normalization procedure through experimentally measured forces is implemented as follows. Instead of scalarization, which requires prior knowledge of appropriate weights, one can minimize the objectives of Eqs.~\eqref{SubSect:DIC:Eq1} and~\eqref{SubSect:DIC:Eq4} sequentially, making use of the fact that the deformation state of the considered (elastic) specimen under the applied load does not depend on any particular material parameter normalization (recall that during the tensile experiment adopted, displacement control with force measurement is used). The numerically computed force, $ \vec{F} $, is therefore known only up to a multiplicative constant~$\alpha$, i.e., the simulated force reads~$ \vec{F}_\mathrm{sim} = \alpha \vec{F} $. The multiplicative constant can be eliminated from the quadratic objective of Eq.~\eqref{SubSect:DIC:Eq4}, providing optimal value denoted by the asterisk
\begin{equation}
\alpha^\star = \frac{ \vec{F}_\mathrm{exp} \cdot \vec{F}(\underline{\lambda}) }{ \vec{F}(\underline{\lambda}) \cdot \vec{F}(\underline{\lambda}) }.
\label{Subsect:DNS:Eq1}
\end{equation}

Although normalization through the measured force in Eq.~\eqref{Subsect:DNS:Eq1} is used for simplicity, we note that other approaches can be used, following for instance the line of reasoning proposed in~\cite{Rahmani:2014}, where homogenized stiffness properties of the microstructure have been fitted to the experimentally observed stiffness, or \citep{Rethore:2010,Rethore:2013}, where equilibrium gap method and the balance between external and internal forces has been adopted to extend the least-squares objective of Eq.~\eqref{SubSect:DIC:Eq1}.

\begin{figure}
	\centering
	% \psfragfig{matfrag/DNSmesh1} 
	% \psfragfig{matfrag/DNSmesh2}
	\subfloat[DNS mesh (image generation)]{\includegraphics[scale=1]{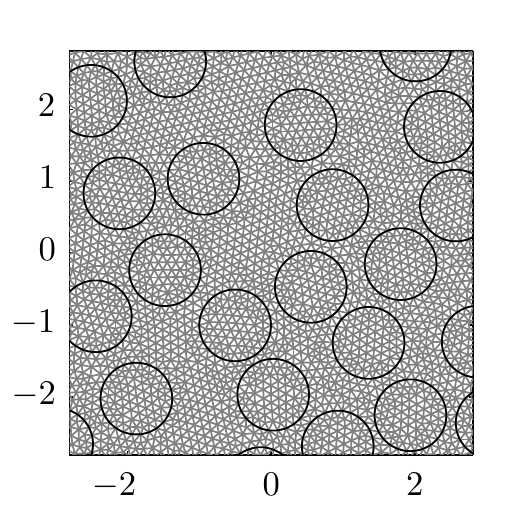}\label{Ex:DNS:Fig1a}}\hspace{1.0em}
	\subfloat[DNS mesh (IDIC)]{\includegraphics[scale=1]{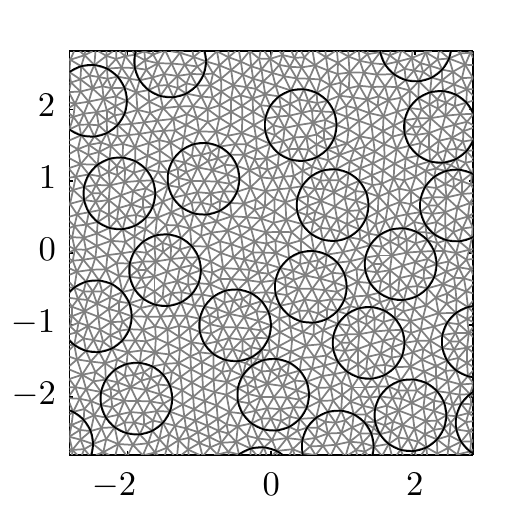}\label{Ex:DNS:Fig1b}}
	\caption{Triangulations associated with two DNS systems used for~(a) generating experimental observations, and~(b) for micro-mechanical identification.}
	\label{Ex:DNS:Fig1}
\end{figure}

\begin{procedure}
	\caption{Full-Scale Direct Numerical Simulation~(DNS) used for micromechanical parameter identification.}
	\label{Sect:DNS:Tab1}
	\vspace{-\topsep}
	\begin{tcolorbox}
		\centering
		\begin{enumerate}[1:]\itemsep0pt
			
			\item Experimental observations:
			\begin{itemize}[\textbf{-}]
				
				\item before the test, record a high-resolution image of the microstructure of the entire specimen~$\Omega$ in undeformed configuration (located inside~$\Omega_\mathrm{fov}^\mathrm{M}$) which typically requires multiple microscopic scans,
				
				\item  at each load step, record a (high-resolution) scan of the microscopic region of interest~$\Omega_\mathrm{roi}^\mathrm{m}$ (not necessarily spanning the entire specimen to save experimental efforts), the applied macroscopic kinematic boundary conditions~$\vec{u}_\mathrm{D}$, and the resulting reaction force~$\vec{F}_\mathrm{exp}$.
			\end{itemize}
			
			\item Construct an underlying IDIC model, taking into account the entire microstructure (i.e., in Eq.~\eqref{SubSect:DIC:Eq3}, $\Omega_\mathcal{M} \equiv \Omega$), and apply the measured kinematic boundary conditions~$\vec{u}_\mathrm{D}$ directly to the model. 
			
			\item Perform IDIC, by correlating the scans of the micro ROI (i.e., in Eq.~\eqref{SubSect:DIC:Eq1}, $\Omega_\mathrm{roi} \equiv \Omega_\mathrm{roi}^\mathrm{m}$), followed by normalization through Eqs.~\eqref{SubSect:DIC:Eq4} and~\eqref{Subsect:DNS:Eq1} to obtain the resulting set of micromechanical parameters~$\underline{\lambda}$.
			
		\end{enumerate}
	\end{tcolorbox}
	\vspace{-\topsep}
\end{procedure}
%
%----------------------------------
%	SUBSECTION DNS Results
%----------------------------------
%
\subsection{Results}
\label{Ex:DNS}
The numerical models used for identification of micromechanical parameters employ the discretization shown in Fig.~\ref{Ex:DNS:Fig1b}, which is $ 1.5 $ times coarser compared to the forward-model mesh used for generating the virtual measurements (shown in Fig.~\ref{Ex:DNS:Fig1a}). Correlation is performed for microscopic images~$f_\mathrm{m}$ and~$g_\mathrm{m}$, using~$\Omega_\mathrm{roi}^\mathrm{m}$ positioned at the specimen's centre. In order to quantify the accuracy of all four identified micromechanical parameters ($G_\beta$ and~$K_\beta$, $\beta \in \{ 1, 2 \}$) through only one simple quantity, the following Root-Mean-Square (RMS) norm is introduced
\begin{equation}
\eta_\mathrm{rms} = \sqrt{\frac{1}{n_{\lambda_\mathrm{mat}}}\sum_{i=1}^{n_{\lambda_\mathrm{mat}}} \left( \frac{\lambda_i}{\lambda_{\mathrm{ex},i}}-1 \right)^2 },
\label{Ex:DNS:Eq1}
\end{equation}
where~$\lambda_{\mathrm{ex},i}$ is the exact value of corresponding $i$-th material parameter~$\lambda_i$. The achieved typical accuracy of DNS is~$\eta_\mathrm{rms} = 0.03\,\%$, as reported in Tab.~\ref{Ex:Comparison:Tab1} of Section~\ref{Ex:Comparison}, where an overall comparison of all methods is presented.

For consistency with other methods, the Rigid Body Motion~(RBM) has been corrected for in IDIC to eliminate related systematic errors by adding the corresponding DOFs to~$\underline{\lambda}$. As expected, the RBM displacements vanish, namely~$ \| \vec{u}_\mathrm{rbm} \|_\mathrm{E} \leq 1 \times 10^{-4} $. Furthermore, various sizes of~$\Omega_\mathrm{roi}^\mathrm{m}$ positioned at the specimen's centre have been considered to test the influence of the micro-image resolution in the range~$ \ell_\mathrm{roi} \in [4, 20]d $, used later for Stress Integration~(SI) and computational homogenization~(FE\textsuperscript{2}) methods described in Sections~\ref{Sect:SI} and~\ref{Sect:SIFE2}. No significant influence on the accuracy of identified parameters has been observed, suggesting that the resolution~$\ell_\mathrm{roi} = 20d$ (i.e., density of~$50 \times 50$ pixels per~$d^2$) still suffices for accurate micromechanical identification (obtained accuracy~$\eta_\mathrm{rms}$ corresponds to~$0.08\,\%$).
%
%-----------------------------------------------------------------------------
%	SECTION Concurrent Multiscale Modelling
%-----------------------------------------------------------------------------
%
\section{Concurrent Multiscale Modelling}
\label{Sect:CMM}
%
%----------------------------------
%	SUBSECTION CMM Methodology
%----------------------------------
%
\subsection{Methodology}
\label{SubSect:CMM}
In order to reduce computational and experimental efforts, Concurrent Multiscale Modelling~(CMM) \cite[see][]{Tian:2010,Gracie:2011,Hutter:2014} takes into account only a small portion of the microstructure (cf. Fig.~\ref{Fig:methods_b}), whereas a homogeneous material with a suitable (effective) constitutive law is assumed elsewhere. The two regions, with either the fully resolved microstructure ($\Omega_\mathrm{full}^\mathrm{m}$) or the homogeneous material (outside~$\Omega_\mathrm{full}^\mathrm{m}$), are seamlessly bridged by a gradually refined mesh. The homogenized material transmits forces applied at the specimen's boundary to the fully resolved microstructural region, inside which the images are correlated (i.e., inside~$\Omega_\mathrm{roi}^\mathrm{m}$). 

The identification itself proceeds in two steps and requires observations at two scales, cf. Proc.~\ref{Sect:CMM:Tab1}. First, macroscopic IDIC is performed using the applied kinematic boundary conditions, measured reaction forces, macroscopically observed images inside~$\Omega_\mathrm{roi}^\mathrm{M}$, and a macroscopic homogeneous model~$\vec{\mathcal{M}}^\mathrm{hom}$. This model spans the entire domain (i.e., $\Omega_\mathcal{M}^\mathrm{hom} \equiv \Omega$), in which homogeneous effective material properties are adopted. In the subsequent identification stage, the actual microstructural material parameters are sought for, which are considered only inside the fully resolved region of which~$\Omega_\mathrm{roi}^\mathrm{m}$ is part. To this end, a multiscale heterogeneous model~$\vec{\mathcal{M}}^\mathrm{het}$ is constructed, which uses the same domain (i.e., $\Omega_\mathcal{M}^\mathrm{het} \equiv \Omega$) and boundary conditions as the homogeneous model~$\vec{\mathcal{M}}^\mathrm{hom}$. The only difference is that the multiscale model~$\vec{\mathcal{M}}^\mathrm{het}$ considers fully refined microstructure in certain part of the domain~$\Omega$ (having there an independent set of micromechanical parameters) and denoted as~$\Omega_\mathrm{full}^\mathrm{m}$ (of size~$\ell_\mathrm{full} \times \ell_\mathrm{full}$), whereas outside this region the previously identified homogeneous material is used.
\begin{procedure}
	\caption{Concurrent Multiscale Modelling~(CMM) used for micromechanical parameter identification.}
	\label{Sect:CMM:Tab1}
	\vspace{-\topsep}
	\begin{tcolorbox}
		\centering
		\begin{enumerate}[1:]
			\item Experimental observations:
			\begin{itemize}[\textbf{-}]
				\item at each load step, record a macroscopic image spanning~$\Omega_\mathrm{fov}^\mathrm{M}$ with low spatial resolution (no microstructure needed), the applied displacements~$\vec{u}_\mathrm{D}$, and the reaction forces~$\vec{F}_\mathrm{exp}$,
				
				\item at each load step, also record a microscopic image inside~$\Omega_\mathrm{fov}^\mathrm{m}$ with a high spatial resolution (capturing micro-structural morphology with sufficient accuracy).
			\end{itemize}
			
			\item Construct two models:
			\begin{itemize}[\textbf{-}]
				\item a homogeneous model, $\mathcal{M}^\mathrm{hom}$, $\Omega_\mathcal{M}^\mathrm{hom} \equiv \Omega$, in which constant effective material parameters are adopted everywhere,
				
				\item and a multiscale heterogeneous model, $\mathcal{M}^\mathrm{het}$, $\Omega_\mathcal{M}^\mathrm{het} \equiv \Omega$, in which effective material parameters are used outside the fully resolved region~$\Omega_\mathrm{full}^\mathrm{m}$, whereas inside this region an independent set of micro-mechanical parameters is used.
				
			\end{itemize}
			
			\item Perform IDIC using the homogeneous model~$\mathcal{M}^\mathrm{hom}$ to identify effective material parameter ratios by correlating inside~$\Omega_\mathrm{roi}^\mathrm{M}$. Normalize obtained parameter ratios by the measured forces~$\vec{F}_\mathrm{exp}$ through Eq.~\eqref{Subsect:DNS:Eq1}.
			
			\item Perform IDIC using the heterogeneous model~$\mathcal{M}^\mathrm{het}$, in which the normalized effective material parameters (Step~$3$) are used and kept fixed outside the fully resolved region. Inside the fully resolved region (of which~$\Omega_\mathrm{roi}^\mathrm{m}$ is part of), an independent set of micro-mechanical parameters is identified by correlating the microscopic images inside~$\Omega_\mathrm{roi}^\mathrm{m}$. (Note that no additional normalization is required, because homogeneous material parameters have already been normalized in Step~$3$.)
			
		\end{enumerate}
	\end{tcolorbox}
	\vspace{-\topsep}
\end{procedure}
%
%----------------------------------
%	SUBSECTION CMM Results
%----------------------------------
%
\subsection{Results}
\label{Ex:CMM}
\begin{figure}
	\centering
	\subfloat[triangulation of the homogeneous model~$\vec{\mathcal{M}}^\mathrm{hom}$]{\includegraphics[width=0.49\linewidth]{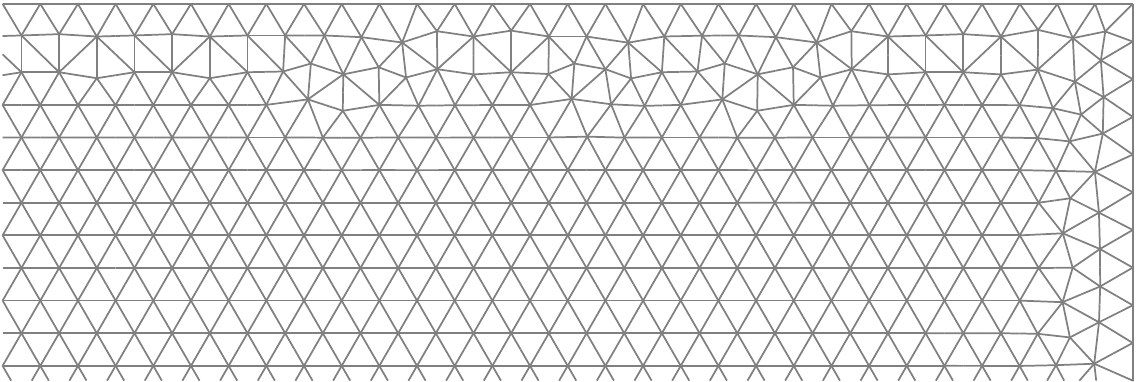}\label{Ex:CMM:Fig0a}}\hspace{0.5em}
	\subfloat[triangulation of the heterogeneous model~$\vec{\mathcal{M}}^\mathrm{het}$]{\includegraphics[width=0.49\linewidth]{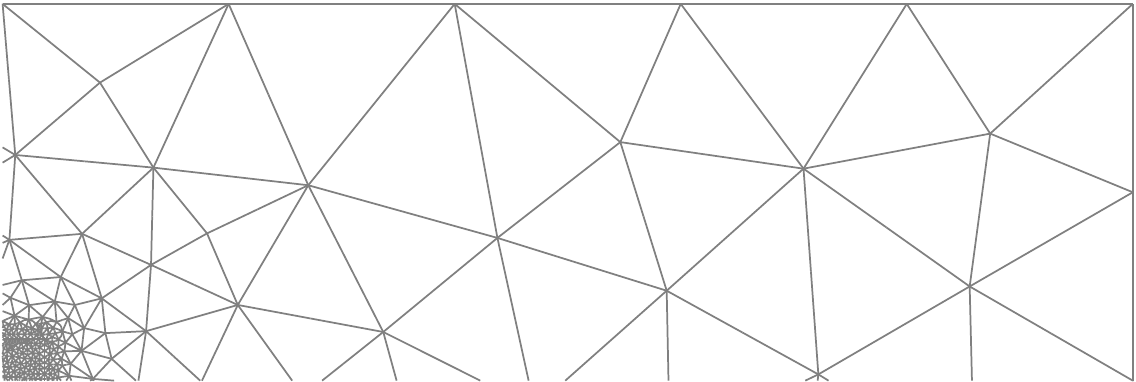}\label{Ex:CMM:Fig0b}}	
	\caption{Two triangulations used for the Concurrent Multiscale Modelling~(CMM). (a)~A triangulation with a homogeneous constitutive model used in the first step for identification of the homogenized properties~$\overline{G}$ and~$\overline{K}$. (b)~A seamless triangulation bridging two scales with heterogeneous constitutive properties used for identification of the micromechanical parameters. A quarter of the entire domain~$\Omega$ is shown in both cases. For this particular case, the size of the fully resolved region, in which the heterogeneous microstructure is considered, corresponds to~$\ell_\mathrm{full} = \ell_\mathrm{roi} = 4d$.}
	\label{Ex:CMM:Fig0}
\end{figure}

\emph{Step~1:} Using the triangulation of Fig.~\ref{Ex:CMM:Fig0a} in the homogeneous model~$\vec{\mathcal{M}}^\mathrm{hom}$ of Step~3, Proc.~\ref{Sect:CMM:Tab1}, the \emph{effective} constitutive parameters are obtained through IDIC by correlating images inside the macroscopic ROI, $\Omega_\mathrm{roi}^\mathrm{M}$, assuming the constitutive law of Eq.~\eqref{Sect:VE:Eq4}. Obtained values are~$\overline{G} = 1.649$ and~$\overline{K} = 4.584$, which are within~$2\,\%$ of relative error compared to the reference homogenized values~$\widetilde{G}_\mathrm{dns}$ and~$\widetilde{K}_\mathrm{dns}$ (obtained for~$\ell_\mathrm{mve} = L = 108d$) reported in Tab.~\ref{Sect:VE:Tab3}. The discrepancy between limit values and IDIC values~$\overline{G}$ and~$\overline{K}$ are explained by different state of the material and its non-linearities. In particular, reference configuration has been used for the identification of~$\widetilde{G}_\mathrm{dns}$ and~$\widetilde{K}_\mathrm{dns}$, which means that the constitutive law effectively corresponds to the Hooke's law. On the contrary, deformed configuration was used for the identification of~$\overline{G}$ and~$\overline{K}$, for which constitutive law does not fulfil Hooke's law anymore due to large strains and hyperelasticity, recall Eq.~\eqref{Sect:VE:Eq4}. The differences in the deformed state then give rise to observed discrepancy.

\emph{Step~2:} Micromechanical parameter identification using a multiscale heterogeneous model $\vec{\mathcal{M}}^\mathrm{het}$ is performed, cf. Step~4, Proc.~\ref{Sect:CMM:Tab1}, using~$\overline{G}$ and~$\overline{K}$ outside the fully resolved region of the heterogeneous microstructure. Typical triangulation is shown in Fig.~\ref{Ex:CMM:Fig0b} for the size of the fully resolved region~$\ell_\mathrm{full} = \ell_\mathrm{roi} = 4d$. Rigid body motion is corrected for, which now does not vanish due to neglected heterogeneities outside~$\Omega_\mathrm{full}^\mathrm{m}$, achieving values as large as~$\| \vec{u}_\mathrm{rbm} \|_\mathrm{E} \approx 1.2 \times 10^{-2}$ (cf. also Fig.~\ref{Ex:SS:Fig1}, where the systematic errors in MVE BCs are compared for the different methods). When~$\vec{u}_\mathrm{rbm}$ is neglected, however, large errors in the identified material parameters result. 

\begin{figure}
	\centering
	\subfloat[accuracy of CMM]{\includegraphics[scale=1]{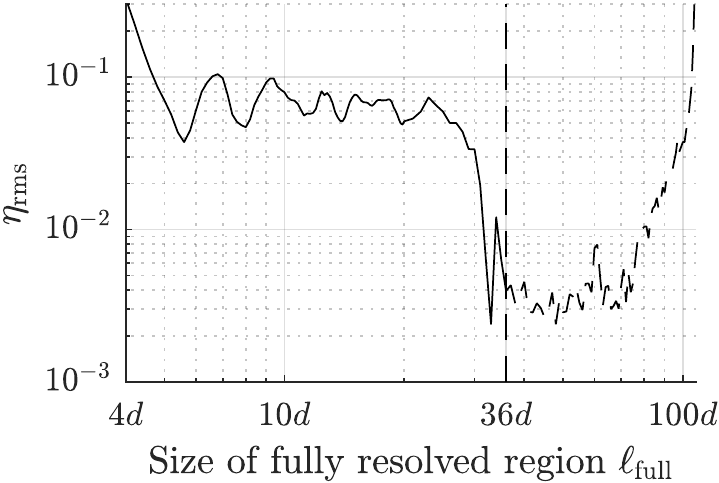}\label{Ex:CMM:Fig1b}}\hspace{1.0em}
	\subfloat[accuracy of BSM]{\includegraphics[scale=1]{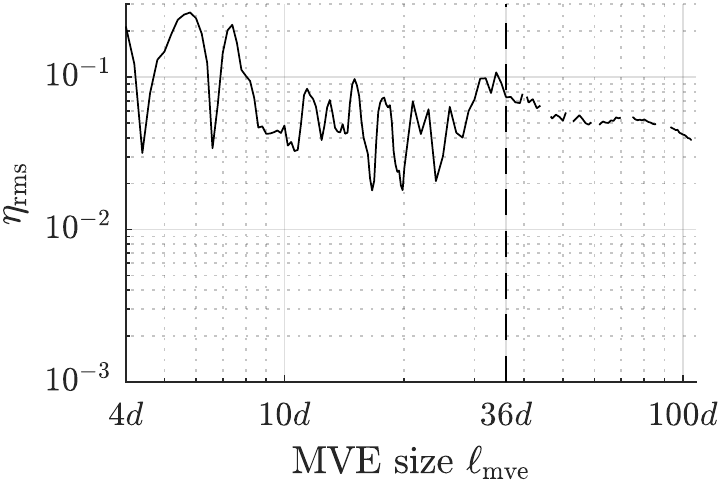}\label{Ex:CMM:Fig1c}}	
	\caption{(a)~Convergence of~$\eta_\mathrm{rms}$ as a function of the size of the fully resolved region~$\ell_\mathrm{full}$ for the Concurrent Multiscale Modelling~(CMM). (b)~Convergence of~$\eta_\mathrm{rms}$ as a function of the MVE size~$\ell_\mathrm{mve}$ for the Bridging Scale Method~(BSM). Micro ROI size is fixed in both cases to~$\ell_\mathrm{roi} = 4d$.}
	\label{Ex:CMM:Fig1}
\end{figure}

In order to evaluate the influence of the size of the fully resolved region on the accuracy of the micromechanical parameter identification, the following convergence study is performed. The microscopic region of interest~$\Omega_\mathrm{roi}^\mathrm{m}$ (of constant size~$\ell_\mathrm{roi} = 4d$) is fixed at the specimen's centre, whereas the size of the fully resolved region is varied as $ \ell_\mathrm{full} \in [4, 108]d $. With increasing~$\ell_\mathrm{full}$, the effect of the homogeneous material fades out (i.e., inaccurate displacements at the interface between the homogeneous and heterogeneous regions, cf. Fig.~\ref{Ex:SS:Fig1}). The effect of the homogeneous region can be observed in Fig.~\ref{Ex:BSM:Fig1a}, where the residual corresponding to~$\ell_\mathrm{full} = 5.6d$ and~$\ell_\mathrm{roi} = 4d$ is shown. In terms of~$\eta_\mathrm{rms}$, shown in Fig.~\ref{Ex:CMM:Fig1b}, two regions can be distinguished:
(i) $\ell_\mathrm{full}$ is smaller than the specimen's width~$W$ (solid line);
and (ii) $\ell_\mathrm{full}$ is larger than the specimen's width (dashed line).
For smaller sizes of the fully resolved regions, the achieved accuracy decreases due to neglected heterogeneities outside the fully resolved region, and achieves on average~$\overline{\eta}_\mathrm{rms} \approx 7\,\%$ of RMS for $4d \leq \ell_\mathrm{full} \leq 20d$. In the second case, the method is no longer in a multiscale regime and hence, these results should be interpreted with caution as they are not representative for cases aimed for in this contribution (for~$\ell_\mathrm{full} = 48d$, $\eta_\mathrm{rms} = 0.2\,\%$, which is an accurate result). For very large fully resolved regions ($\ell_\mathrm{full} > 80d$), the accuracy decreases again, which is caused by too narrow homogenized regions positioned between the fully resolved region and the two fixed vertical boundaries. Note that the limit in which the entire specimen is fully resolved cannot be reached as no homogenized material outside~$\Omega^\mathrm{m}_\mathrm{full}$, needed for normalization, is included. In this case, one would obviously opt for the full DNS model instead.

%
%-----------------------------------------------------------------------------
%	SECTION Bridging Scale Method
%-----------------------------------------------------------------------------
%
\section{Bridging Scale Method}
\label{Sect:BSM}
%
%----------------------------------
%	SUBSECTION BSM Methodology
%----------------------------------
%
\subsection{Methodology}
\label{SubSect:BSM}
The Bridging Scale Method~(BSM) \cite[see, e.g.,][]{Tang:2005,Kaye:2013}, similarly to the CMM, fully resolves the microstructure inside a small region of interest, as indicated in Fig.~\ref{Fig:methods_c}. It differs, however, in the fact that there is no seamless transition between the micro- and macro-scale by a gradually refined mesh, but the microscale model is coupled to the macroscale one. Although multiple options exist, a kinematic coupling is adopted in what follows.

Two separate macro- and microscopic models are introduced in this approach, which are solved sequentially in two steps, cf. Proc.~\ref{Sect:BSM:Tab1}. In the first step, IDIC correlating on~$\Omega_\mathrm{roi}^\mathrm{M}$ is performed through a macroscopic model~$\vec{\mathcal{M}}^\mathrm{M}$, containing no microstructure and spanning the entire specimen ($\Omega_\mathcal{M}^\mathrm{M} \equiv \Omega$). This step provides homogenized material properties, and displacement and stress fields inside~$\Omega_\mathcal{M}^\mathrm{M}$. Next, boundary conditions are sampled from the macroscopic model and prescribed to a MVE, represented by a microscopic model~$\vec{\mathcal{M}}^\mathrm{m}$ (in which~$\Omega_\mathcal{M}^\mathrm{m} \equiv \Omega_\mathrm{mve}^\mathrm{m}$), where the full microstructure is considered. Using this model, a second IDIC is performed, by correlating the images inside~$\Omega_\mathrm{roi}^\mathrm{m} \subseteq \Omega_\mathrm{mve}^\mathrm{m}$. Because Dirichlet BCs are applied along the entire boundary~$\partial\Omega_\mathrm{mve}^\mathrm{m}$, only material parameter ratios can be extracted. In order to normalize them, the difference between the macroscopic~$\bs{P}_\mathrm{M}$ and effective MVE stress~$\overline{\bs{P}}_\mathrm{m}$ tensors is minimized in analogy to Eq.~\eqref{SubSect:DIC:Eq4}, i.e.
\begin{equation}
\alpha^\star = \underset{\alpha \in \mathbb{R}}{\mbox{arg min}}\ \| \bs{P}_\mathrm{M} - \alpha \overline{\bs{P}}_\mathrm{m}(\underline{\lambda}) \|_\mathrm{T}^2, 
\quad \mbox{providing} \quad
\alpha^\star = \frac{ \bs{P}_\mathrm{M} : \overline{\bs{P}}_\mathrm{m}^\mathsf{T}(\underline{\lambda}) }{ \overline{\bs{P}}_\mathrm{m}(\underline{\lambda}) : \overline{\bs{P}}_\mathrm{m}^\mathsf{T}(\underline{\lambda}) },
\label{SubSect:BSM:Eq1}
\end{equation}
where
\begin{equation}
\overline{\bs{P}}_\mathrm{m}(\underline{\lambda}) = \frac{1}{|\Omega_\mathrm{mve}^\mathrm{m}|}\int_{\Omega_\mathrm{mve}^\mathrm{m}} \bs{P}_\mathrm{m}(\vec{X}_\mathrm{m}, \underline{\lambda})\,\mathrm{d}\vec{X}_\mathrm{m}
\label{SubSect:BSM:Eq2}
\end{equation}
denotes the volume average of the microscopic first Piola--Kirchhoff stress tensor, $\| \bs{A} \|_\mathrm{T}^2 = \bs{A} : \bs{A}^\mathsf{T}$ is the square of the tensor norm induced by the corresponding scalar product, and~$\alpha^\star$ the optimal normalization constant. In practice, the average stress of Eq.~\eqref{SubSect:BSM:Eq2} computes as
\begin{equation}
\overline{\underline{P}}_\mathrm{m} = \frac{1}{|\Omega_\mathrm{mve}^\mathrm{m}|}\bs{\mathsf{B}}^\mathsf{T}\underline{\mu},
\label{SubSect:BSM:Eq3}
\end{equation}
where~$\overline{\underline{P}}_\mathrm{m}$ is a column storing components of the homogenized microscopic stress tensor~$\overline{\bs{P}}_\mathrm{m}$, $\underline{\mu}$ is the Lagrange multiplier enforcing the periodicity of Eq.~\eqref{eq:weak_periodicity}, and~$\bs{\mathsf{B}}$ is the coefficient matrix, recall Eq.~\eqref{SubSect:FE2:Eq1} and the discussion therein.

A potential drawback of this approach is that boundary conditions prescribed to the MVE model are sampled from the meso- or macroscopic model~$\Omega_\mathcal{M}^\mathrm{M}$, and hence they do not contain any micro-structural features and fluctuations, which are known to be essential for accurate microstructural IDIC material identification without stress normalization, cf.~\cite{Rokos:2018}.
\begin{procedure}
	\caption{Bridging Scale Method~(BSM) used for micromechanical parameter identification.}
	\label{Sect:BSM:Tab1}
	\vspace{-\topsep}
	\begin{tcolorbox}
		\centering
		\begin{enumerate}[1:]
			\item Experimental observations:
			\begin{itemize}[\textbf{-}]
				
				\item at each load step, record a macroscopic image spanning~$\Omega_\mathrm{fov}^\mathrm{M}$ with low spatial resolution (no microstructure is needed), the applied displacements~$\vec{u}_\mathrm{D}$, and reaction forces~$\vec{F}_\mathrm{exp}$,
				
				\item at each load step, also record a microscopic image~$\Omega_\mathrm{fov}^\mathrm{m}$ with a high spatial resolution (capturing the microstructural morphology with sufficient accuracy).
			\end{itemize}
			
			\item Construct a homogeneous macroscopic model~$\vec{\mathcal{M}}^\mathrm{M}$ of the entire specimen ($\Omega_\mathcal{M}^\mathrm{M} \equiv \Omega$); perform IDIC inside~$\Omega_\mathrm{roi}^\mathrm{M}$ to obtain the effective homogeneous material parameters.
			
			\item Construct a heterogeneous microscopic model~$\vec{\mathcal{M}}^\mathrm{m}$ inside the MVE (i.e., $\Omega_\mathcal{M}^\mathrm{m} \equiv \Omega_\mathrm{mve}^\mathrm{m}$), taking into account the microstructural morphology. Sample the kinematic MVE boundary conditions along~$\partial\Omega_\mathrm{mve}^\mathrm{m}$ and the corresponding average macroscopic stress~$\bs{P}_\mathrm{M}$ from the macroscopic model~$\vec{\mathcal{M}}^\mathrm{M}$. Perform IDIC on the MVE model, correlating the microscopic images inside~$\Omega_\mathrm{roi}^\mathrm{m}$, to obtain the micromechanical parameter ratios, while simultaneously enforcing normalization according to Eq.~\eqref{SubSect:BSM:Eq1}.
			
		\end{enumerate}
	\end{tcolorbox}
	\vspace{-\topsep}
\end{procedure}
%
%----------------------------------
%	SUBSECTION DNS Results
%----------------------------------
%
\subsection{Results}
\label{Ex:BSM}
The first identification step towards homogenized material parameters in BSM is identical to that of CMM, and hence provides the same results. The second step consists in prescribing kinematic boundary conditions, sampled from the homogeneous model, along the entire MVE boundary~$\partial\Omega_\mathrm{mve}^\mathrm{m}$. Finally, micromechanical parameters are normalized by comparing homogenized stresses of the MVE model to stresses of the homogeneous model. The MVE rigid body motion has been corrected for ($\| \vec{u}_\mathrm{rbm} \|_\mathrm{E} \approx 1.1 \times 10^{-2}$), which is again significant due to inaccuracies in the applied boundary conditions.

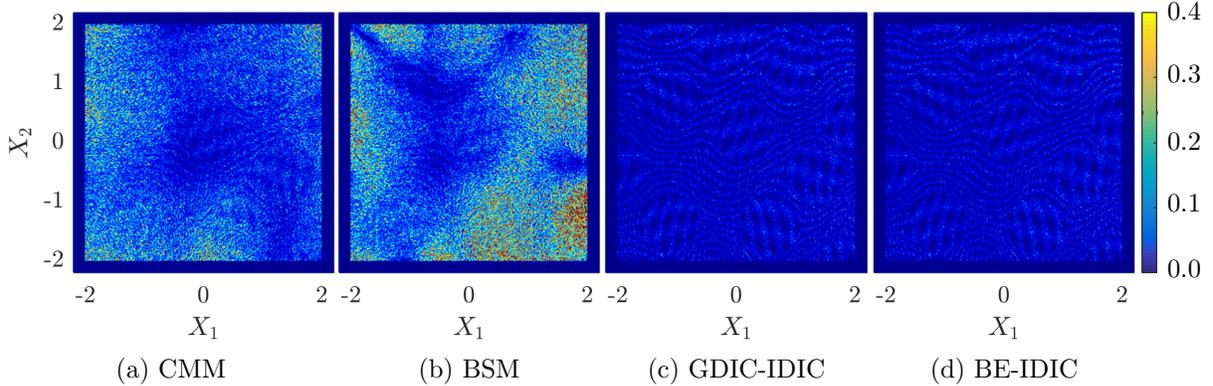
\begin{figure}
	\centering
	\subfloat[CMM]{\includegraphics[scale=0.9]{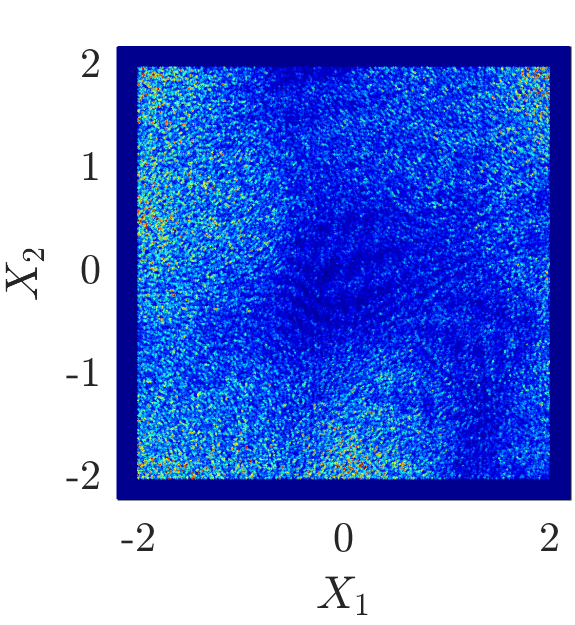}\label{Ex:BSM:Fig1a}}
	\subfloat[BSM]{\includegraphics[scale=0.9]{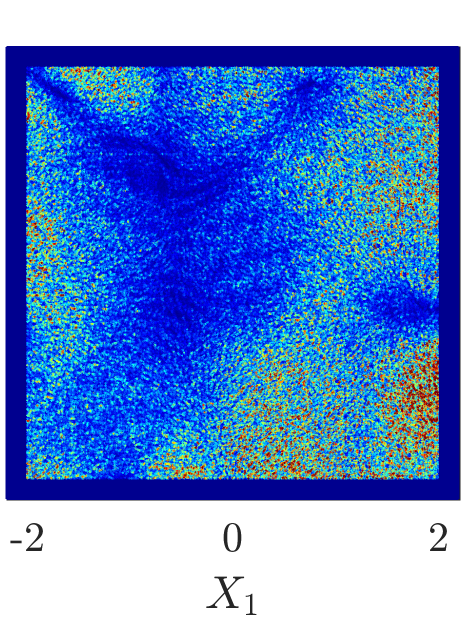}\label{Ex:BSM:Fig1b}}
	\subfloat[GDIC-IDIC]{\includegraphics[scale=0.9]{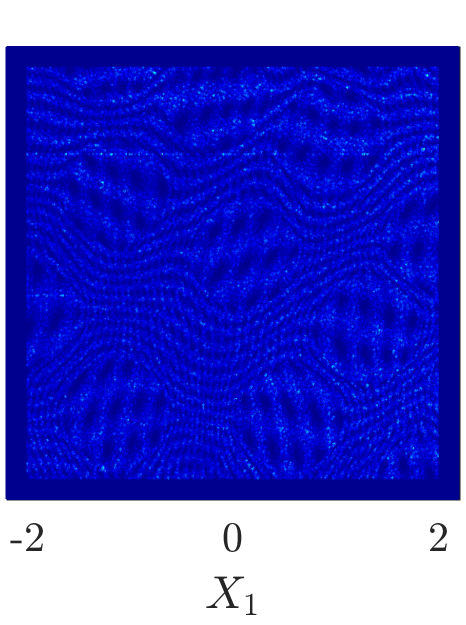}\label{Ex:BSM:Fig1c}}
	\subfloat[BE-IDIC]{\includegraphics[scale=0.9]{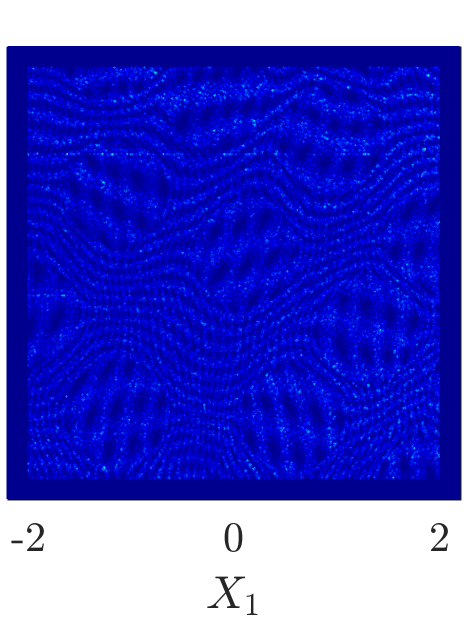}\label{Ex:BSM:Fig1d}}
	\def\svgwidth{2.175em}%% Creator: Inkscape inkscape 0.92.3, www.inkscape.org
%% PDF/EPS/PS + LaTeX output extension by Johan Engelen, 2010
%% Accompanies image file 'mresiduals_colorbar.pdf' (pdf, eps, ps)
%%
%% To include the image in your LaTeX document, write
%%   \input{<filename>.pdf_tex}
%%  instead of
%%   \includegraphics{<filename>.pdf}
%% To scale the image, write
%%   \def\svgwidth{<desired width>}
%%   \input{<filename>.pdf_tex}
%%  instead of
%%   \includegraphics[width=<desired width>]{<filename>.pdf}
%%
%% Images with a different path to the parent latex file can
%% be accessed with the `import' package (which may need to be
%% installed) using
%%   \usepackage{import}
%% in the preamble, and then including the image with
%%   \import{<path to file>}{<filename>.pdf_tex}
%% Alternatively, one can specify
%%   \graphicspath{{<path to file>/}}
%% 
%% For more information, please see info/svg-inkscape on CTAN:
%%   http://tug.ctan.org/tex-archive/info/svg-inkscape
%%
\begingroup%
  \makeatletter%
  \providecommand\color[2][]{%
    \errmessage{(Inkscape) Color is used for the text in Inkscape, but the package 'color.sty' is not loaded}%
    \renewcommand\color[2][]{}%
  }%
  \providecommand\transparent[1]{%
    \errmessage{(Inkscape) Transparency is used (non-zero) for the text in Inkscape, but the package 'transparent.sty' is not loaded}%
    \renewcommand\transparent[1]{}%
  }%
  \providecommand\rotatebox[2]{#2}%
  \newcommand*\fsize{\dimexpr\f@size pt\relax}%
  \newcommand*\lineheight[1]{\fontsize{\fsize}{#1\fsize}\selectfont}%
  \ifx\svgwidth\undefined%
    \setlength{\unitlength}{28.34645669bp}%
    \ifx\svgscale\undefined%
      \relax%
    \else%
      \setlength{\unitlength}{\unitlength * \real{\svgscale}}%
    \fi%
  \else%
    \setlength{\unitlength}{\svgwidth}%
  \fi%
  \global\let\svgwidth\undefined%
  \global\let\svgscale\undefined%
  \makeatother%
  \begin{picture}(1,5)%
    \lineheight{1}%
    \setlength\tabcolsep{0pt}%
    \put(0.39055281,2.88190002){\color[rgb]{0,0,0}\makebox(0,0)[lt]{\lineheight{1.25}\smash{\begin{tabular}[t]{l}{\footnotesize\setlength{\fboxsep}{1pt}\colorbox{white}{$0.2$}}\end{tabular}}}}%
    \put(0.39055281,1.97442627){\color[rgb]{0,0,0}\makebox(0,0)[lt]{\lineheight{1.25}\smash{\begin{tabular}[t]{l}{\footnotesize\setlength{\fboxsep}{1pt}\colorbox{white}{$0.1$}}\end{tabular}}}}%
    \put(0.39055281,3.88882751){\color[rgb]{0,0,0}\makebox(0,0)[lt]{\lineheight{1.25}\smash{\begin{tabular}[t]{l}{\footnotesize\setlength{\fboxsep}{1pt}\colorbox{white}{$0.3$}}\end{tabular}}}}%
    \put(0.39055281,0.99555359){\color[rgb]{0,0,0}\makebox(0,0)[lt]{\lineheight{1.25}\smash{\begin{tabular}[t]{l}{\footnotesize\setlength{\fboxsep}{1pt}\colorbox{white}{$0.0$}}\end{tabular}}}}%
    \put(0.39055281,4.81315308){\color[rgb]{0,0,0}\makebox(0,0)[lt]{\lineheight{1.25}\smash{\begin{tabular}[t]{l}{\footnotesize\setlength{\fboxsep}{1pt}\colorbox{white}{$0.4$}}\end{tabular}}}}%
    \put(0,0){\includegraphics[width=\unitlength,page=1]{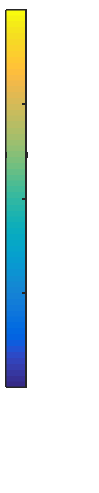}}%
  \end{picture}%
\endgroup%

	\caption{Microscopic image residuals~$| f_\mathrm{m} - g_\mathrm{m} |(\vec{X}_\mathrm{m})/2^8 $ (normalized by dynamic contrast~$2^8$) corresponding to (a)~Concurrent Multiscale Method~(CMM) with~$\ell_\mathrm{full} = 5.6d$ (RMS grey value of~$0.06$), (b)~Bridging Scale Method~(BSM) with~$\ell_\mathrm{mve} = 4.4d$ (RMS grey value of~$0.09$), (c)~uncoupled GDIC-IDIC with~$\ell_\mathrm{mve} = 4d$ (RMS grey value of~$0.02$), and~(d) uncoupled BE-IDIC with~$\ell_\mathrm{mve} = 4d$ (RMS grey value of~$0.02$). In all cases, $\ell_\mathrm{roi} = 4d$ has been used.}
	\label{Ex:BSM:Fig1}
\end{figure}

A convergence study is carried out, which consists in enlarging the employed MVE ($\Omega_\mathrm{mve}^\mathrm{m}$, with~$ \ell_\mathrm{mve} \in [4, 108]d $) while using a ROI with a fixed size~$\ell_\mathrm{roi} = 4d$. The results are shown in Fig.~\ref{Ex:CMM:Fig1c}, where the best accuracy, within multiscale regime (i.e., for~$4d \leq \ell_\mathrm{mve} \leq 20d$), is achieved for~$\eta_\mathrm{rms}(\ell_\mathrm{mve} = 16.6d) = 1.8\,\%$. The average performance corresponds approximately to~$\overline{\eta}_\mathrm{rms} \approx 8\,\%$ of RMS. The image residual corresponding to~$\ell_\mathrm{mve} = 4.4d$ is shown in Fig.~\ref{Ex:BSM:Fig1b}, which indicates a relatively poor correlation.

%
%-----------------------------------------------------------------------------
%	SECTION Novel Methods
%-----------------------------------------------------------------------------
%
\section{Normalization by Newly Proposed Methods}
\label{Sect:NM}
Two novel methodologies for the identification of micro-mechanical parameters are described below in this section. They both utilize the same ingredient, uncoupled microscale methods described first in Section~\ref{Sect:SM}, which serve to identify micromechanical parameter \underline{ratios} at the microscale. Proper normalization constants are obtained through Stress Integration~(SI) or computational homogenization~(FE\textsuperscript{2}) methods, described subsequently in Sections~\ref{Sect:SI} and~\ref{Sect:SIFE2}.

Although both methodologies are presented in a simplified context, i.e., focusing on conservative hyperelastic systems and considering only two configurations (reference and deformed), they can both be extended to time-evolving and history-dependent problems. To this end, simultaneous correlations of multiple images in a time-integrated scheme would need to be considered, see~\cite{Neggers:2015}. In the case of SI (Section~\ref{Sect:SI}), the displacement~$\vec{u}_\mathrm{M}$ and strain~$\bs{B}$ fields would need to be considered incrementally at each of the time steps (to approximate virtual test fields), effectively minimizing the difference between the internal and external forces over all time steps tested with the incremental displacement and strain fields, cf. also \citep[Chapter~4]{VFM}. For FE\textsuperscript{2} (Section~\ref{Sect:SIFE2}), time-integrated difference between the observed~$\vec{F}_\mathrm{exp}$ and simulated~$\vec{F}_\mathrm{sim}$ forces would need to be minimized. Since microstructural simulations are run only with Dirichlet boundary conditions, the parameters of an elastic-plastic constitutive law are expected to have a small influence on the displacement field, see, e.g., \cite{Rethore:2013}. For that reason, low accuracy of history-dependent variables might occur. Moreover, based on history evolution of the resulting macroscopic forces, it might be difficult to distinguish between non-linearities stemming from elastic and plastic deformations, cf., e.g., \cite{Rappel2020}. Additional investigations are thus needed to obtain quantitative as well as qualitative performance of the newly proposed methods for history-dependent materials. This is outside of the scope of the current manuscript, and will be considered in future studies.
%
%-----------------------------------------------------------------------------
%	SECTION Uncoupled Methods
%-----------------------------------------------------------------------------
%
\subsection{Uncoupled Methods}
\label{Sect:SM}
%
%----------------------------------
%	SUBSECTION SM Methodology
%----------------------------------
%
\subsubsection{Methodology}
\label{SubSect:SS}
\begin{figure}
	\centering
	% \psfragfig{matfrag/MVEmesh} 
	% \psfragfig{matfrag/GDICmesh}
	\subfloat[GDIC-IDIC]{\def\svgwidth{0.25\textwidth}%% Creator: Inkscape inkscape 0.92.3, www.inkscape.org
%% PDF/EPS/PS + LaTeX output extension by Johan Engelen, 2010
%% Accompanies image file 'GDICmesh.pdf' (pdf, eps, ps)
%%
%% To include the image in your LaTeX document, write
%%   \input{<filename>.pdf_tex}
%%  instead of
%%   \includegraphics{<filename>.pdf}
%% To scale the image, write
%%   \def\svgwidth{<desired width>}
%%   \input{<filename>.pdf_tex}
%%  instead of
%%   \includegraphics[width=<desired width>]{<filename>.pdf}
%%
%% Images with a different path to the parent latex file can
%% be accessed with the `import' package (which may need to be
%% installed) using
%%   \usepackage{import}
%% in the preamble, and then including the image with
%%   \import{<path to file>}{<filename>.pdf_tex}
%% Alternatively, one can specify
%%   \graphicspath{{<path to file>/}}
%% 
%% For more information, please see info/svg-inkscape on CTAN:
%%   http://tug.ctan.org/tex-archive/info/svg-inkscape
%%
\begingroup%
  \makeatletter%
  \providecommand\color[2][]{%
    \errmessage{(Inkscape) Color is used for the text in Inkscape, but the package 'color.sty' is not loaded}%
    \renewcommand\color[2][]{}%
  }%
  \providecommand\transparent[1]{%
    \errmessage{(Inkscape) Transparency is used (non-zero) for the text in Inkscape, but the package 'transparent.sty' is not loaded}%
    \renewcommand\transparent[1]{}%
  }%
  \providecommand\rotatebox[2]{#2}%
  \newcommand*\fsize{\dimexpr\f@size pt\relax}%
  \newcommand*\lineheight[1]{\fontsize{\fsize}{#1\fsize}\selectfont}%
  \ifx\svgwidth\undefined%
    \setlength{\unitlength}{368.50393701bp}%
    \ifx\svgscale\undefined%
      \relax%
    \else%
      \setlength{\unitlength}{\unitlength * \real{\svgscale}}%
    \fi%
  \else%
    \setlength{\unitlength}{\svgwidth}%
  \fi%
  \global\let\svgwidth\undefined%
  \global\let\svgscale\undefined%
  \makeatother%
  \begin{picture}(1,1)%
    \lineheight{1}%
    \setlength\tabcolsep{0pt}%
    \put(0,0){\includegraphics[width=\unitlength,page=1]{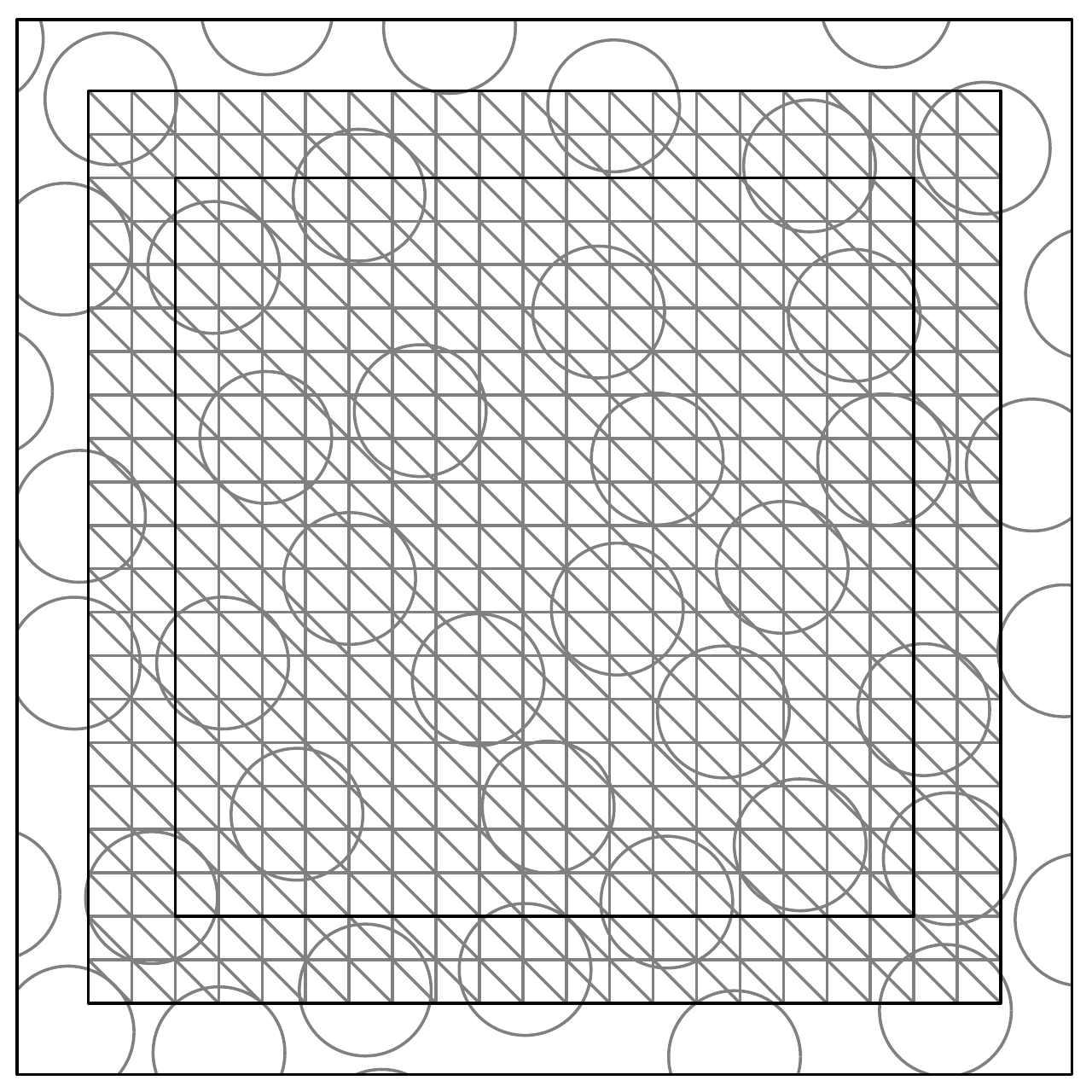}}%
    \put(0.30378171,0.74548786){\color[rgb]{0,0,0}\makebox(0,0)[lt]{\lineheight{1.25}\smash{\begin{tabular}[t]{l}{\setlength{\fboxsep}{1pt}\colorbox{white}{$\Omega_\mathrm{roi}^\mathrm{m} \equiv \Omega_\mathrm{mve}^\mathrm{m}$}}\end{tabular}}}}%
    \put(0.57243582,0.89744943){\color[rgb]{0,0,0}\makebox(0,0)[lt]{\lineheight{1.25}\smash{\begin{tabular}[t]{l}{\setlength{\fboxsep}{1pt}\colorbox{white}{$\Omega_\mathrm{roi,gdic}^\mathrm{m}$}}\end{tabular}}}}%
    \put(0.7839312,0.05316092){\color[rgb]{0,0,0}\makebox(0,0)[lt]{\lineheight{1.25}\smash{\begin{tabular}[t]{l}{\setlength{\fboxsep}{1pt}\colorbox{white}{$\Omega_\mathrm{fov}^\mathrm{m}$}}\end{tabular}}}}%
    \put(0,0){\includegraphics[width=\unitlength,page=2]{GDICmesh.pdf}}%
  \end{picture}%
\endgroup%
\label{Sect:SS:Fig1b}}\hspace{2.0em}
	\subfloat[BE-IDIC]{\def\svgwidth{0.25\textwidth}%% Creator: Inkscape inkscape 0.92.3, www.inkscape.org
%% PDF/EPS/PS + LaTeX output extension by Johan Engelen, 2010
%% Accompanies image file 'MVEmesh.pdf' (pdf, eps, ps)
%%
%% To include the image in your LaTeX document, write
%%   \input{<filename>.pdf_tex}
%%  instead of
%%   \includegraphics{<filename>.pdf}
%% To scale the image, write
%%   \def\svgwidth{<desired width>}
%%   \input{<filename>.pdf_tex}
%%  instead of
%%   \includegraphics[width=<desired width>]{<filename>.pdf}
%%
%% Images with a different path to the parent latex file can
%% be accessed with the `import' package (which may need to be
%% installed) using
%%   \usepackage{import}
%% in the preamble, and then including the image with
%%   \import{<path to file>}{<filename>.pdf_tex}
%% Alternatively, one can specify
%%   \graphicspath{{<path to file>/}}
%% 
%% For more information, please see info/svg-inkscape on CTAN:
%%   http://tug.ctan.org/tex-archive/info/svg-inkscape
%%
\begingroup%
  \makeatletter%
  \providecommand\color[2][]{%
    \errmessage{(Inkscape) Color is used for the text in Inkscape, but the package 'color.sty' is not loaded}%
    \renewcommand\color[2][]{}%
  }%
  \providecommand\transparent[1]{%
    \errmessage{(Inkscape) Transparency is used (non-zero) for the text in Inkscape, but the package 'transparent.sty' is not loaded}%
    \renewcommand\transparent[1]{}%
  }%
  \providecommand\rotatebox[2]{#2}%
  \newcommand*\fsize{\dimexpr\f@size pt\relax}%
  \newcommand*\lineheight[1]{\fontsize{\fsize}{#1\fsize}\selectfont}%
  \ifx\svgwidth\undefined%
    \setlength{\unitlength}{368.50393701bp}%
    \ifx\svgscale\undefined%
      \relax%
    \else%
      \setlength{\unitlength}{\unitlength * \real{\svgscale}}%
    \fi%
  \else%
    \setlength{\unitlength}{\svgwidth}%
  \fi%
  \global\let\svgwidth\undefined%
  \global\let\svgscale\undefined%
  \makeatother%
  \begin{picture}(1,1)%
    \lineheight{1}%
    \setlength\tabcolsep{0pt}%
    \put(0,0){\includegraphics[width=\unitlength,page=1]{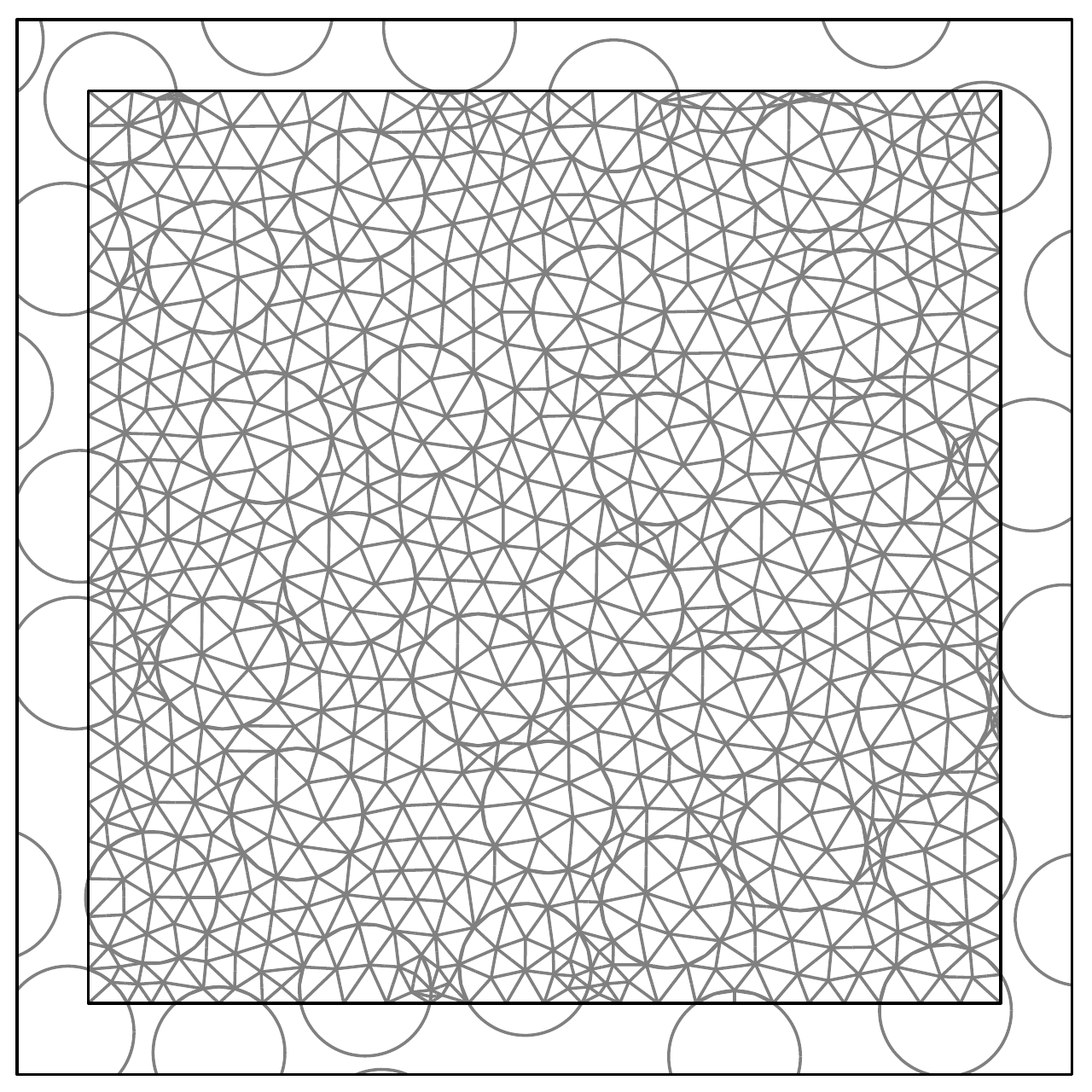}}%
    \put(0.38926397,0.83232187){\color[rgb]{0,0,0}\makebox(0,0)[lt]{\lineheight{1.25}\smash{\begin{tabular}[t]{l}{\setlength{\fboxsep}{1pt}\colorbox{white}{$\Omega_\mathrm{roi}^\mathrm{m}\equiv \Omega_\mathrm{mve}^\mathrm{m}$}}\end{tabular}}}}%
    \put(0,0){\includegraphics[width=\unitlength,page=2]{MVEmesh.pdf}}%
    \put(0.78392956,0.05316444){\color[rgb]{0,0,0}\makebox(0,0)[lt]{\lineheight{1.25}\smash{\begin{tabular}[t]{l}{\setlength{\fboxsep}{1pt}\colorbox{white}{$\Omega_\mathrm{fov}^\mathrm{m}$}}\end{tabular}}}}%
  \end{picture}%
\endgroup%
\label{Sect:SS:Fig1a}}
	\caption{A sketch of uncoupled methods. (a) The GDIC-IDIC method identifies in the first step kinematic boundary conditions by means of a GDIC mesh constructed over~$\Omega_\mathrm{roi,gdic}^\mathrm{m} \subset \Omega_\mathrm{fov}^\mathrm{m}$, and subsequently performs IDIC on a smaller region, $\Omega_\mathrm{roi}^\mathrm{m}$. (b) BE-IDIC uses material parameters as well as kinematic displacements at the MVE boundary as IDIC DOFs ($\Omega_\mathrm{roi}^\mathrm{m} \equiv \Omega_\mathrm{mve}^\mathrm{m} \subset \Omega_\mathrm{fov}^\mathrm{m}$).}
	\label{Sect:SS:Fig1}
\end{figure}

Uncoupled methods rely on IDIC being carried out inside an MVE only, as sketched in Fig.~\ref{Sect:SS:Fig1}, requiring microstructural scans of~$\Omega_\mathrm{roi}^\mathrm{m}$ and a corresponding IDIC MVE model~$\vec{\mathcal{M}}^\mathrm{m}$. The MVE model necessitates kinematic boundary conditions, applied along the entire MVE boundary~$\partial\Omega_\mathrm{mve}^\mathrm{m}$, which are generally unknown and highly irregular due to material heterogeneities. Furthermore, because Dirichlet BCs are prescribed, only material parameter ratios can be extracted. The key question here is how sufficiently accurate microscopic boundary conditions along the MVE can be obtained, as discussed below.

Two alternatives are considered, as summarized in Proc.~\ref{Sect:SS:Tab1} and~\ref{Sect:SS:Tab2}: (i)~sequential identification of microscopic displacement fields by GDIC followed by material parameter identification through IDIC (introduced by~\citealt{Hild:2016} and~\citealt{Shakoor:2017}), abbreviated as GDIC-IDIC in what follows; and (ii) simultaneous identification of boundary conditions and material parameters, abbreviated as Boundary-Enriched IDIC (BE-IDIC), cf.~\cite{Fedele:2012} and \cite{Rokos:2018}. Because the GDIC-IDIC approach is subject to the well-known GDIC compromise (i.e., lack of kinematic freedom for coarse GDIC meshes and random errors due to image noise for fine meshes, cf.~\citealt{Leclerc:2009}), the identified kinematic boundary conditions (and hence identified micromechanical parameters) suffer from inaccuracies for highly heterogeneous microstructures~\citep{Rokos:2018}.

The second alternative (i.e., BE-IDIC) is used to eliminate the GDIC step. Material parameters as well as all kinematic displacements at the MVE boundary are considered as DOFs in Eq.~\eqref{SubSect:DIC:Eq1}, i.e.,
\begin{equation}
\underline{\lambda} = [\underline{\lambda}_\mathrm{mat}^\mathsf{T},\underline{\lambda}_\mathrm{kin}^\mathsf{T}]^\mathsf{T},
\label{SubSect:SS:Eq1}
\end{equation}
where
\begin{equation}
\begin{aligned}
\underline{\lambda}_\mathrm{mat} &= [G_1, K_1, G_2, K_2]^\mathsf{T}, \\
\underline{\lambda}_\mathrm{kin} &= \underline{u}_\mathrm{mve}(\vec{X}), \quad \vec{X} \in \partial\Omega_\mathrm{mve}^\mathrm{m}.
\end{aligned}
\label{SubSect:SS:Eq2}
\end{equation}
In Eq.~\eqref{SubSect:SS:Eq2}, $\underline{u}_\mathrm{mve}(\vec{X})$, $\vec{X} \in \partial\Omega_\mathrm{mve}^\mathrm{m}$, denotes a column of all DOFs situated at the MVE boundary. A standard IDIC procedure with an extended set of DOFs is then conducted.

Note that for both uncoupled methods the MVE chosen does not have to be representative~\citep{Rokos:2018}, meaning that it should merely contain all relevant microstructural constituents. This is in contrast with normalization approaches which require representative effective stresses (typically evaluated at macroscopic Gauss integration points). Accuracy of the uncoupled methods can be further improved by considering multiple MVE realizations, and averaged over the ensemble of identified material parameter ratios to eliminate random scatter (cf.~\citealt{Rokos:2018}, where it was shown that the identified micromechanical parameter ratios show a low systematic bias for the BE-IDIC case).

\begin{procedure}
	\caption{Uncoupled method GDIC-IDIC for identification of micromechanical parameter ratios.}
	\label{Sect:SS:Tab1}
	\vspace{-\topsep}
	\begin{tcolorbox}
		\centering
		\begin{enumerate}[1:]
			\item Experimental observations:
			\begin{itemize}[\textbf{-}]
				\item at each load step, record a microscopic image inside~$\Omega_\mathrm{fov}^\mathrm{m}$ with a high spatial resolution.
			\end{itemize}
			
			\item Construct a GDIC triangulation and perform GDIC on~$\Omega_\mathrm{roi,gdic}^\mathrm{m}$, $ \Omega_\mathrm{roi}^\mathrm{m} \subset \Omega_\mathrm{roi,gdic}^\mathrm{m} \subset \Omega_\mathrm{fov}^\mathrm{m}$ (cf. Fig.~\ref{Sect:SS:Fig1b}).
			
			\item Construct a microscopic MVE model~$\vec{\mathcal{M}}^\mathrm{m}$ over~$ \Omega_\mathrm{mve}^\mathrm{m} \subseteq \Omega_\mathrm{roi,gdic}^\mathrm{m} $ (cf. Fig.~\ref{Sect:SS:Fig1b} where~$\Omega_\mathrm{mve}^\mathrm{m} \equiv \Omega_\mathrm{roi}^\mathrm{m}$), applying the MVE BCs along~$\partial\Omega_\mathrm{mve}^\mathrm{m}$ sampled from the GDIC data of Step~2.
			
			\item Perform IDIC to obtain micromechanical parameter ratios, correlating on~$\Omega_\mathrm{roi}^\mathrm{m} \subseteq \Omega_\mathrm{roi,gdic}^\mathrm{m}$.
			
		\end{enumerate}
	\end{tcolorbox}
	\vspace{-\topsep}
\end{procedure}

\begin{procedure}
	\caption{Uncoupled method BE-IDIC for identification of micromechanical parameter ratios.}
	\label{Sect:SS:Tab2}
	\vspace{-\topsep}
	\begin{tcolorbox}
		\centering
		\begin{enumerate}[1:]
			\item Experimental observations:
			\begin{itemize}[\textbf{-}]
				\item at each time step, record a microscopic image inside~$\Omega_\mathrm{fov}^\mathrm{m}$ with a high spatial resolution.
			\end{itemize}
			
			\item Construct a microscopic MVE model~$\vec{\mathcal{M}}^\mathrm{m}$ over~$\Omega_\mathrm{mve}^\mathrm{m} \equiv \Omega_\mathrm{roi}^\mathrm{m}$ (cf. Fig.~\ref{Sect:SS:Fig1a}).
			
			\item Perform IDIC, correlating on~$\Omega_\mathrm{roi}^\mathrm{m}$ while optimizing all material and kinematic DOFs (cf. Eqs.~\eqref{SubSect:SS:Eq1} and~\eqref{SubSect:SS:Eq2}).
			
		\end{enumerate}
	\end{tcolorbox}
	\vspace{-\topsep}
\end{procedure}

%
%----------------------------------
%	SUBSECTION SM Results
%----------------------------------
%
\subsubsection{Results}
\label{Ex:SS}
The uncoupled methods are designed to identify accurately material parameters at the microscale, assuming that one of the material parameters is a priori known to provide normalization. The micromechanical boundary conditions are accounted for accurately, whereby BE-IDIC outperforms GDIC-IDIC~\citep[cf.][]{Rokos:2018}. The achieved overall precision in terms of identified material parameter ratios is typically~$\eta_\mathrm{rms} \approx 0.3\,\%$ (results not shown), which is one order of magnitude more accurate compared to CMM and BSM in the multiscale regime, recall Fig.~\ref{Ex:CMM:Fig1}. This result is obtained by normalizing all parameters~$G_1$, $K_1$, and~$G_2$, relative to a fixed inclusions' bulk modulus~$K_2$. Corresponding micro-residual images are shown in Figs.~\ref{Ex:BSM:Fig1c} and~\ref{Ex:BSM:Fig1d}, resulting in a RMS grey value of~$0.02$.

\begin{figure}
	\centering
	\includegraphics[scale=1]{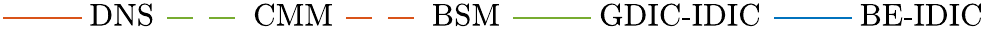}\\
	\vspace{-0.75em}
	\subfloat[horizontal component~$u_1(\xi)$]{\def\svgwidth{0.455\textwidth}%% Creator: Inkscape inkscape 0.92.3, www.inkscape.org
%% PDF/EPS/PS + LaTeX output extension by Johan Engelen, 2010
%% Accompanies image file 'displacements_ux.pdf' (pdf, eps, ps)
%%
%% To include the image in your LaTeX document, write
%%   \input{<filename>.pdf_tex}
%%  instead of
%%   \includegraphics{<filename>.pdf}
%% To scale the image, write
%%   \def\svgwidth{<desired width>}
%%   \input{<filename>.pdf_tex}
%%  instead of
%%   \includegraphics[width=<desired width>]{<filename>.pdf}
%%
%% Images with a different path to the parent latex file can
%% be accessed with the `import' package (which may need to be
%% installed) using
%%   \usepackage{import}
%% in the preamble, and then including the image with
%%   \import{<path to file>}{<filename>.pdf_tex}
%% Alternatively, one can specify
%%   \graphicspath{{<path to file>/}}
%% 
%% For more information, please see info/svg-inkscape on CTAN:
%%   http://tug.ctan.org/tex-archive/info/svg-inkscape
%%
\begingroup%
  \makeatletter%
  \providecommand\color[2][]{%
    \errmessage{(Inkscape) Color is used for the text in Inkscape, but the package 'color.sty' is not loaded}%
    \renewcommand\color[2][]{}%
  }%
  \providecommand\transparent[1]{%
    \errmessage{(Inkscape) Transparency is used (non-zero) for the text in Inkscape, but the package 'transparent.sty' is not loaded}%
    \renewcommand\transparent[1]{}%
  }%
  \providecommand\rotatebox[2]{#2}%
  \newcommand*\fsize{\dimexpr\f@size pt\relax}%
  \newcommand*\lineheight[1]{\fontsize{\fsize}{#1\fsize}\selectfont}%
  \ifx\svgwidth\undefined%
    \setlength{\unitlength}{212.5984252bp}%
    \ifx\svgscale\undefined%
      \relax%
    \else%
      \setlength{\unitlength}{\unitlength * \real{\svgscale}}%
    \fi%
  \else%
    \setlength{\unitlength}{\svgwidth}%
  \fi%
  \global\let\svgwidth\undefined%
  \global\let\svgscale\undefined%
  \makeatother%
  \begin{picture}(1,0.69333333)%
    \lineheight{1}%
    \setlength\tabcolsep{0pt}%
    \put(0,0){\includegraphics[width=\unitlength,page=1]{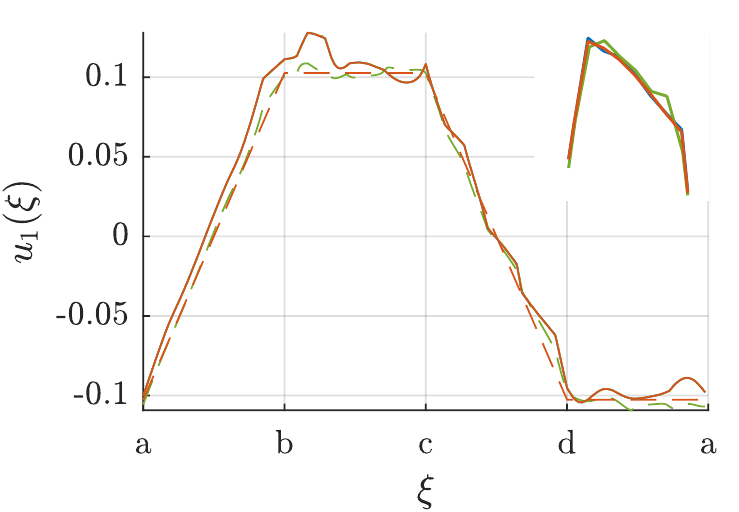}}%
    \put(0.78749191,0.38741577){\color[rgb]{0,0,0}\makebox(0,0)[lt]{\lineheight{1.25}\smash{\begin{tabular}[t]{l}{\footnotesize\setlength{\fboxsep}{1pt}\colorbox{white}{zoom}}\end{tabular}}}}%
    \put(0,0){\includegraphics[width=\unitlength,page=2]{displacements_ux.pdf}}%
  \end{picture}%
\endgroup%
\label{Ex:SS:Fig1a}}\hspace{1em}
	\subfloat[vertical component~$u_2(\xi)$]{\includegraphics[scale=1]{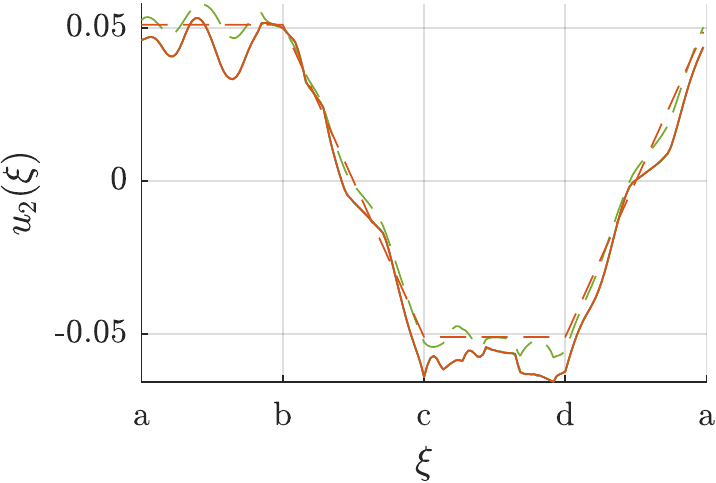}\label{Ex:SS:Fig1b}}
	\caption{Line traces of the displacement field~$\vec{u}(\xi) = u_1(\xi)\vec{e}_1 + u_2(\xi)\vec{e}_2$, as obtained for individual methods, compared against the exact DNS result. The parametric coordinate~$\xi$ sweeps along the ROI boundary~$\partial\Omega_\mathrm{roi}^\mathrm{m}$. The chosen ROI is positioned at the specimen's centre (cf. Fig.~\ref{Sect:VE:Fig1}) and has a size~$\ell_\mathrm{roi} = 4d$. The horizontal components~$u_1(\xi)$ are shown in~(a), whereas the vertical components~$u_2(\xi)$ in~(b).}
	\label{Ex:SS:Fig1}
\end{figure}

An example of displacement components~$u_1(\xi)$ and~$u_2(\xi)$ along the micro ROI boundary~$\partial\Omega_\mathrm{roi}^\mathrm{m}$, parametrized by~$\xi \in [0, 16]d$ for~$\ell_\mathrm{roi} = 4d$, is shown in Fig.~\ref{Ex:SS:Fig1}. The results show that, in contrast to CMM and BSM methods, uncoupled approaches (especially BE-IDIC) achieve practically indistinguishable profiles compared to the DNS results. Even the peak displacement~$u_1 \approx 0.1d$ of~$20$ pixels is captured accurately, see zoom inset in Fig.~\ref{Ex:SS:Fig1a}. Note that in contrast to other methods, the rigid body motion is accounted for automatically.

%
%-----------------------------------------------------------------------------
%	SECTION Stress Integration
%-----------------------------------------------------------------------------
%
\subsection{Stress Integration}
\label{Sect:SI}
%
%----------------------------------
%	SUBSECTION SI Methodology
%----------------------------------
%
\subsubsection{Methodology}
\label{SubSect:SI}
In sequential multiscale modelling, the microscale information, such as the first Piola--Kirchhoff stress tensor obtained from the uncoupled methods of Section~\ref{Sect:SM}, is computed and passed to the macroscale. By integrating homogenized stresses at the macroscale over~$\Omega$, the resulting macroscopic reaction forces are assembled and compared against experimentally observed data to provide proper normalization. In particular, the stress Integration~(SI) method, summarized in Proc.~\ref{Sect:SI:Tab1} and sketched in Fig.~\ref{Fig:methods_d}, employs experimental observations at both scales. At the microscale, images with sufficiently high resolution capturing all morphological features are observed (at specific locations, as detailed below). Microstructural IDIC identification is subsequently performed on~$\Omega_\mathrm{roi}^\mathrm{m}$ for all MVEs simultaneously to yield micromechanical parameter ratios by means of one of the uncoupled methods. This step also provides homogenized stresses for each MVE. At the macroscale, the reaction forces, applied displacements, and coarse images are exploited to provide normalization and strain fields.

Normalization of micromechanical parameters is achieved through six steps, as follows:
\begin{enumerate}[(i)]
	\item The Principle of Virtual Work~(PVW) is first employed, i.e.,
	\begin{equation}
	\underbrace{\int_{\Omega} \bs{P}_\mathrm{M}(\vec{X}_\mathrm{M}):\delta\bs{F}^\mathsf{T}_\mathrm{M}(\vec{X}_\mathrm{M})\,\mathrm{d}\vec{X}_\mathrm{M}}_
	{\delta W^\mathrm{int}}
	=
	\underbrace{\int_{\Gamma_\mathrm{D}} \vec{t}_\mathrm{M}(\vec{X}_\mathrm{M}) \cdot \delta\vec{u}_\mathrm{M}(\vec{X}_\mathrm{M}) \, \mathrm{d}\vec{X}_\mathrm{M}}_
	{\delta W^\mathrm{ext}},
	\label{eq:vw}
	\end{equation}
	where~$\Gamma_\mathrm{D} \subseteq \partial\Omega$ is the fixed part of the domain boundary along which the traction~$\vec{t}_\mathrm{M}$ is determined from the measured force (which can be done for a specific loading configuration such as a uni-axial tensile test), $\bs{P}_\mathrm{M}$ is the macroscopic first Piola--Kirchhoff stress tensor, $\delta\vec{u}_\mathrm{M}$ macroscopic virtual displacement field, and~$\delta\bs{F}_\mathrm{M}^\mathsf{T} = \delta(\bs{I}+\vec{\nabla}_\mathrm{M}\vec{u}_\mathrm{M}) = \vec{\nabla}_\mathrm{M}\delta\vec{u}_\mathrm{M}$ its gradient; note that $\vec{\nabla}_\mathrm{M} = \partial/\partial X_{\mathrm{M},i}$ denotes the gradient operator with respect to the macroscopic coordinates~$X_{\mathrm{M},i}$ in the reference configuration.
	
	\item The macroscopic stress field~$\bs{P}_\mathrm{M}$, entering the internal virtual work~$\delta W^\mathrm{int}$, is next approximated by its homogenized counterpart (available only point-wise, typically in Gauss integration points, and computed at the microscale through one of the uncoupled methods), i.e.,
	\begin{equation}
	\bs{P}_\mathrm{M}(\vec{X}_\mathrm{M}) 
	\approx
	\alpha\,\overline{\bs{P}}_\mathrm{m}(\vec{X}_\mathrm{M})
	=
	\frac{\alpha}{|\Omega_\mathrm{mve}^\mathrm{m}|}\int_{\Omega_\mathrm{mve}^\mathrm{m}} \bs{P}_\mathrm{m}(\vec{X}_\mathrm{M},\vec{X}_\mathrm{m}, \underline{\lambda})\,\mathrm{d}\vec{X}_\mathrm{m},
	\label{eq:stress}
	\end{equation}
	which is known only at the positions of the measured MVEs (denoted~$\vec{X}_\mathrm{M}$), and up to a yet to be identified normalization constant~$\alpha$.	
	
	\item The macroscopic virtual displacement~$\delta\vec{u}_\mathrm{M}$ and strain~$\delta\bs{F}_\mathrm{M}$ fields in~$\delta W^\mathrm{int}$ are approximated by the kinematics experimentally identified at the macroscale, i.e.,
	\begin{equation}
	\delta\bs{F}^\mathsf{T}_\mathrm{M}(\vec{X}_\mathrm{M}) \approx \vec{\nabla}_\mathrm{M}\vec{u}_\mathrm{M}(\vec{X}_\mathrm{M}).
	\label{eq:mode}
	\end{equation}
	The displacement field~$\vec{u}_\mathrm{M}$ is typically obtained by GDIC with globally or locally supported polynomials, or by IDIC with homogeneous material in which only the strain field is of interest.
	
	\item The continuous integration over~$\Omega$ in~$\delta W^\mathrm{int}$ of Eq.~\eqref{eq:vw} is approximated numerically considering a set of macroscopic Gauss integration points:
	\begin{equation}
	\begin{aligned}
	\delta W^\mathrm{int}
	= 
	\int_{\Omega} \bs{P}_\mathrm{M}(\vec{X}_\mathrm{M}) : \delta\bs{F}_\mathrm{M}^\mathsf{T}(\vec{X}_\mathrm{M})\,\mathrm{d}\vec{X}_\mathrm{M}
	&\approx
	\alpha\int_{\Omega} \overline{\bs{P}}_\mathrm{m}(\vec{X}_\mathrm{M}) : \underbrace{\vec{\nabla}_\mathrm{M}\vec{u}_\mathrm{M}(\vec{X}_\mathrm{M})}_{\bs{B}(\vec{X}_\mathrm{M})}\,\mathrm{d}\vec{X}_\mathrm{M} \\
	&\approx \alpha\underbrace{\sum_{i=1}^{n_\mathrm{g}}w_iJ_i \overline{\underline{P}}_{\mathrm{m},i}^\mathsf{T} \underline{B}_i }_{\delta\widetilde{W}^\mathrm{int}}.
	\end{aligned}
	\label{eq:integration}
	\end{equation}
	In Eq.~\eqref{eq:integration}, $\overline{\underline{P}}_{\mathrm{m},i}$ and~$\underline{B}_i$ are columns storing components of~$\overline{\bs{P}}_\mathrm{m}(\vec{X}_{\mathrm{M},i})$ and~$\bs{B}(\vec{X}_{\mathrm{M},i})$ at positions~$\vec{X}_{\mathrm{M},i}$ of~$n_\mathrm{g}$ adopted integration points with weights~$w_i$ and Jacobians~$J_i$. It is clear that the spatial positions of individual MVEs need to match with the particular integration scheme chosen, influencing the experimental set-up. Therefore, an integrated experimental--computational scheme results. Note that the approximation steps of Eqs.~\eqref{eq:stress}--\eqref{eq:integration} resemble a Reduced Order Modelling~(ROM) approach~\citep[see, e.g.,][]{Yvonnet:2007}, in which~$\bs{B}(\vec{X}_\mathrm{M})$ corresponds to a strain snapshot (or a mode) of the considered specimen with domain~$\Omega$. This also suggests that more efficient ROM integration techniques (compared to the Gauss integration scheme suggested here) can be employed~\citep[see, e.g.,][]{An:2008,Ryckelynck:2009,HERNANDEZ:2017,YANO:2019}. Some affinity with the Virtual Fields Method~(VFM) can be observed as well (cf., e.g., \citealt{VFM}), in which~$\bs{B}(\vec{X}_\mathrm{M})$ represents a single virtual field. This means that the sensitivity of the VFM on the choice of multiple virtual fields~\citep[as discussed, e.g., by][]{Rahmani:2014} is relaxed, as only one parameter needs to be identified (i.e., the normalization parameter~$\alpha$). 

	\item In analogy to the internal virtual work~$\delta W^\mathrm{int}$, the external virtual work~$\delta W^\mathrm{ext}$ is estimated based on available data (that is, experimentally measured applied macroscopic displacements and reaction forces), i.e.,
	\begin{equation}
	\delta W^\mathrm{ext}
	=
	\int_{\Gamma_\mathrm{D}} \vec{t}_\mathrm{M}(\vec{X}_\mathrm{M}) \cdot \delta\vec{u}_\mathrm{M}(\vec{X}_\mathrm{M}) \, \mathrm{d}\vec{X}_\mathrm{M}
	\approx
	\underbrace{\sum_{i=1}^{n_\mathrm{f}}\underline{F}_{\mathrm{exp},i}^\mathsf{T}\underline{u}_{\mathrm{D},i}}_{\delta\widetilde{W}^\mathrm{ext}}.
	\label{eq:external}
	\end{equation}
	Note that accurate experimental observations of tractions and displacements are not straightforward, if even possible. The formulation is, nevertheless, introduced here in the context of PVW for completeness in the form of the left-hand-side of Eq.~\eqref{eq:external}. In practice, force resultants are considered instead, see, e.g., \citep[Section~3.4]{Rethore:2013} or \citep[Section 1.2 and Chapter~2]{VFM}, whereas displacements are obtained through DIC or by using boundary enrichments (i.e., using virtual boundaries as described in Section~\ref{Sect:SM}, Eqs.~\eqref{SubSect:SS:Eq1}--\eqref{SubSect:SS:Eq2}). Within an experimental setting considered hereafter, the knowledge of~$\vec{t}_\mathrm{M}$ and~$\Gamma_\mathrm{D}$ under specific considerations reduces to a set of reaction force resultants~$\underline{F}_{\mathrm{exp},i}$ along with their corresponding displacements~$\underline{u}_{\mathrm{D},i}$, $i = 1, \dots, n_\mathrm{f}$.
	
	\item The optimal normalization constant is finally obtained by equating the approximate virtual internal and external work, providing
	\begin{equation}
	\alpha^\star = \frac{\delta\widetilde{W}^\mathrm{ext}}{\delta\widetilde{W}^\mathrm{int}}.
	\label{eq:normalization:SI}
	\end{equation}
\end{enumerate}

Unlike the uncoupled methods introduced in Section~\ref{Sect:SM}, it may furthermore become clear that individual MVEs should be representative in order to provide accurate homogenized macroscopic properties~$\overline{\bs{P}}_\mathrm{m}$. This may entail an increase in experimental and computational costs because larger portions of the microstructure need to be accurately scanned and modelled.
\begin{procedure}
	\caption{Stress Integration~(SI) method used for micromechanical parameter identification.}
	\label{Sect:SI:Tab1}
	\vspace{-\topsep}
	\begin{tcolorbox}
		\centering
		\begin{enumerate}[1:]
			\item Experimental observations:
			\begin{itemize}[\textbf{-}]
				
				\item at each load step, record a macroscopic image spanning~$\Omega_\mathrm{fov}^\mathrm{M}$ with low spatial resolution (no microstructure is needed), applied displacements~$\vec{u}_\mathrm{D}$, and reaction forces~$\vec{F}_\mathrm{exp}$,
				
				\item at each load step, also record a microscopic image inside~$\Omega_\mathrm{fov}^\mathrm{m}$ with high spatial resolution for all MVEs; individual MVEs are positioned close to Gauss integration points of the entire specimen~$\Omega$.
				
			\end{itemize}
			
			\item Use one of the uncoupled methods of Section~\ref{Sect:SM} to identify micromechanical parameter ratios (simultaneously for all MVEs to achieve the best accuracy).
			
			\item Determine all macroscopic quantities to be able to compute~$\delta\widetilde{W}^\mathrm{int}$ in Eq.~\eqref{eq:integration} and~$\delta\widetilde{W}^\mathrm{ext}$ in Eq.~\eqref{eq:external}. Compute the optimal normalization constant~$\alpha^\star$ from Eq.~\eqref{eq:normalization:SI}.
			
		\end{enumerate}
	\end{tcolorbox}
	\vspace{-\topsep}
\end{procedure}

The above description suggests that the main advantages of the SI are as follows:
\begin{itemize}\setlength{\itemsep}{0pt}\setlength{\parskip}{0pt}\setlength{\parsep}{0pt}
	
	\item No constraints are implied for MVE kinematic boundary conditions; all fluctuations are captured accurately (up to image noise present and MVE discretization used);
	
	\item The homogenized stresses~$\overline{\bs{P}}_\mathrm{m}$ correspond to the true deformation state of the underlying microstructure, as observed experimentally (no assumption on homogeneous deformation);
	
	\item The macroscopic strain mode~$\bs{B}$ results directly from the underlying physics at the corresponding scale, as it has been obtained experimentally.
	
\end{itemize}
On the contrary, several disadvantages need to be emphasized as well:
\begin{itemize}	\setlength{\itemsep}{0pt}\setlength{\parskip}{0pt}\setlength{\parsep}{0pt}
	
	\item The method requires substantial experimental effort, because~$n_\mathrm{g}$ MVEs need to be observed at a priori chosen positions (e.g., Gauss integration points);
	
	\item The experimentally observed strain~$\bs{B}$ and displacement~$\vec{u}_\mathrm{M}$ fields may introduce excessive inaccuracies due to image noise and systematic errors (as a result of, e.g., too coarse or fine GDIC mesh);
	
	\item The virtual displacement and strain fields should ideally be incremental to maintain a high accuracy (i.e., infinitesimally small), which is not always guaranteed, but can be improved using a time-incremental scheme.
	
\end{itemize}
Although the effect of noise and systematic error in GDIC can be mitigated, the number of required microstructural observations~$n_\mathrm{g}$ and force measurements may be too large for complex geometries to be experimentally feasible. To this end, an alternative methodology, described in Section~\ref{Sect:SIFE2} below, is proposed as well.
%
%----------------------------------
%	SUBSECTION SI Results
%----------------------------------
%
\subsubsection{Results}
\label{Ex:SI}
For the adopted simple tensile test example shown in Fig.~\ref{Sect:VE:Fig1}, the resulting experimental forces have only horizontal components~$ \vec{F}_\mathrm{exp} = \pm F_\mathrm{exp}\vec{e}_1 $, induced through prescribed displacements at both ends, $ \vec{u}_\mathrm{D} = \pm u_\mathrm{D}\vec{e}_1 $, where the displacements are constant along the two vertical edges. Substituting these assumptions into Eq.~\eqref{eq:external} yields for the variation of the external virtual work~$ \delta W^\mathrm{ext} = 2 F_\mathrm{exp} u_\mathrm{D}$, which eventually gives through Eq.~\eqref{eq:normalization:SI} the normalization constant
\begin{equation}
\alpha^\star = \frac{2 F_\mathrm{exp} u_\mathrm{D}}{\delta\widetilde{W}^\mathrm{int}},
\label{eq:normalization}
\end{equation}
where Eq.~\eqref{eq:integration} is used to compute the approximate variation of the internal virtual work, i.e., $\delta\widetilde{W}^\mathrm{int} = \sum_{i=1}^{n_\mathrm{g}}w_iJ_i \overline{\underline{P}}_{\mathrm{m},i}^\mathsf{T} \underline{B}_i$, Eq.~\eqref{SubSect:BSM:Eq3} to compute homogenized stresses~$\overline{\bs{P}}_{\mathrm{m},i}$, and~$w_i$ and~$J_i$ result from the adopted integration rule.

\begin{figure}
	\centering
	\subfloat[SI\textsubscript{idic}]{\includegraphics[scale=1]{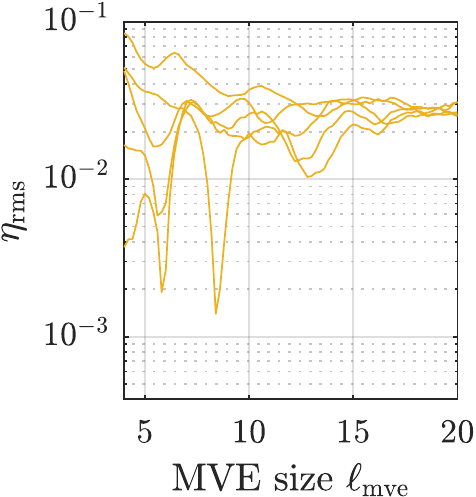}\label{Ex:SI:Fig3a}}\hspace{0.2em}
	\subfloat[SI\textsubscript{fe}]{\includegraphics[scale=1]{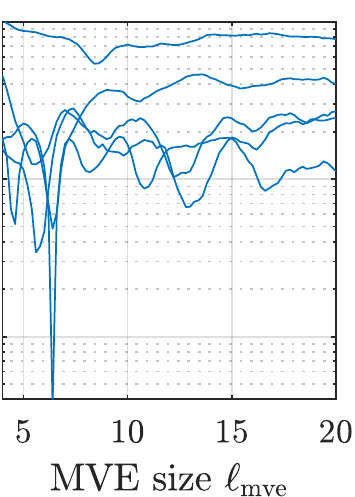}\label{Ex:SI:Fig3b}}\hspace{0.2em}
	\subfloat[SI\textsubscript{poly}]{\includegraphics[scale=1]{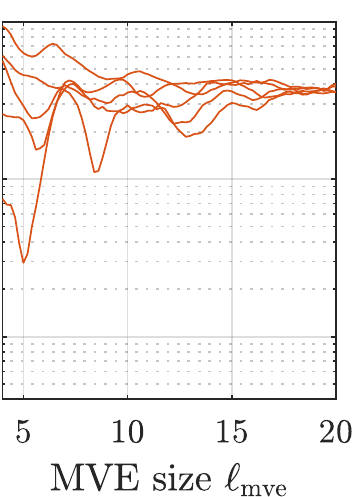}\label{Ex:SI:Fig3c}}\hspace{0.2em}
	\subfloat[SI\textsubscript{vf}]{\includegraphics[scale=1]{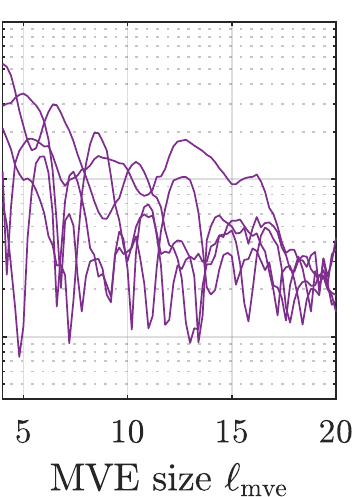}\label{Ex:SI:Fig3d}}
	\caption{Accuracy of the Stress Integration~(SI) method with respect to the size of the MVE~$ \ell_\mathrm{mve} \in [4, 20]d $. Four strain fields~$\bs{B}$ for the normalization have been used, obtained through (a)~IDIC with homogeneous material properties, (b)~GDIC with finite element interpolation functions (cf. employed mesh in Fig.~\ref{Ex:SI:Fig1e}), (c)~GDIC with polynomial functions of fifth order, and~(d) spatially constant virtual strain field of Eq.~\eqref{eq:vf}. Individual lines in each of the plots correspond to various MVE spatial positioning, as indicated in Fig.~\ref{Ex:SI:Fig1a}.}
	\label{Ex:SI:Fig3}
\end{figure}

\begin{figure}
	\centering
	\subfloat[$n_\mathrm{g} = 1$]{\def\svgwidth{0.284\textwidth}%% Creator: Inkscape inkscape 0.92.3, www.inkscape.org
%% PDF/EPS/PS + LaTeX output extension by Johan Engelen, 2010
%% Accompanies image file 'integration_SI_1g.pdf' (pdf, eps, ps)
%%
%% To include the image in your LaTeX document, write
%%   \input{<filename>.pdf_tex}
%%  instead of
%%   \includegraphics{<filename>.pdf}
%% To scale the image, write
%%   \def\svgwidth{<desired width>}
%%   \input{<filename>.pdf_tex}
%%  instead of
%%   \includegraphics[width=<desired width>]{<filename>.pdf}
%%
%% Images with a different path to the parent latex file can
%% be accessed with the `import' package (which may need to be
%% installed) using
%%   \usepackage{import}
%% in the preamble, and then including the image with
%%   \import{<path to file>}{<filename>.pdf_tex}
%% Alternatively, one can specify
%%   \graphicspath{{<path to file>/}}
%% 
%% For more information, please see info/svg-inkscape on CTAN:
%%   http://tug.ctan.org/tex-archive/info/svg-inkscape
%%
\begingroup%
  \makeatletter%
  \providecommand\color[2][]{%
    \errmessage{(Inkscape) Color is used for the text in Inkscape, but the package 'color.sty' is not loaded}%
    \renewcommand\color[2][]{}%
  }%
  \providecommand\transparent[1]{%
    \errmessage{(Inkscape) Transparency is used (non-zero) for the text in Inkscape, but the package 'transparent.sty' is not loaded}%
    \renewcommand\transparent[1]{}%
  }%
  \providecommand\rotatebox[2]{#2}%
  \newcommand*\fsize{\dimexpr\f@size pt\relax}%
  \newcommand*\lineheight[1]{\fontsize{\fsize}{#1\fsize}\selectfont}%
  \ifx\svgwidth\undefined%
    \setlength{\unitlength}{176.31496279bp}%
    \ifx\svgscale\undefined%
      \relax%
    \else%
      \setlength{\unitlength}{\unitlength * \real{\svgscale}}%
    \fi%
  \else%
    \setlength{\unitlength}{\svgwidth}%
  \fi%
  \global\let\svgwidth\undefined%
  \global\let\svgscale\undefined%
  \makeatother%
  \begin{picture}(1,0.33762057)%
    \lineheight{1}%
    \setlength\tabcolsep{0pt}%
    \put(0,0){\includegraphics[width=\unitlength,page=1]{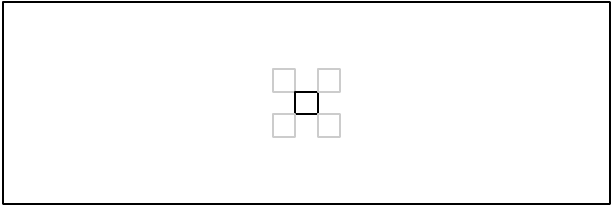}}%
    \put(0.57416627,0.14061734){\color[rgb]{0,0,0}\makebox(0,0)[lt]{\lineheight{1.25}\smash{\begin{tabular}[t]{l}{\footnotesize\setlength{\fboxsep}{1pt}\colorbox{white}{MVE~1}}\end{tabular}}}}%
  \end{picture}%
\endgroup%
\label{Ex:SI:Fig1a}}\hspace{2em}
	\subfloat[FE-GDIC mesh]{\includegraphics[scale=0.75]{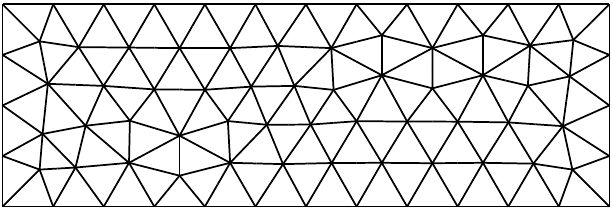}\label{Ex:SI:Fig1e}}\hspace{2em}
	\subfloat[FE\textsuperscript{2} mesh]{\includegraphics[scale=0.75]{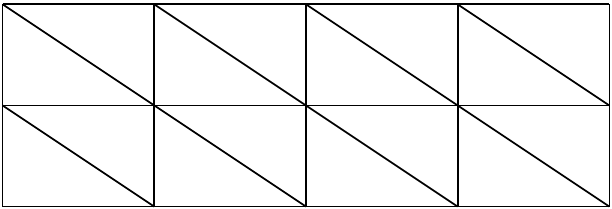}\label{Ex:SI:Fig1f}}
	\caption{(a)~Position of Gauss integration point as employed in the Stress Integration~(SI) method, MVE size~$\ell_\mathrm{mve} = 4d$ (four shifted RVEs are shown in grey). Macroscopic triangulation~(b) used in FE-GDIC for identification of the strain field~$\bs{B}(\vec{X}_\mathrm{M})$, using quadratic triangular elements, and~(c) triangulation used in FE\textsuperscript{2} with~$n_\mathrm{e} = 16$ elements.}
	\label{Ex:SI:Fig1}
\end{figure}

The results are summarized in terms of~$\eta_\mathrm{rms}$ in Fig.~\ref{Ex:SI:Fig3}, where three aspects of the SI methodology have been tested for the case of only a single integration point positioned at the specimen's centre, i.e., at $\vec{X}_\mathrm{M} = \vec{0}$. First, the influence of the macroscopic strain field~$\bs{B}(\vec{X}_\mathrm{M})$ is investigated by considering four options: 
\begin{enumerate}[(a)]
	\item Strain field obtained from macroscopic IDIC, which correlates inside~$\Omega_\mathrm{roi}^\mathrm{M}$ with the non-linear homogeneous hyper-elastic Neo--Hookean constitutive law of Eq.~\eqref{Sect:VE:Eq4} and the mesh of Fig.~\ref{Ex:CMM:Fig0a}, abbreviated SI\textsubscript{idic};
		
	\item Macroscopic GDIC with FE hat-shaped functions, employing quadratic triangular elements and the mesh shown in Fig.~\ref{Ex:SI:Fig1e}, referred to as SI\textsubscript{fe};
		
	\item Macroscopic GDIC with globally-supported smooth polynomials of fifth order, abbreviated as SI\textsubscript{poly};
		
	\item And a macroscopic homogeneous virtual strain field of the form
	\begin{equation}
		\bs{B}(\vec{X}) = \frac{2u_\mathrm{D}}{L}\vec{e}_1\vec{e}_1,
		\label{eq:vf}
	\end{equation}
	which in this simple case effectively provides normalization by the experimentally applied homogenized stress field~$\sigma_\mathrm{exp} = F_\mathrm{exp}/W$, i.e.,
	\begin{equation}
	\alpha^\star = \frac{\sigma_\mathrm{exp}}{\overline{P}_{\mathrm{m},11}},
	\label{eq:normalization:stress}
	\end{equation}
	referred to as SI\textsubscript{vf} (recall also the discussion of point (iv), Eq.~\eqref{eq:integration}, on the relation of SI to VFM).
\end{enumerate}
The results corresponding to the normalization through individual strain fields (a)--(d) are presented in Figs.~\ref{Ex:SI:Fig3a}--\ref{Ex:SI:Fig3d}. The best normalization in terms of~$\eta_\mathrm{rms}$ is achieved by the virtual strain field, i.e., SI\textsubscript{vf}, whereas the other options provide a lower accuracy. Note that results presented for the strain fields (a)--(c) have been obtained for only one time increment, which is not an optimal choice. A higher accuracy could be reached with a time incrementation scheme. 

The second aspect investigated is the influence of the MVE size~$\ell_\mathrm{mve}$, which has been assessed by considering~$ \ell_\mathrm{mve} \in [4, 20 ]d $. As can be observed in Fig.~\ref{Ex:SI:Fig3}, for small MVE sizes significant fluctuations in the achieved accuracy are present for all four strain fields (due to the lack of MVE representativeness, cf.~also Fig.~\ref{Sect:VE:Fig3}), for larger MVE sizes ($ \ell_\mathrm{mve} \geq 16 $) fluctuations in~$\eta_\mathrm{rms}$ rapidly decay.

Finally, sensitivity of spatial positioning of the MVE is investigated by considering five different positions depicted in Fig.~\ref{Ex:SI:Fig1a} (with shifts~$\Delta\vec{X} = \pm 4d\vec{e}_1 \pm 4d\vec{e}_2$), which might yield different results due to mild inhomogeneity of the considered deformation state of the specimen. Corresponding results are presented in Fig.~\ref{Ex:SI:Fig3} through individual lines which, except for the SI\textsubscript{fe} case that is sensitive to local GDIC fluctuations, reveal a typical error level with increasing~$\ell_\mathrm{mve}$.

In general we can conclude that the SI method provides a typical accuracy below~$1\,\%$ when the normalization through a virtual field is used, approaching for large MVEs a lower bound emanating from the adopted uncoupled methods (i.e., $0.3\,\%$ RMS). On the contrary, normalization through a virtual field can only be performed reliably for simple geometries, in which an appropriate virtual field can be established. This may not be possible for complex geometries, for which inhomogeneous stress and strain fields result. As an alternative, a normalization through a strain field obtained from the macroscopic IDIC scheme can be used instead, providing a typical accuracy of~$\eta_\mathrm{rms} = 3\,\%$.

Although the choice of a single integration point might seem overly simplified, it minimizes the experimental efforts required to perform identification. Multiple integration points could in principle be introduced for more accurate predictions, which is especially important in cases in which significantly varying macroscopic stress and strain fields are present. The positions and weights of integration points should be chosen optimally with respect to the integrand, i.e., the product~$\bs{P}_\mathrm{M}(\vec{X}_\mathrm{M}) : \bs{B}_\mathrm{M}(\vec{X}_\mathrm{M})$, see Eq.~\eqref{eq:integration}. However, for the simple geometry chosen here for the demonstration purposes (cf. Fig.~\ref{Sect:VE:Fig1}), a single integration point positioned at the centre of the specimen has already proven to be sufficient because of approximately constant strain field. For completeness, multiple macroscopic integration points have been considered (not reported here in full detail for brevity), using Gauss integration rule with~$n_\mathrm{g} \in \{4, 9\}$ integration points for a rectangular domain. The results are in both cases of higher accuracy compared to those of a single integration point, since in case of multiple integration points the identified forces are a weighted average over stresses evaluated at those integration points (or arithmetic average in the case of SI\textsubscript{vf} because the strain field is constant). This also leads to higher robustness of the identification in addition to higher accuracy. In particular, the highest error for~$n_\mathrm{g} = 9$ and~$\ell_\mathrm{mve} = 4d$ was $0.7\,\%$ RMS for SI\textsubscript{poly}, whereas the lowest error was $0.1\,\%$ RMS for SI\textsubscript{vf}. In both cases, this is a significant improvement, cf. Fig.~\ref{Ex:SI:Fig3}, although at the cost of additional experimental effort.

%
%-----------------------------------------------------------------------------
%	SECTION Computational Homogenization
%-----------------------------------------------------------------------------
%
\subsection{Computational Homogenization}
\label{Sect:SIFE2}
%
%----------------------------------
%	SUBSECTION SIFE2 Methodology
%----------------------------------
%
\subsubsection{Methodology}
\label{SubSect:FE2}
Normalization by FE\textsuperscript{2} method, sketched in Fig.~\ref{Fig:methods_e} and summarized in Proc.~\ref{Sect:FE2:Tab1}, requires in contrast to the SI method only a single experimental MVE observation. This is possible because deformation state of the microstructure associated with each macroscopic quadrature point is computed numerically rather than observed experimentally. This in turn allows through a macroscale model to estimate macroscopic reaction forces to be compared against experimental observations, and thus provides normalization of micromechanical parameters. Overall, the methodology proceeds as follows. First, material parameter ratios are extracted from a single MVE, using one of the uncoupled methods. Subsequently, the same MVE is assigned to all macroscopic Gauss integration points of a macroscopic model~$\vec{\mathcal{M}}^\mathrm{M}$ (assuming, however, that microstructural properties do not significantly vary in space). Periodic MVEs are adopted by using periodic MVE meshes, although the microstructural morphology is not necessarily periodic. Next, an FE\textsuperscript{2} simulation is performed in order to obtain the macroscopic displacement field and reaction forces, which provide normalization of the microstructural material parameter ratios upon comparison with~$\vec{F}_\mathrm{exp}$ through Eq.~\eqref{Subsect:DNS:Eq1}. Boundary conditions according to Fig.~\ref{Sect:VE:Fig1} are directly applied to the macroscopic model, by prescribing horizontal displacements at the two vertical edges to a given value while fixing all vertical displacements to zero. Note that when normalization through Eq.~\eqref{Subsect:DNS:Eq1} is not used, unscaled material parameters are identified, which typically means that one of the microstructural parameters is fixed to an arbitrary value.

Unlike the stress integration method, where the micro--macro loop is open and no iterations take place (cf. Fig.~\ref{Fig:methods_d}), computational homogenization requires equilibration of the macroscale system, and hence knowledge of the homogenized stresses~$\overline{\bs{P}}_\mathrm{m}$ (recall Eq.~\eqref{SubSect:BSM:Eq3}) as well as homogenized constitutive tangents~$\overline{\mathbb{C}}_\mathrm{m}$, if a Newton solver is used (recall Eq.~\eqref{SubSect:FE2:Eq1}). Due to the assumption on periodicity, normalization through FE\textsuperscript{2} may suffer from a higher inaccuracy compared to the stress integration method because of inexact displacements at the MVE boundary. Assumption on separation of scales adopted in the first-order FE\textsuperscript{2}, i.e., periodicity along with homogeneity of the underlying deformation~$\bs{F}_\mathrm{M}$ over each MVE, may thus limit the accuracy of the method. On the other hand, as the displacement field at the macroscale is now available, the macroscopic image residual can be constructed and the quality of the entire multiscale model can be assessed. When the adopted MVE is not representative enough, or the microstructural constitutive models are inadequate, one expects these deficiencies to emerge at the macroscale level through the macroscopic image residuals.

Note that although the first-order computational homogenization scheme has been adopted in this work as an example, any other multiscale approach (which may be more relevant for a given microstructure) can be used as well, including second-order computational homogenization or micro-morphic schemes \citep[see, e.g.,][]{Kouznetsova:2004,Biswas:2017,Rokos:2018:micromorphic}.
\begin{procedure}
	\caption{Computational homogenization~(FE\textsuperscript{2}) used for micromechanical parameter identification.}
	\label{Sect:FE2:Tab1}
	\vspace{-\topsep}
	\begin{tcolorbox}
		\centering
		\begin{enumerate}[1:]
			\item Obtain microscopic images inside~$\Omega_\mathrm{fov}^\mathrm{m}$ with high spatial resolution (for a single MVE only).
			
			\item Use one of the uncoupled methods to identify material parameter ratios.
			
			\item Build a mesh of the macroscopic specimen~$\Omega$ and set up a macroscale model~$\vec{\mathcal{M}}^\mathrm{M}$. With each Gauss integration point, $i = 1, \dots, n_\mathrm{g}$, associate the experimentally obtained MVE and perform FE\textsuperscript{2} simulation (using periodic BCs).
			
			\item Minimize the difference between macroscopic forces predicted by FE\textsuperscript{2} through Eq.~\eqref{Subsect:DNS:Eq1} to normalize material parameter ratios.
			
			\item If macroscopic observations inside~$\Omega_\mathrm{fov}^\mathrm{M}$ are available in addition to Step~$1$, macroscopic residual images may help to reveal any shortcomings of the multiscale model used.
			
		\end{enumerate}
	\end{tcolorbox}
	\vspace{-\topsep}
\end{procedure}
%
%----------------------------------
%	SUBSECTION SIFE2 Results
%----------------------------------
%
\subsubsection{Results}
\label{Ex:FE2}
The performance of the computational homogenization scheme is tested for various numbers of macroelements~$n_\mathrm{e} \in \{2,4,16,24,48\}$, and for two types of shape interpolation functions: linear and quadratic. In the case of linear triangles, one integration point per element is used, whereas in the case of quadratic triangles, three Gauss integration points are considered. Regular triangulations with isoparametric elements are employed, as shown in Fig.~\ref{Ex:SI:Fig1f} for~$n_\mathrm{e} = 16$. Finer or coarser discretizations use a similar mesh with right-angled triangles. The displacement field~$\vec{u}_\mathrm{M}(\vec{X}_M)$ is obtained using these discretizations which solve for the equilibrium of the macroscopic system, while all nodes on the two vertical lateral edges are considered with known prescribed displacements (for both linear as well as quadratic elements).

\begin{figure}
	\centering
	\subfloat[$\eta_\mathrm{rms}$, linear elements]{\includegraphics[scale=1]{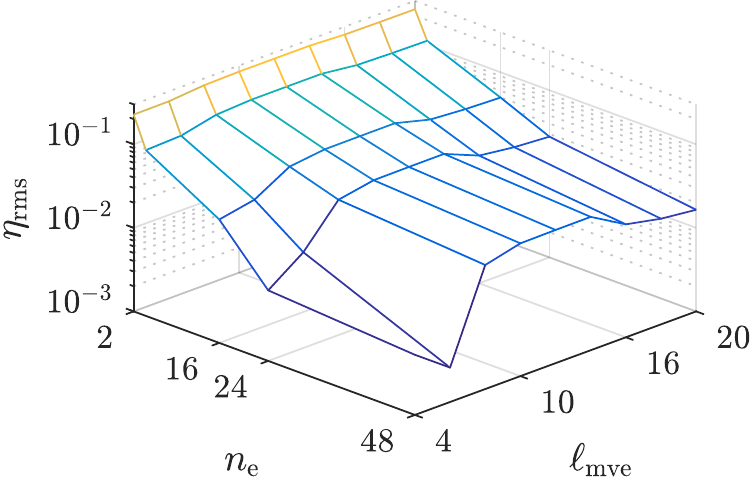}\label{Ex:FE2:Fig1a}}\hspace{1em}
	\subfloat[$\eta_\mathrm{rms}$, quadratic elements]{\includegraphics[scale=1]{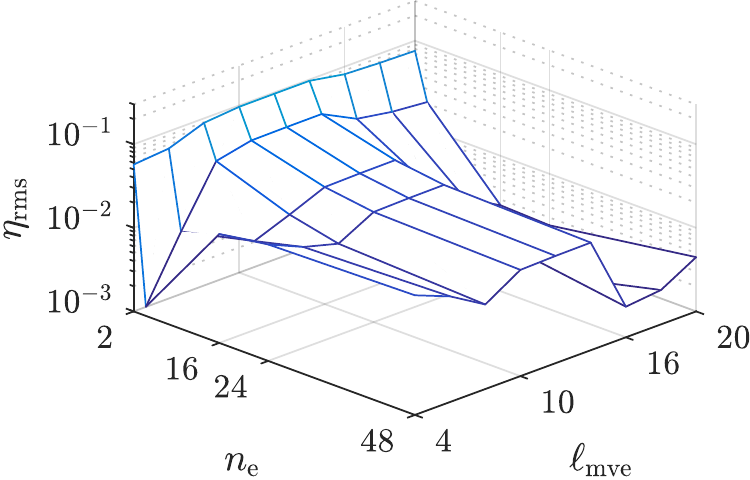}\label{Ex:FE2:Fig1b}}\\
	\subfloat[macroscopic grey residual, linear elements]{\includegraphics[scale=1]{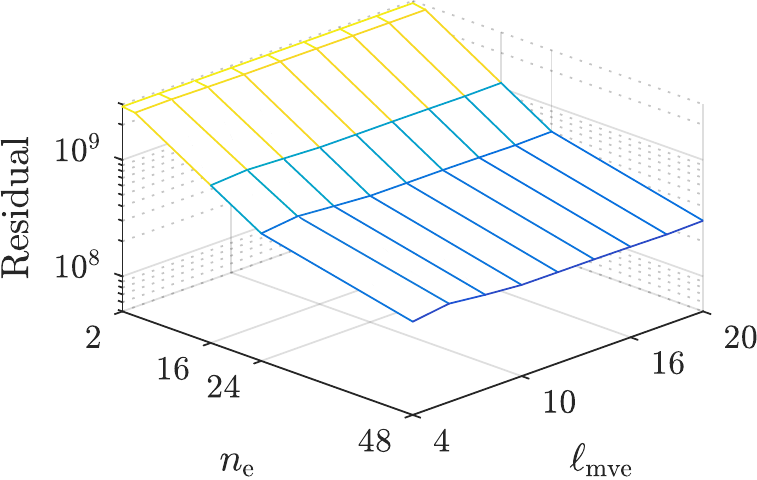}\label{Ex:FE2:Fig1c}}\hspace{1em}
	\subfloat[macroscopic grey residual, quadratic elements]{\includegraphics[scale=1]{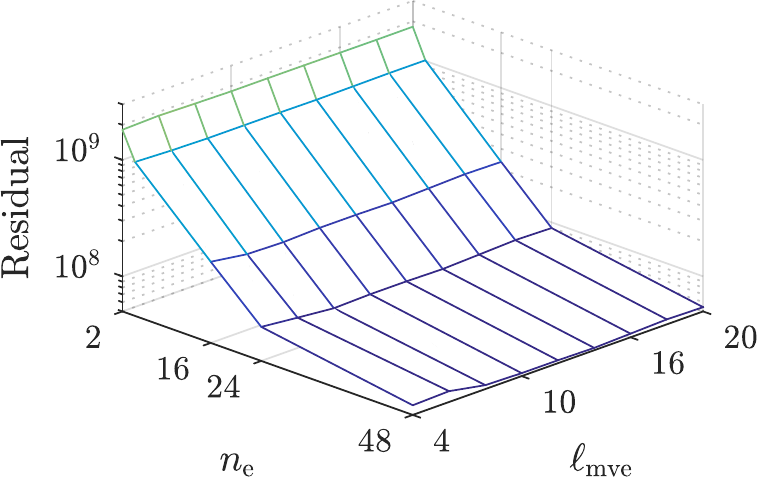}\label{Ex:FE2:Fig1d}}
	\caption{Results for the computational homogenization~(FE\textsuperscript{2}) scheme in connection with BE-IDIC for micromechanical parameter identification as a function of the number of macro-elements~$n_\mathrm{e} \in \{ 2, 4, 16, 24, 48 \}$ and size of the MVE~$\ell_\mathrm{mve} \in \{ 4, 6, 8, 10, 12, 14, 16, 18, 20 \}d$. RMS error~$\eta_\mathrm{rms}$ for~(a) linear, and~(b) quadratic triangular macro-elements. Macro residual of Eq.~\eqref{eq:Mresidual} for~(c) linear, and~(d) quadratic triangular macro-elements.}
	\label{Ex:FE2:Fig1}
\end{figure}

The resulting accuracy is shown in Figs.~\ref{Ex:FE2:Fig1a} and~\ref{Ex:FE2:Fig1b} in terms of~$\eta_\mathrm{rms}$, where a clear correlation exists between the non-monotonic trend of~$\eta_\mathrm{rms}$ data along the~$\ell_\mathrm{mve}$-axis and the MVE representativeness of Fig.~\ref{Sect:VE:Fig3a}. The relatively large errors indicate that the normalization step through the force measurements is responsible for most of the errors observed, as the BE-IDIC used at the microscale level is again within~$0.3\,\%$ RMS error in terms of material parameter ratios, even for only a single MVE. The results further show that a relatively large kinematic freedom needs to be provided to the macroscopic system (quadratic interpolation with relatively large number of elements) in order to achieve good accuracy and to capture inhomogeneity of the overall deformation. Average error for the finest macroscopic triangulations correspond to~$2$ and~$3\,\%$ or RMS error for~$48$ linear and quadratic elements.

\begin{figure}
	\centering
	\scalebox{0.75}{
		\def\svgwidth{1.1\textwidth}
		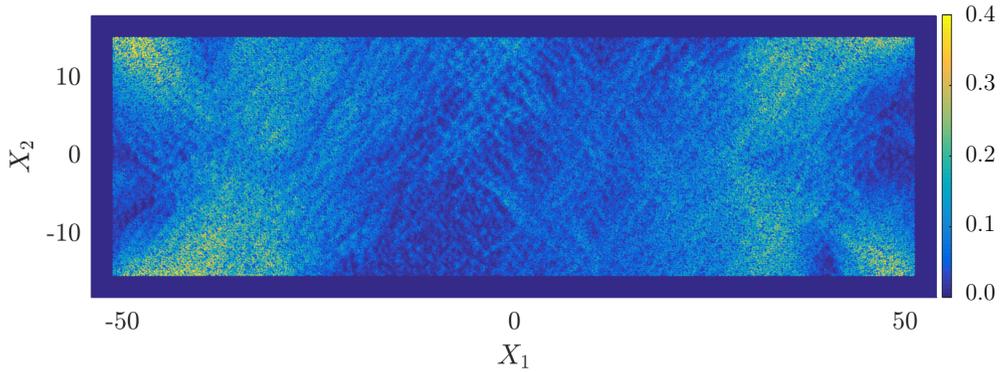}
	\caption{Macroscopic image residual~$| f_\mathrm{M}-g_\mathrm{M} |(\vec{X}_\mathrm{M})/2^8$ (normalized by dynamic contrast~$2^8$) corresponding to computational homogenization~(FE\textsuperscript{2}) with~$n_\mathrm{e} = 16$ quadratic triangular macro-elements and~$\ell_\mathrm{mve} = 6d$ ($0.07$ RMS in grey values).}
	\label{Ex:FE2:Fig2}
\end{figure}

Because for FE\textsuperscript{2} the displacement field~$\vec{u}_\mathrm{M}(\vec{X}_\mathrm{M})$ is available at the macroscale, the image residual inside~$\Omega_\mathrm{roi}^\mathrm{M}$ can be computed. This is shown for~$n_\mathrm{e} = 16$ and quadratic elements in Fig.~\ref{Ex:FE2:Fig2}, indicating that close to the corner points the residual is relatively poor (values are as high as~$40\,\%$). Finally, the macro-residual, i.e.,
\begin{equation}
\frac{1}{2}\| f_\mathrm{M}(\vec{X}_\mathrm{M})-g_\mathrm{M}(\vec{X}_\mathrm{M}+\vec{u}_\mathrm{M}(\vec{X}_\mathrm{M})) \|^2_{\Omega^\mathrm{M}_\mathrm{roi}},
\label{eq:Mresidual}
\end{equation}
is reported for each FE\textsuperscript{2} configuration in Figs.~\ref{Ex:FE2:Fig1c} and~\ref{Ex:FE2:Fig1d}, revealing a clear decrease with increasing number of macro-elements~$n_\mathrm{e}$. However, except for this general trend, no correlation of the image residual (Figs.~\ref{Ex:FE2:Fig1c} and~\ref{Ex:FE2:Fig1d}) with the accuracy of the FE\textsuperscript{2} method (Figs.~\ref{Ex:FE2:Fig1a} and~\ref{Ex:FE2:Fig1b}) can be observed, meaning that exact reproduction of the macroscopic displacement field does not necessarily guarantee accurate micromechanical material parameter identification for finite separation of scales in random heterogeneous microstructures.

%
%-----------------------------------------------------------------------------
%	SECTION Comparison
%-----------------------------------------------------------------------------
%
\section{Comparison}
\label{Sect:comparison}
%
%----------------------------------
%	SUBSECTION Comparison of Individual Methods
%----------------------------------
%
\subsection{Comparison of Individual Methods}
\label{Ex:Comparison}
A short overview of all methods is compiled in Tab.~\ref{Ex:Comparison:Tab1} along with parameters such as size of the microscopic ROI ($\ell_\mathrm{roi}$), fully resolved region ($\ell_\mathrm{full}$), MVE size used ($\ell_\mathrm{mve}$), and the number of Gauss integration points ($n_\mathrm{g}$), or number of elements ($n_\mathrm{e}$). These results have been chosen such that they reflect the typical results of each of the multiscale methods, which are compared to the DNS results serving as the reference. In three cases (DNS, CMM, and BSM), the same size of the ROI region has been used ($\ell_\mathrm{roi} = 4d$), whereas for SI and FE\textsuperscript{2} the size of the ROI is the same as the MVE size, i.e.~$\ell_\mathrm{roi} = \ell_\mathrm{mve}$.

The Concurrent Multiscale Method~(CMM) provides a relatively stable accuracy of approximately~$7\,\%$ for the entire range of the fully resolved region~$4d \leq \ell_\mathrm{full} \leq 20d$, which is still in a multiscale regime. Beyond this, once the entire height of the specimen is fully resolved (i.e., $\ell_\mathrm{full} \geq W = 36d$), a much better accuracy of~$0.2\,\%$ can be achieved.

The Bridging Scale Method~(BSM) has a comparable performance to CMM in spite of the fact that all fluctuations at the MVE boundary~$\partial\Omega_\mathrm{mve}^\mathrm{m}$ are neglected (recall Fig.~\ref{Ex:SS:Fig1}). The overall accuracy corresponds to approximately~$\overline{\eta}_\mathrm{rms} \approx 8\,\%$ of RMS error for~$4d \leq \ell_\mathrm{mve} \leq 20d$, and the best result is~$\eta_\mathrm{rms}(\ell_\mathrm{mve} = 16.6d) = 1.8\,\%$ RMS. In contrast to the CMM method, $\ell_\mathrm{mve} \geq 36d$ does not seem to yield a higher accuracy of identification. BSM is generally highly sensitive to~$\ell_\mathrm{mve}$, which makes it impossible to choose an optimal~$\ell_\mathrm{mve}$, and hence errors as high as~$20\,\%$ should be expected. It is worth mentioning that BSM is quite similar to SI\textsubscript{idic}, neglecting, however, heterogeneities at the MVE boundary~$\partial\Omega_\mathrm{mve}^\mathrm{m}$. This seems to be at the root of its lower accuracy, as SI\textsubscript{idic} performs almost three times better.

The sequential multiscale method performs adequately, especially in combination with a virtual strain field~(SI\textsubscript{vf}), which achieves an accuracy as low as~$1 \,\%$, although the material parameter \underline{ratios} can be identified more accurately through uncoupled methods (error as low as~$0.3\,\%$ of RMS). The normalization step is responsible for the increase in the error, mainly due to the lack of MVE representativeness. The presented result is therefore achievable under the condition that MVEs used are representative. The high accuracy is explained by the fact that the maximum of experimentally observed data is used (micro- as well as macro-images, and force and displacement measurements at the macroscale), which reflect the actual underlying physics. When the strain field obtained through a macroscopic IDIC is used, the overall error increases to~$3\,\%$ of RMS.

Computational homogenization~(FE\textsuperscript{2}) provides a relatively good accuracy while saving experimental efforts compared to SI, as it only requires a single MVE even for complex structures. On the contrary, relatively large computational efforts (fine macro-discretization) need to be used to provide an accurate normalization. Similarly to the SI method, high accuracy can be achieved only with a representative MVE. Note that in that case (i.e., $\ell_\mathrm{mve} \geq 16d$ according to Fig.~\ref{Ex:Noise:Fig1a}), combined area of all MVEs considered is comparable to the area of the entire specimen, when considering~$n_\mathrm{e} = 16$ and linear triangular elements as shown in Fig.~\ref{Ex:SI:Fig1f}. Although computational effort scales directly with the number of macroscopic Gauss integration points~$n_\mathrm{g}$, this is still an efficient approach, since only one MVE needs to be observed experimentally, thus substantially saving experimental efforts in comparison to SI. Overall result achieved corresponds to approximately~$2\,\%$ of RMS.

\begin{table}
	\centering
	\caption{Typical RMS errors (corresponding to Eq.~\eqref{Ex:DNS:Eq1}) in the identified micromechanical parameters for all tested methodologies (as detailed in Sections~\ref{Sect:DNS}--\ref{Sect:NM}) considered for~$ \ell_\mathrm{mve}, \ell_\mathrm{full} \in [4, 20]d $, and for two levels of image noise~$\zeta$.}
	\renewcommand{\arraystretch}{1.1}
	\begin{tabular}{c|r@{}l|r@{}l|r@{}l|r@{}l|r@{}l}
		Method  &  \multicolumn{2}{c|}{DNS}   & \multicolumn{2}{c|}{CMM} & \multicolumn{2}{c|}{BSM} & \multicolumn{2}{c|}{SI\textsubscript{vf}} & \multicolumn{2}{c}{FE\textsuperscript{2}} \\
		$\zeta\ [\%]$  &  \multicolumn{1}{c}{$0$} & \multicolumn{1}{c|}{$2$}&  \multicolumn{1}{c}{$0$} & \multicolumn{1}{c|}{$2$}&  \multicolumn{1}{c}{$0$} & \multicolumn{1}{c|}{$2$}&  \multicolumn{1}{c}{$0$} & \multicolumn{1}{c|}{$2$}&  \multicolumn{1}{c}{$0$} & \multicolumn{1}{c}{$2$}\\\hline
		$ \eta_\mathrm{rms} $ [\%] & \multicolumn{1}{c}{$0.03$} & \multicolumn{1}{c|}{$0.05$} &\multicolumn{1}{c}{$7.0$} & \multicolumn{1}{c|}{$7.1$} &\multicolumn{1}{c}{$8.0$} & \multicolumn{1}{c|}{$8.0$} &\multicolumn{1}{c}{$1.0$} & \multicolumn{1}{c|}{$1.1$} &\multicolumn{1}{c}{$2.0$} &\multicolumn{1}{c}{$2.1$}\\
		\multirow{3}{*}{Parameters} & \multicolumn{2}{c|}{---} & $\ell_\mathrm{full}$ & $= 5d$ & $\ell_\mathrm{mve}$ & $ = 6.8d$ & $\ell_\mathrm{mve}$ & $ = 6.4d$ & $\ell_\mathrm{mve}$ & $ = 10d$ \\
		&  $\ell_\mathrm{roi}$ & $ = 4d$   & $\ell_\mathrm{roi}$ & $ = 4d$ & $\ell_\mathrm{roi}$ & $ = 4d$ & $\ell_\mathrm{roi}$ & $ = 6.4d$ & $\ell_\mathrm{roi} $ & $= 10d$ \\
		& \multicolumn{2}{c|}{---} & \multicolumn{2}{c|}{---} & \multicolumn{2}{c|}{---} & $n_\mathrm{g}$ & $ = 1$ & $n_\mathrm{e}^\mathrm{quad}$ & $ = 48$
	\end{tabular}
	\label{Ex:Comparison:Tab1}
\end{table}
%
%----------------------------------
%	SUBSECTION Image Noise Study
%----------------------------------
%
\subsection{Image Noise Study}
\label{Ex:Noise}
\begin{figure}
	\centering
	\includegraphics[scale=1]{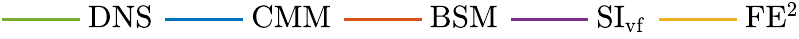}\vspace{-0.5em}\\
	\subfloat[comparison in multiscale regime]{\includegraphics[scale=1]{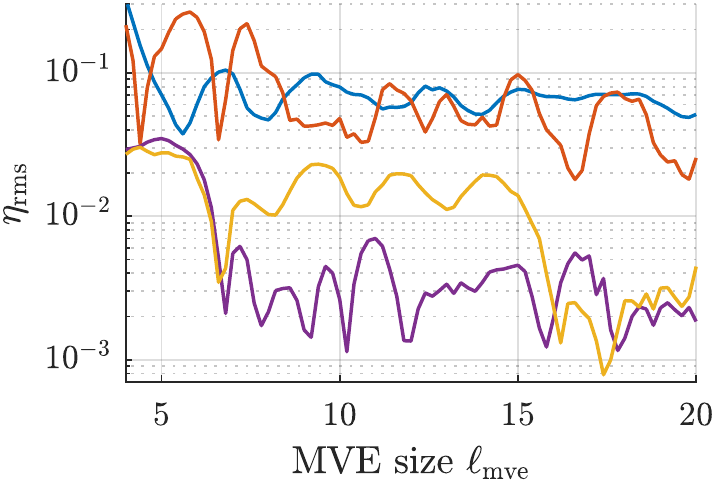}\label{Ex:Noise:Fig1a}}\hspace{2em}
	\subfloat[image noise study]{\includegraphics[scale=1]{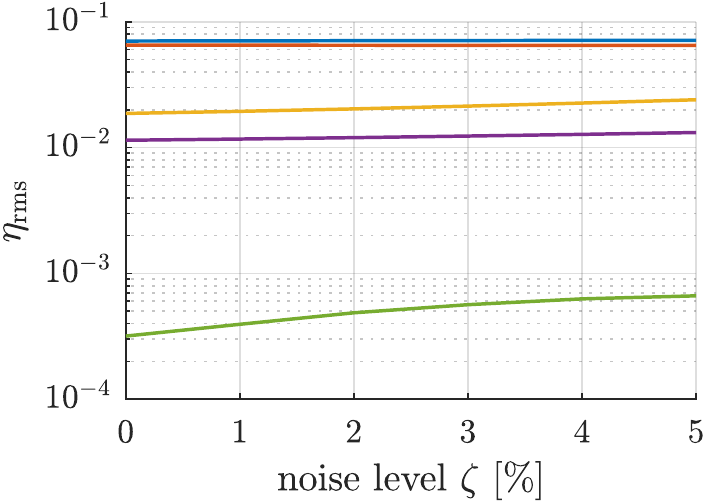}\label{Ex:Noise:Fig1b}}
	\caption{(a) Comparison of individual methods in the multiscale regime ($\ell_\mathrm{mve} \leq 20d$). (b) Accuracy in terms of the RMS error of Eq.~\eqref{Ex:DNS:Eq1} achieved by the individual multiscale identification methods, expressed as a function of added image noise~$\zeta \in [0, 0.05]\,\%$ of the dynamic image contrast~$2^8$ for the configurations of Tab.~\ref{Ex:Comparison:Tab1}. The force measurements are assumed noise-free.}
	\label{Ex:Noise:Fig1}
\end{figure}

The robustness of individual multiscale methods is evaluated through an image noise study by adding Gaussian white noise with a standard deviation~$\zeta \in [0, 0.05]\,\%$ of the image dynamic range to all (virtual) experimental observations. The force measurements are, however, assumed noise-free. The obtained results are summarized in Fig.~\ref{Ex:Noise:Fig1b} for the configurations reported in Tab.~\ref{Ex:Comparison:Tab1}. The results show that whereas CMM and BSM methods are practically insensitive to image noise 
(relative to their overall accuracy), DNS, SI\textsubscript{vf}, and FE\textsuperscript{2} reveal a mild dependence on~$\zeta$.

%
%-----------------------------------------------------------------------------
%	SECTION Summary and Conclusion
%-----------------------------------------------------------------------------
%
\section{Summary and Conclusions}
\label{Sect:Conclusion}
In this paper, a comparison of various multiscale parameter identification methodologies has been performed, aiming for micromechanical parameter identification with Integrated Digital Image Correlation~(IDIC). A theoretical overview of the adopted techniques with their description was given, combined with numerical assessments on the basis of virtual experiments. The convergence and performance of individual methods was investigated, including an image noise study to test the robustness. The main results of this contribution can be summarized as follows:
\begin{enumerate}
	\item The Concurrent Multiscale Method is robust with respect to the size of the fully resolved region and image noise. However, its accuracy is lower ($7\,\%$ RMS) than other multiscale techniques.

	\item The Bridging Scale Method suffers from the assumed homogeneous boundary conditions along the adopted Microstructural Volume Element~(MVE) boundary. This results in an accuracy that slowly improves with increasing size of the MVE reaching, nevertheless, a stable accuracy of~$8\,\%$ RMS error.
	
	\item The uncoupled methods provide only micromechanical parameter \underline{ratios} with an accuracy that is within~$0.3\,\%$ RMS error.
		
	\item The normalization of micromechanical parameters by Stress Integration provides an overall RMS error of~$1\,\%$. A high correlation with the representativeness of the MVEs as well as a mild sensitivity to image noise has been observed.
	
	\item Computational homogenization is of comparable accuracy to Stress Integration, i.e., of~$2\,\%$ RMS error. High requirements for the multiscale model in terms of kinematic freedom need to be satisfied to achieve such a result (i.e., a fine macroscopic mesh), although less experimental effort is required. Some dependence on MVE representativeness and sensitivity with respect to image noise is also observed.

\end{enumerate}

Throughout this contribution, a 2D plane-strain configuration was considered, neglecting any inaccuracies due to out-of-plane motion and subsurface effects. Furthermore, the micro-structural morphology and constitutive model were assumed to be known exactly, while in reality they may introduce significant additional errors. Considerably lower accuracies of all tested methodologies may, therefore, be expected in real in-situ experiments compared to the results reported here for virtual experiments. We suggest that in such cases the most accurate multiscale methodology would be used, although approaches capable of addressing respective sources of inaccuracies should be considered as well, including digital height correlation or digital volume correlation in case of 3D or subsurface effects, for instance.
%
%-----------------------------------------------------------------------------
%	APPENDIX A
%-----------------------------------------------------------------------------
% \appendix
%
% \section{Equivalence of Approach~I and~II}
% \label{Sect:A}
%
%-----------------------------------------------------------------------------
%	ACKNOWLEDGEMENTS
%-----------------------------------------------------------------------------
%
\section*{Acknowledgements}
The research leading to these results has received funding from the European Research Council under the European Union's Seventh Framework Programme (FP7/2007-2013)/ERC grant agreement \textnumero~[339392].
%
%-----------------------------------------------------------------------------
%	REFERENCES
%-----------------------------------------------------------------------------
%
% \section*{References}
\bibliography{mybibfile}
\end{document}

\centering
\def\svgwidth{0.7\textwidth}
%% Creator: Inkscape inkscape 0.92.3, www.inkscape.org
%% PDF/EPS/PS + LaTeX output extension by Johan Engelen, 2010
%% Accompanies image file 'sketch_SI_ink.pdf' (pdf, eps, ps)
%%
%% To include the image in your LaTeX document, write
%%   \input{<filename>.pdf_tex}
%%  instead of
%%   \includegraphics{<filename>.pdf}
%% To scale the image, write
%%   \def\svgwidth{<desired width>}
%%   \input{<filename>.pdf_tex}
%%  instead of
%%   \includegraphics[width=<desired width>]{<filename>.pdf}
%%
%% Images with a different path to the parent latex file can
%% be accessed with the `import' package (which may need to be
%% installed) using
%%   \usepackage{import}
%% in the preamble, and then including the image with
%%   \import{<path to file>}{<filename>.pdf_tex}
%% Alternatively, one can specify
%%   \graphicspath{{<path to file>/}}
%% 
%% For more information, please see info/svg-inkscape on CTAN:
%%   http://tug.ctan.org/tex-archive/info/svg-inkscape
%%
\begingroup%
  \makeatletter%
  \providecommand\color[2][]{%
    \errmessage{(Inkscape) Color is used for the text in Inkscape, but the package 'color.sty' is not loaded}%
    \renewcommand\color[2][]{}%
  }%
  \providecommand\transparent[1]{%
    \errmessage{(Inkscape) Transparency is used (non-zero) for the text in Inkscape, but the package 'transparent.sty' is not loaded}%
    \renewcommand\transparent[1]{}%
  }%
  \providecommand\rotatebox[2]{#2}%
  \newcommand*\fsize{\dimexpr\f@size pt\relax}%
  \newcommand*\lineheight[1]{\fontsize{\fsize}{#1\fsize}\selectfont}%
  \ifx\svgwidth\undefined%
    \setlength{\unitlength}{375bp}%
    \ifx\svgscale\undefined%
      \relax%
    \else%
      \setlength{\unitlength}{\unitlength * \real{\svgscale}}%
    \fi%
  \else%
    \setlength{\unitlength}{\svgwidth}%
  \fi%
  \global\let\svgwidth\undefined%
  \global\let\svgscale\undefined%
  \makeatother%
  \begin{picture}(1,0.6)%
    \lineheight{1}%
    \setlength\tabcolsep{0pt}%
    \put(0,0){\includegraphics[width=\unitlength,page=1]{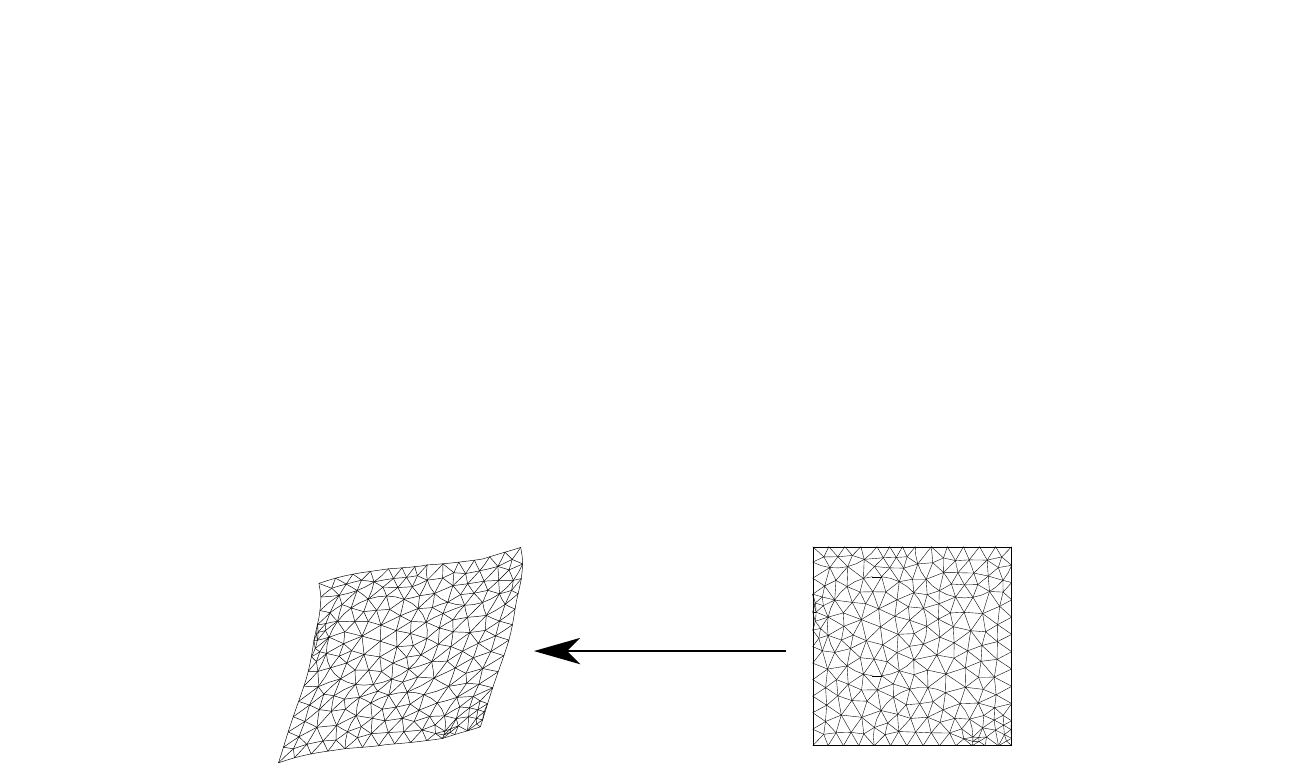}}%
    \put(0.454,0.10999999){\color[rgb]{0,0,0}\makebox(0,0)[lt]{\lineheight{0}\smash{\begin{tabular}[t]{l}BE-IDIC\end{tabular}}}}%
    \put(0.624,0.18999999){\color[rgb]{0,0,0}\makebox(0,0)[lt]{\lineheight{0}\smash{\begin{tabular}[t]{l}image $f_{\mathrm{m},i}$\end{tabular}}}}%
    \put(0.244,0.18999999){\color[rgb]{0,0,0}\makebox(0,0)[lt]{\lineheight{0}\smash{\begin{tabular}[t]{l}image $g_{\mathrm{m},i}$\end{tabular}}}}%
    \put(0,0){\includegraphics[width=\unitlength,page=2]{sketch_SI_ink.pdf}}%
    \put(0.33,0.33999999){\color[rgb]{0,0,0}\makebox(0,0)[lt]{\lineheight{0}\smash{\begin{tabular}[t]{l}$\overline{\bs{P}}_\mathrm{m}$\end{tabular}}}}%
    \put(0,0){\includegraphics[width=\unitlength,page=3]{sketch_SI_ink.pdf}}%
    \put(0.04,0.36999999){\color[rgb]{0,0,0}\makebox(0,0)[lt]{\lineheight{0}\smash{\begin{tabular}[t]{l}$\vec{u}_\mathrm{D}$\end{tabular}}}}%
    \put(0.9,0.36999999){\color[rgb]{0,0,0}\makebox(0,0)[lt]{\lineheight{0}\smash{\begin{tabular}[t]{l}$\vec{u}_\mathrm{D}$\end{tabular}}}}%
    \put(0.04,0.44999999){\color[rgb]{0,0,0}\makebox(0,0)[lt]{\lineheight{0}\smash{\begin{tabular}[t]{l}$\vec{F}_\mathrm{sim}$\end{tabular}}}}%
    \put(0.9,0.44999999){\color[rgb]{0,0,0}\makebox(0,0)[lt]{\lineheight{0}\smash{\begin{tabular}[t]{l}$\vec{F}_\mathrm{sim}$\end{tabular}}}}%
    \put(0.45,0.56999999){\color[rgb]{0,0,0}\makebox(0,0)[lt]{\lineheight{0}\smash{\begin{tabular}[t]{l}image $f_\mathrm{M}$\end{tabular}}}}%
    \put(0.45,0.41999999){\color[rgb]{0,0,0}\makebox(0,0)[lt]{\lineheight{0}\smash{\begin{tabular}[t]{l}image $g_\mathrm{M}$\end{tabular}}}}%
    \put(0.6,0.49999999){\color[rgb]{0,0,0}\makebox(0,0)[lt]{\lineheight{0}\smash{\begin{tabular}[t]{l}$\Omega$\end{tabular}}}}%
    \put(0.794,0.08999999){\color[rgb]{0,0,0}\makebox(0,0)[lt]{\lineheight{0}\smash{\begin{tabular}[t]{l}$\Omega_\mathrm{roi}^\mathrm{m} \equiv \Omega_\mathrm{mve}^\mathrm{m}$\end{tabular}}}}%
    \put(0.25788403,0.48803991){\color[rgb]{0,0,0}\makebox(0,0)[lt]{\lineheight{0}\smash{\begin{tabular}[t]{l}$f_{\mathrm{m},1}$\end{tabular}}}}%
    \put(0,0){\includegraphics[width=\unitlength,page=4]{sketch_SI_ink.pdf}}%
    \put(0.51195532,0.34601123){\color[rgb]{0,0,0}\makebox(0,0)[lt]{\lineheight{0}\smash{\begin{tabular}[t]{l}$f_{\mathrm{m},i}$\end{tabular}}}}%
    \put(0.7660293,0.34605944){\color[rgb]{0,0,0}\makebox(0,0)[lt]{\lineheight{0}\smash{\begin{tabular}[t]{l}{\setlength{\fboxsep}{1pt}\colorbox{white}{$f_{\mathrm{m},{n_\mathrm{g}}}$}}\end{tabular}}}}%
  \end{picture}%
\endgroup%